\documentclass[reprint,prl,twocolumn,twoside,aps,superscriptaddress]{revtex4-2}

\usepackage{pdfpages} 
\usepackage{pgffor} 

\makeatletter
\AtBeginDocument{\let\LS@rot\@undefined}
\makeatother

\def\supplementfilename{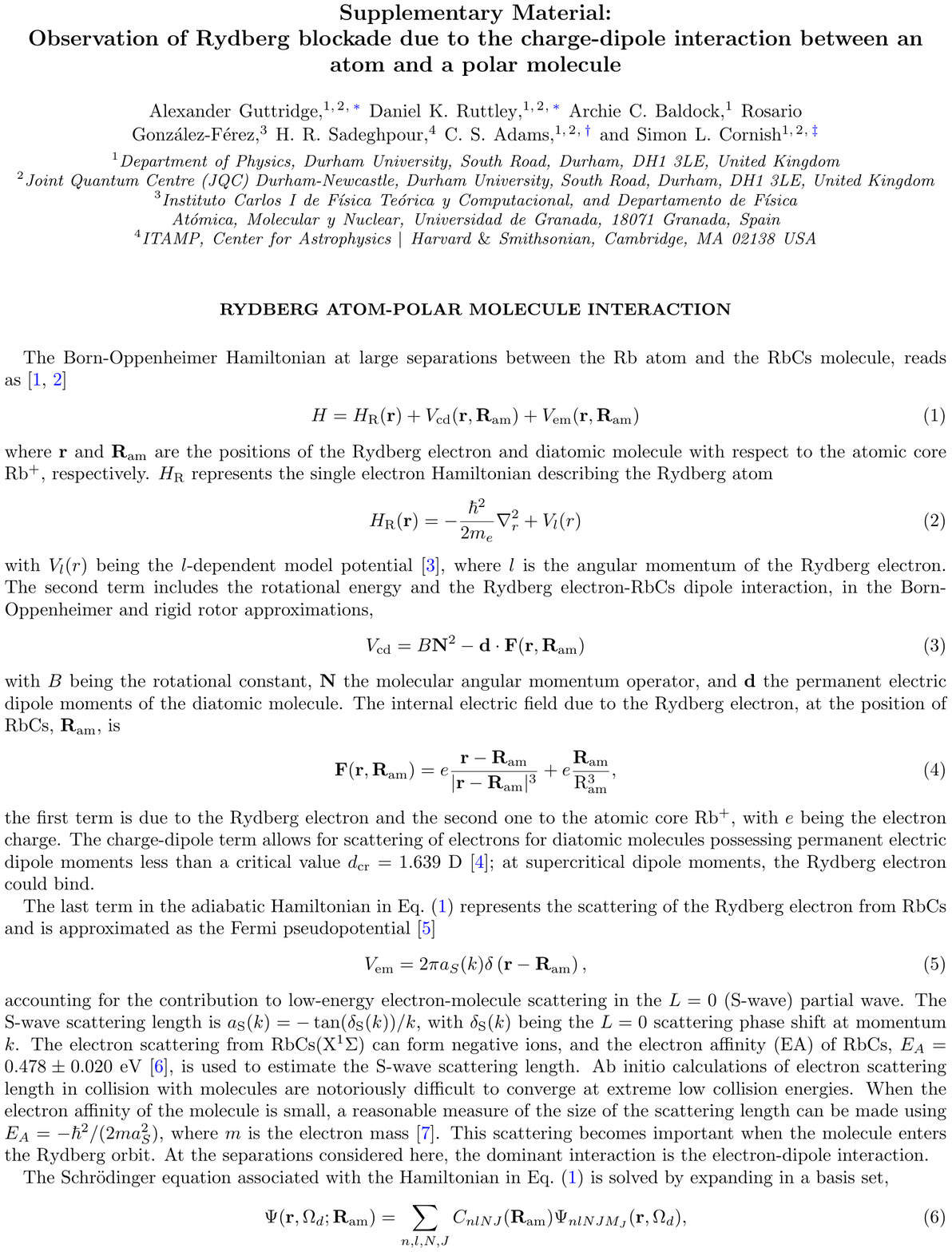}

\pdfximage{\supplementfilename}
\def\numbersupplementpages{\the\pdflastximagepages}

\newif\ifarXiv
\arXivtrue

\usepackage{amsmath,amssymb,amsfonts}
\usepackage{graphicx}
\usepackage{siunitx}
\usepackage{natbib}
\usepackage{braket}
\usepackage{multirow}
\usepackage[normalem]{ulem}
\usepackage{xspace}
\usepackage{url}
\usepackage[breaklinks]{hyperref}
\usepackage{braket}
\usepackage{lipsum}

\hypersetup{
    unicode=true,
    colorlinks=true,
    linkcolor=blue,
    citecolor=blue,
    urlcolor=blue
  }
\usepackage{breakurl}

\newcommand{\feshbach}{$\ket{F}$\xspace}
\newcommand{\ground}{$\ket{G}$\xspace}
\newcommand{\grounddetail}{$\ket{G} = \ket{\mathrm{X}^{1}\mathrm{\Sigma}^{+},v=0,N=0}$\xspace}
\newcommand{\intermediatedetail}{$\ket{E} = \ket{^{3}\Pi_{1}, v'=29, J'=1}$\xspace}
\newcommand{\intermediate}{$\ket{E}$\xspace}
\newcommand{\atomground}{$\ket{g}$\xspace}
\newcommand{\atomgrounddetail}{$\ket{g} = \ket{5\mathrm{s}_{1/2},f=1,m_f=1}$\xspace}
\newcommand{\rydberg}{$\ket{r}$\xspace}
\newcommand{\rydbergdetail}{$\ket{r} = \ket{52 \mathrm{s}}$\xspace}
\newcommand{\atompairgrounddetail}{${\ket{f\mathrm=1,m_f=1}}_\mathrm{Rb}+{\ket{f\mathrm=3,m_f=3}\xspace}_\mathrm{Cs}$}
\newcommand{\sep}{$\mathrm{R}_{\mathrm{am}}$\xspace}

\usepackage{xr}
\begin{document}
\title{Observation of Rydberg blockade due to the charge-dipole interaction between an atom and a polar molecule}

\newcommand{\physics}{Department of Physics, Durham University, South Road, Durham, DH1 3LE, United Kingdom}
\newcommand{\jqc}{Joint Quantum Centre Durham-Newcastle, Durham University, South Road, Durham, DH1 3LE, United Kingdom}
\newcommand{\grenada}{Instituto Carlos I de F\'{i}sica Te\'{o}rica y Computacional, and Departamento de F\'{i}sica At\'{o}mica, Molecular y Nuclear, Universidad de Granada, 18071 Granada, Spain}
\newcommand{\itamp}{ITAMP, Center for Astrophysics~$|$~Harvard $\&$ Smithsonian, Cambridge, MA 02138 USA} 

\author{Alexander Guttridge}
\thanks{A. G. and D. K. R. contributed equally to this work.}
\affiliation{\physics}
\affiliation{\jqc}
\author{Daniel K. Ruttley}
\thanks{A. G. and D. K. R. contributed equally to this work.}
\affiliation{\physics}
\affiliation{\jqc}
\author{Archie C. Baldock}
\affiliation{\physics}
\author{Rosario~Gonz\'{a}lez-F\'{e}rez}
\affiliation{\grenada}
\author{H. R. Sadeghpour}
\affiliation{\itamp}
\author{C. S. Adams}
\email{c.s.adams@durham.ac.uk}
\affiliation{\physics}
\affiliation{\jqc}
\author{Simon L. Cornish}
\email{s.l.cornish@durham.ac.uk}
\affiliation{\physics}
\affiliation{\jqc}

\begin{abstract}
We demonstrate Rydberg blockade due to the charge-dipole interaction between a single Rb atom and a single RbCs molecule confined in optical tweezers.
The molecule is formed by magnetoassociation of a Rb+Cs atom pair and subsequently transferred to the rovibrational ground state 
with an efficiency of 91(1)\%. 
Species-specific tweezers are used to control the separation between the atom and molecule. The charge-dipole interaction causes blockade of the transition to the Rb(52s) Rydberg state, when the atom-molecule separation is set to $310(40)$~nm. The observed excitation dynamics are in good agreement with simulations using calculated interaction potentials. Our results open up the prospect of a hybrid platform where quantum information is transferred between individually trapped molecules using Rydberg atoms.
\end{abstract}
\date{\today}

\maketitle
Ultracold dipolar systems, such as Rydberg atoms 
and polar molecules, are 
promising platforms for quantum simulation and computation \cite{Barnett2006,Gorshkov2011,Baranov2012,Lechner2013,Wall2015,Bohn2017,Yao2018,Carr2009,Saffman2016,Adams2020,Browaeys2020,Morgado2021,Wu2021,Kaufman2021}. Rydberg atoms exhibit strong, long-range interactions that can be exploited to engineer quantum entanglement and multi-qubit gates \cite{Jaksch2000,Lukin2001,Isenhower2010,Wilk2010,Saffman2010,Levine2019,Morgado2021}. 
This approach exploits the Rydberg blockade mechanism, where strong van der Waals interactions between neighbouring Rydberg atoms prevent simultaneous excitation of multiple atoms within a certain radius. Ultracold polar molecules also exhibit long-range interactions and possess a rich manifold of long-lived rotational states which can be coupled using microwave fields \cite{Park2017,Gregory2021,Burchesky2021} to realise high-fidelity quantum operations \cite{DeMille2002,Yelin2006,Ni2018,Hughes2020,Sawant2020,Albert2020}. Recent advances in optical tweezer arrays of Rydberg atoms \cite{Adams2020,Browaeys2020,Morgado2021,Wu2021} and ultracold molecules \cite{Liu2018,Anderegg2019,He2020,Cairncross2021,Zhang2022a,Ruttley2023} provide the foundation to develop hybrid atom-molecule systems.

A hybrid system composed of polar molecules and Rydberg atoms trapped in optical tweezer arrays offers a way to combine the advantages of both platforms. For example, quantum information can be encoded in the 
internal states of the molecule, and gates can be performed utilising the strong 
interactions of Rydberg atoms \cite{Kuznetsova2011,Kuznetsova2016,Wang2022,Zhang2022}.
This combines the fast high-fidelity interactions and readout possible with Rydberg atoms \cite{Levine2019,Graham2019} with the long coherence times and 
lifetimes of polar molecules \cite{Park2017,Gregory2021,Burchesky2021}. In addition, this hybrid system offers new capabilities, such as nondestructive readout of the molecular state~\cite{Jarisch2018,Gawlas2020,Patsch2022},  cooling of molecules using Rydberg atoms~\cite{Zhao2012,Huber2012}, and photoassociation of giant polyatomic Rydberg molecules \cite{Rittenhouse2010,Rittenhouse2011,GonzalezFerez2020}. 

Realising controlled interactions between molecules and Rydberg atoms remains an outstanding challenge. These interactions extend beyond the van der Waals and dipole-dipole interactions which have been widely used in single-species Rydberg systems \cite{Browaeys2020} and the dipole-dipole interactions recently observed between polar molecules \cite{Yan2013,Christakis2023,Holland2022,Bao2022}. The long-range interaction of the Rydberg electron with the permanent dipole, {\bf d}, of the polar molecule takes, in first order, the form of a charge-dipole interaction \cite{Rittenhouse2010,Shaffer2018}. 
The interaction arises when the internal field due to the Rydberg electron and atomic core polarizes the molecular dipole, $V_{\mathrm{cd}}({\bf r} , {\bf R}_{\mathrm{am}}) = {B{\bf N}^2 - {\bf d} \cdot {\bf F ({\bf r, {\bf R}_{\mathrm{am}})}}}$. Here {\bf r} is the electron position, ${\bf R}_{\mathrm{am}}$ is the dipole position with respect to the atomic core, and ${\bf N}$ and $B$ are the quantum operators for molecular rotation and the associated rotational constant respectively. The internal electric field is
${\bf F} =\frac{e ({\bf r} - {\bf R}_{\mathrm{am}})}{|{\bf r} - {\bf R}_{\mathrm{am}}|^3}+\frac{e\mathbf{R}_{\mathrm{am}}}{\mathrm{R}_{\mathrm{am}}^3}$, 
leading 
to an anisotropic $1/\mathrm{R}_{\mathrm{am}}^2$ interaction. For micron-scale separations, achievable in optical lattices and optical tweezers, these interactions are predicted to be strong enough to preclude the excitation of an atom to a Rydberg state in the presence of a molecule 
\cite{Kuznetsova2011,Kuznetsova2016}. 

In this Letter, we demonstrate Rydberg blockade due to the charge-dipole interaction in a hybrid platform composed of a single $^{87}$Rb atom and a single $^{87}$Rb$^{133}$Cs molecule confined in separate optical tweezers. The molecule is prepared in the rovibrational ground state using a combination of magnetoassociation and coherent optical transfer. 
We use species-specific optical tweezers to control the atom-molecule separation down to $\sim300$\,nm without significant collisional loss. At this separation, we find that excitation of the Rb atom to a Rydberg state is suppressed.

\begin{figure}[t]
\includegraphics[width=\hsize]{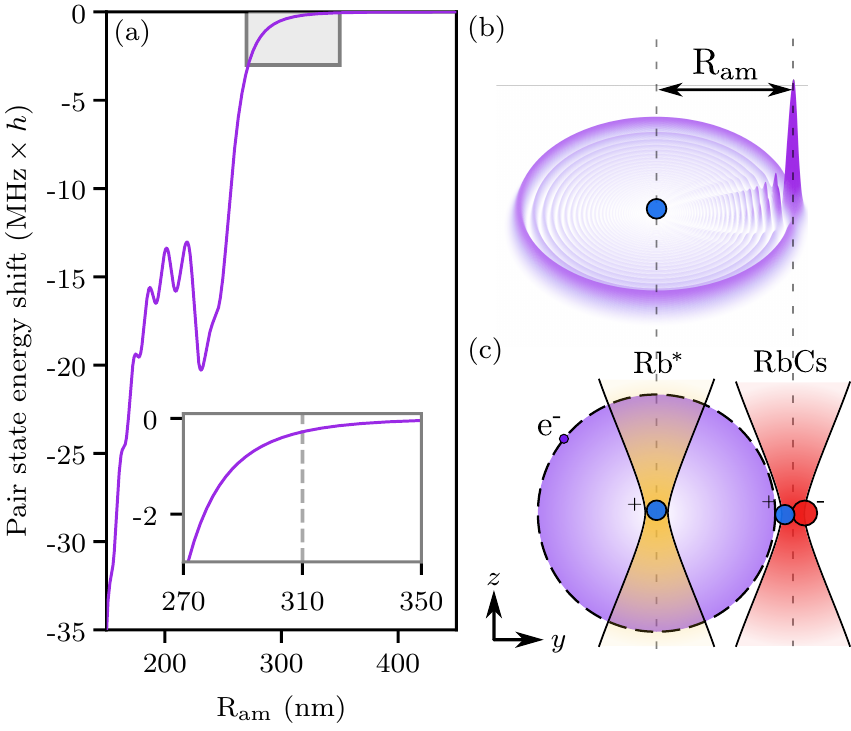}
\caption{(a) Pair state energy shift 
as a function of the separation \sep between a Rb 
atom in state \rydbergdetail and a RbCs molecule in state \grounddetail. {The shaded region (shown inset) highlights the region relevant to the blockade measurements; the dashed line shows our best estimate of \sep and the range reflects the associated uncertainty.} 
(b) A surface plot of the radial electron density of the Rydberg electron for {\sep~$= 230$~nm}. 
(c) Schematic of the experiment showing the atom and molecule trapped in species-specific optical tweezers separated along the $y$-axis.
\label{fig:theory}}
\end{figure}

The adiabatic Hamiltonian for the hybrid Rydberg atom-molecule system at large separations contains the charge induced-dipole interaction~\cite{Rittenhouse2010},
 and the S-wave scattering of the slow electron from the molecule~\cite{Shaffer2018,AMOHandbook2023,supplement}. 
 The results described here hold for $d<d_{\textrm{cr}} = 1.639$~Debye~\cite{fermi47}, 
 the Fermi-Teller critical dipole,  to ensure that the electron only scatters from the molecule.
The trapping potentials due to the optical tweezers and the effect of the magnetic field are neglected in the theoretical description. 

In Fig.~\ref{fig:theory}(a), we show the resulting energy shift as a function of the separation \sep between a Rb atom in Rydberg state \rydbergdetail
and a RbCs molecule in the rovibrational ground state \grounddetail 
with  
${d}=1.225$ Debye~\cite{Molony2014} and 
$B=0.490$~GHz~\cite{Gregory2016}. Here, $v$ and $N$ are the vibrational and rotational quantum numbers, respectively.
At \sep~$\sim300$\,nm, we see the onset of a large shift arising from the charge-dipole interaction. The modulations in the energy arise from the oscillatory nature of the Rydberg electron wavefunction. For our choice of states the interaction is non-resonant and van der Waals interactions are $\simeq$~1\,kHz at these distances.
Figure~\ref{fig:theory}(b) shows the Rydberg electron density for our system with {\sep~$= 230$~nm}, highlighting the perturbation due to the polar molecule. 
The outermost minimum of the Rydberg electron wavefunction sets the range of interactions; this occurs at 220~nm for the state \rydberg. 

The experimental geometry is shown in Fig.~\ref{fig:theory}(c).
The atom and molecule are prepared in species-specific tweezers. Both particles predominantly occupy the motional ground state of their respective traps. 
The tweezer separation, $\mathrm{R}_{\rm t}$, is set by controlling the relative tweezer alignment in all three spatial dimensions. The atom-molecule separation, $\mathrm{R}_{\mathrm{am}}$, is determined from the difference in the resulting potential minima. When the tweezers overlap, the atom-molecule separation is reduced compared to the tweezer separation ($\mathrm{R}_{\mathrm{am}} <\mathrm{R}_{\rm t}$) due to the effect of each potential on the other species. 
For each measurement, we repeat an experimental sequence many times. Fluctuations in the relative alignment of the tweezers occur from shot to shot with an estimated standard deviation of 50\,nm in each coordinate.
Atomic fluorescence images are taken at the start and end of each sequence to determine the occupancy of each tweezer; molecules are detected by reversing the association procedure and imaging the resulting atom pair in separate tweezers.
{We apply various post-selection criteria on the tweezer occupancies to obtain values and their associated confidence intervals from typically 200 -- 1000 runs for different experimental scenarios \cite{supplement}.}

\begin{figure*}[t]
\includegraphics[width=\hsize]{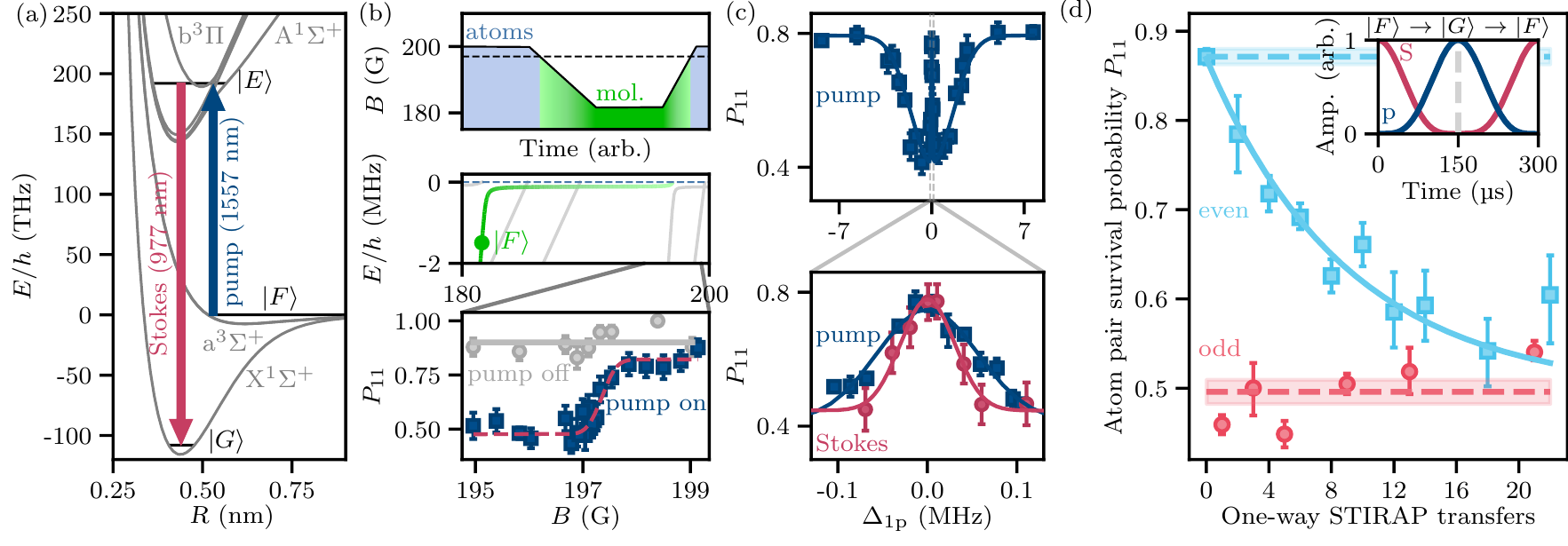}
\caption{Formation of 
ground state RbCs molecules in optical tweezers. (a) Electronic potential curves for RbCs showing the pump and Stokes 
transitions that couple states \feshbach, \intermediate and \ground. 
(b) Formation of weakly bound molecules by magnetoassociation of atom pairs using a Feshbach resonance at 197~G. 
The top panel shows the magnetic field ramps used to form molecules and then navigate the near-threshold bound states shown in the middle panel. 
STIRAP is performed at 181.6~G when the molecule occupies the state \feshbach (indicated point). 
The lower panel shows the pump-induced loss of weakly bound molecules. The atom-pair survival probability $P_{11}$ is measured after reversing the field ramp to dissociate any remaining molecules.
{(c) Atom-pair survival probability for a round trip \feshbach$\rightarrow$\xspace\ground$\rightarrow$\xspace\feshbach as a function of the one-photon detuning $\Delta_\mathrm{1p}$ of either the pump (from the transition \feshbach$\rightarrow$\xspace\intermediate) or the Stokes (from the transition \intermediate$\rightarrow$\xspace\ground) when the other laser is on resonance.}
(d) Repeated STIRAP transfers between states \feshbach and \ground with a one-way efficiency of 91(1)\%. 
The dashed lines show the experimental contrast; the shaded regions are the uncertainties.
The inset shows the pulse profiles for a round-trip transfer.
\label{fig:molecule_formation}}
\end{figure*}

Our experiments begin by loading single $^{87}$Rb and $^{133}$Cs atoms into species-specific optical tweezers \cite{Brooks21}. 
After determining the trap occupations and performing rearrangement, the atoms are further cooled using Raman sideband cooling \cite{Kaufman2012,Thompson2013,Spence22} and transferred to the hyperfine states \atompairgrounddetail. To produce a molecule, we must prepare a Rb+Cs atom pair in the ground state of relative motion in a single tweezer. 
We achieve this by merging a 
817~nm tweezer containing a Rb atom into a 1065~nm tweezer containing a Cs atom.
{This protocol prepares a Rb+Cs atom pair in the ground state of relative motion in 56(5)\% of runs \cite{Spence22}.}

The electronic potential energy curves for RbCs are shown in Fig.~\ref{fig:molecule_formation}(a). Weakly bound RbCs molecules in state \feshbach are formed using magnetoassociation on an interspecies Feshbach resonance at 197~G \cite{Pilch2009,Takekoshi2012,Koeppinger2014,Ruttley2023}.
The magnetic field ramps used to associate and later dissociate the atom pair are shown in the upper panel of Fig.~\ref{fig:molecule_formation}(b); the central panel shows the energy levels that these ramps navigate to access the state \feshbach at 181.6~G. 
The formation of weakly bound RbCs molecules is detected using pump-induced loss 
\cite{supplement,Gregory2015,Molony2016_2,Takekoshi2014} which precludes atom-pair recovery when the association and merging steps are reversed (Fig.~\ref{fig:molecule_formation}(b) lower panel).

We transfer the weakly bound molecule to the rovibrational ground state \ground 
using two-photon stimulated Raman adiabatic passage (STIRAP) \cite{Bergmann1998, Vitanov2017}, as previously demonstrated for bulk gases of RbCs molecules \cite{Takekoshi2014,Molony2014,Gregory2015,Molony2016}.
In Fig.~\ref{fig:molecule_formation}(c) we show the probability of recovering the atom pair after a round trip \feshbach$\rightarrow$\xspace\ground$\rightarrow$\xspace\feshbach as a function of the single-photon detuning of either the pump or Stokes lasers, with the other laser held on single-photon resonance. 
{When the Stokes laser is not resonant, the pump laser causes loss to other molecular states via the intermediate molecular state \intermediatedetail.}
When the pump laser is far from resonant, the molecules remain in state \feshbach throughout the transfer sequence. 

In Fig.~\ref{fig:molecule_formation}(d), we characterise the STIRAP  efficiency using repeated transfers back and forth between states \feshbach and \ground. 
An odd number of successful one-way transfers results in the molecule occupying state \ground, whereas an even number 
returns it to state \feshbach.
Only molecules that occupy state \feshbach at the end of the sequence are dissociated back into atom pairs for detection.
The offset of the odd points indicates the combined efficiency of the cooling, merging, and magnetoassociation stages; 
in 50(1)\% of runs we do not form a molecule, and thus reimage the atom pair independent of the STIRAP pulses. 
The maximum contrast between the odd and even points is limited primarily by the 35(5)\,ms  lifetime of molecules in state \feshbach in the trap and the need to allow the magnetic field to stabilise before STIRAP \cite{supplement}.
 We measure a one-way transfer efficiency of 91(1)\%, consistent with the best reported efficiencies for RbCs in bulk gases \cite{Takekoshi2014,Molony2016}. 

\begin{figure}[t]
\includegraphics[width=\hsize]{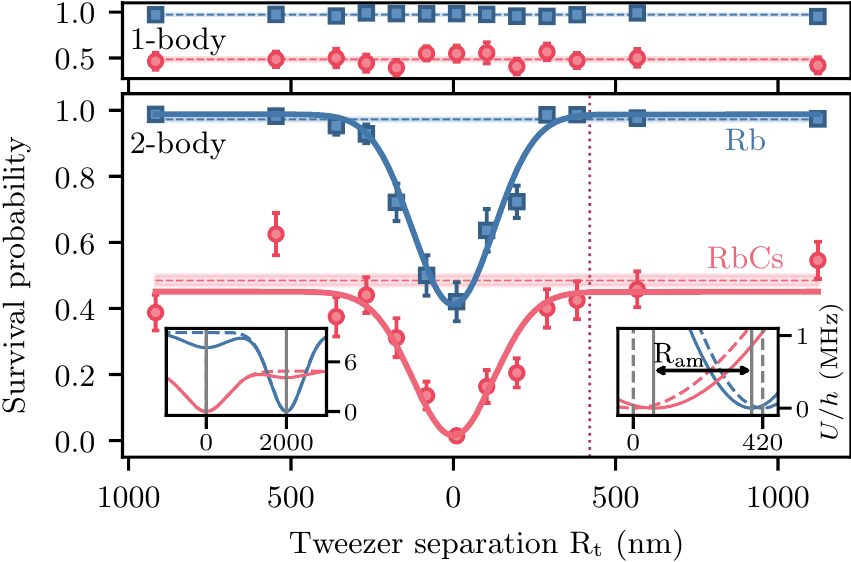}
\caption{Collisions between ground state RbCs molecules and Rb atoms held in separate species-specific optical tweezers. Particle survival probabilities are plotted as a function of the tweezer separation, $\mathrm{R}_{\mathrm{t}}$. For the molecule we report the atom-pair survival probability, post selected on cases where a weakly bound molecule was formed \cite{supplement}.
Upper panel: experimental runs where either a single Rb atom (blue squares) or a single RbCs molecule (red circles) is present.
Lower panel: runs where both the atom and the molecule are present. The dashed lines (and shaded regions) correspond to the mean values (and errors) from the 1-body cases.
The purple dotted line at $\mathrm{R}_{\mathrm{t}}=420$~nm shows the tweezer separation for the measurement in Fig.\ \ref{fig:blockade}(a).
Insets: The potential energy of the atom (blue) and molecule (red) resulting from their own tweezer (dashed lines) and both tweezers (solid lines) for $\mathrm{R}_{\mathrm{t}}=2000$~nm (left) and $\mathrm{R}_{\mathrm{t}}=420$~nm (right).
\label{fig:separation}}
\end{figure}

{To observe blockade, the charge-dipole interaction between the Rydberg atom and the molecule must be greater than the power-broadened transition linewidth.
For our system, this is set by the Rabi frequency of $500(3)$~kHz; 
blockade therefore requires interactions shifts $\gtrsim 1$~MHz. Our calculations in Fig.~\ref{fig:theory} predict the atom-molecule distance must be below a blockade radius $\sim 300$~nm to observe this effect,} a distance smaller than the beam radii of the individual tweezers ($\sim 1$~$\mu$m). We cannot acheive sub-micron separations by loading both species into the same tweezer, as the expected lifetime due to collisional loss is $< 1$~ms \cite{Gregory2021a}. 
Instead we utilise species-specific tweezers at wavelengths of 1065~nm for the molecule and 817~nm for the atom. For the Rb atom, the ratio of polarisabilities for these wavelengths is $ \alpha_{817}^\mathrm{Rb}/\alpha_{1065}^\mathrm{Rb} \sim 6.3$ \cite{UDportal}, so that it is confined predominantly in the 817~nm tweezer (the ``atom tweezer''). Conversely, for the RbCs molecule $ \alpha_{1065}^\mathrm{RbCs}/\alpha_{817}^\mathrm{RbCs} \sim 4.5$ \cite{Vexiau2017} so that it is confined predominantly in the 1065~nm tweezer (the ``molecule tweezer''). Typical trap potentials are illustrated in the insets of Fig.~\ref{fig:separation}.

We investigate loss due to collisions between a Rb atom 
in state \atomgrounddetail and a RbCs molecule by sweeping the position of the atom tweezer to a variable distance $\mathrm{R}_{\mathrm{t}}$ from the molecule tweezer. The particles are held at this separation 
for 9.5~ms before the sweep 
is reversed and the particle survival probabilities are measured. The results are presented in Fig.~\ref{fig:separation}. For the molecule, we report the atom-pair survival probability, post-selected on cases where a weakly bound molecule was formed \cite{supplement}. 
The upper panel shows the one-body survival probabilities from 
runs where \emph{either} the Rb atom \emph{or} the RbCs molecule is present. 
The atomic survival probability is 97.2(4)\%.
For the molecule signal, we observe a survival probability of 48(2)\%, primarily caused by loss prior to STIRAP due to the short lifetime of the state \feshbach in the trap. 
By compensating for the return STIRAP efficiency, we predict that a molecule in \ground is present in 53(3)\% of runs in which a weakly bound molecule is created.
The lower panel in  Fig.~\ref{fig:separation} shows the two-body survival probabilities for runs in which \emph{both} an atom and a weakly bound molecule are initially prepared.
When the tweezers are brought together, the wavefunctions of the particles begin to overlap and collisions cause loss of both the molecule and atom. 
We observe a reduction in the atom survival probability by 58(6)\%, commensurate with the probability a molecule in state \ground is present.
From a Gaussian fit 
we find the loss falls to  $1/e^2$ of its maximum value at $\mathrm{R}_{\mathrm{t}}=250(20)$~nm. 

\begin{figure}[t]
\includegraphics[width=\hsize]{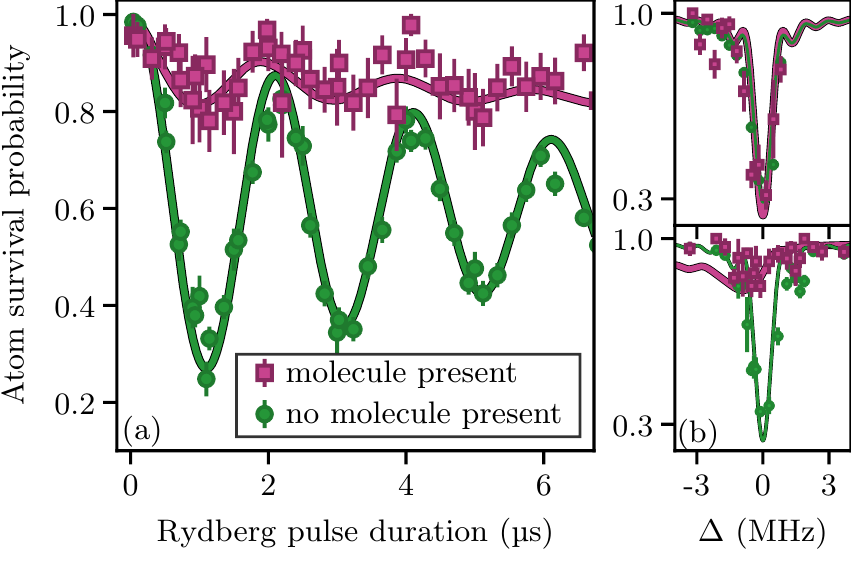}
\caption{(a) Survival probability of the Rb atom as a function of the Rydberg pulse duration for \sep~$= 310(40)$~nm. 
Atoms excited to \rydberg are ejected from the trap and lost. 
Events are post-selected on the detection of a molecule in \ground (purple squares) or unsuccessful formation of a molecule (green circles). The solid lines show the results of simulations using the Lindblad master equation \cite{supplement} using our estimated atom-molecule separation.
(b) Rb atom survival probability as a function of the two-photon detuning, $\Delta$, using a 1~$\mu$s pulse for \sep~$= 700(40)$~nm (upper panel) and \sep~$= 310(40)$~nm (lower panel). 
The detuning is defined relative to the transition centre in the absence of a molecule. Symbols are as in (a) and solid lines show the results of simulations using the estimated atom-molecule separations.
\label{fig:blockade}}
\end{figure} 

To demonstrate blockade, we repeat the routine used to measure collisional loss, but use a shorter hold time of 3~ms when the tweezers are close together.  
{Two-photon excitation of the Rb atom $\ket{g}\rightarrow\ket{6 \mathrm{p}_{3/2}}\rightarrow\ket{r}$} is performed during the hold time with the trapping light still present \cite{supplement}. Atoms excited to state \rydberg are anti-trapped and ejected from the tweezers, mapping Rydberg excitation onto atom loss.
To suppress collisional loss we hold the tweezers at a separation $\mathrm{R}_{\rm t} = 420(40)$~nm, shown by the dotted line in Fig.~\ref{fig:separation}. 
Here the error represents the systematic uncertainty from the alignment calibrations. 
As shown inset in Fig.~\ref{fig:separation}, this equates to an atom-molecule separation of \sep~$= 310(40)$~nm \cite{supplement}.

In Fig.~\ref{fig:blockade}, we demonstrate the blockade of the Rydberg transition of the Rb atom when a RbCs molecule in state \ground is present. Figure~\ref{fig:blockade}(a) shows the survival probability of the Rb atom as the Rydberg pulse duration is varied.
For experimental runs where the molecule tweezer is empty (green circles), {we observe Rabi oscillations between states \atomground and \rydberg~ with a fitted frequency of $500(3)$~kHz.}
{The observed damping is caused by laser frequency noise.}
In contrast, for runs where a molecule in state \ground is present (purple squares), we observe a suppression of the excitation to state \rydberg. Here, the presence of the molecule shifts the energy of state \rydberg through the charge-dipole interaction and thus blockades excitation during the Rydberg pulse. The frequency of the residual Rabi oscillations is almost identical to that for the unblockaded case. This is due to the sharp onset of the interaction shown in Fig.~\ref{fig:theory}(a) combined with shot-to-shot variations in the relative alignment of the tweezers. For runs with the largest separations, the energy shift is smaller than the {Rabi frequency} of the Rydberg transition leading to a signal at the unshifted Rabi frequency.

To simulate the expected excitation dynamics, we solve the Lindblad master equation \cite{supplement}. We use the pair-state energy shifts shown in Fig.~\ref{fig:theory}(a) to include a distance-dependent energy shift. We account for the fact that the atom and molecule are {predominantly} prepared in the motional ground state of their respective tweezers by averaging the interaction over the ground-state wavefunction of relative motion. We also include experimental imperfections such as dephasing from laser frequency noise and shot-to-shot fluctuations in the relative alignment of the tweezers. 
Using our best estimates of the separation, we find good agreement between the results of the simulation and the experiment, as shown by the solid lines in Fig.~\ref{fig:blockade}(a).

Figure~\ref{fig:blockade}(b) shows the effect of changing the atom-molecule separation on the Rydberg blockade. In this experiment, we fix the pulse duration to approximate a $\pi$-pulse and scan the two-photon detuning of the light driving the Rydberg transition.
For \sep~$= 700(40)$~nm, shown in the upper panel, 
the charge-dipole interaction is negligible. Here, the dominant interaction is van der Waals leading to a shift of $\sim 0.1$~kHz \cite{Olaya2020}.  Consequently, the presence of a molecule does not affect the the Rydberg excitation.
However, for \sep~$= 310(40)$~nm, shown in the lower panel, the presence of a molecule leads to an observed shift of the Rydberg transition to lower energy, as expected. The transition is significantly broadened due to the sensitivity of the charge-dipole interaction to the atom-molecule separation. The broadening causes a concomitant reduction in the signal amplitude. Both these effects are reproduced by simulations using the same parameters as in Fig.~\ref{fig:blockade}(a) {with the exception of the appearance of a shoulder in the lower panel of Fig.~\ref{fig:blockade}(b) which is highly sensitive to fluctuations in \sep.}

In conclusion, we have demonstrated blockade of the transition to the Rb(52s) Rydberg state due to the charge-dipole interaction with a RbCs molecule in the rovibrational ground state. This represents the first observation of a charge-dipole induced shift in an ultracold 
setting and opens up many new research directions. 
The blockade we have observed provides a mechanism for 
non-destructive state readout of the molecule \cite{Kuznetsova2011,Kuznetsova2016}.
A single Rydberg atom can also mediate effective spin-spin interactions between a pair of molecular dipoles \cite{Kuznetsova2018}.
For molecules prepared in the $N=2$ rotational state, our calculations for the Rb(52s) Rydberg state predict that resolvable, deeply bound states exist for separations of  $\sim 220$~nm. This offers the possibility to photoassociate giant
polyatomic Rydberg molecules \cite{Rittenhouse2010,Rittenhouse2011,gonzalez15,GonzalezFerez2020}. By selecting Rydberg and molecular states which interact via resonant dipole-dipole interactions, the Rydberg blockade radius can be increased to several microns, enabling high-fidelity entangling gates between molecules mediated by strong interactions with neighbouring Rydberg atoms \cite{Zhang2022,Wang2022}. This presents the tantalising prospect of a hybrid platform where quantum information is transferred between individually trapped molecules using Rydberg atoms.


\begin{acknowledgments}

We thank S. Spence for earlier experimental work, A. L. Tao for assistance in the setup of the STIRAP laser system, X. Yang for informative simulations of the atom and molecule system, and H. J. Williams and S. A. Gardiner for helpful discussions. 
We acknowledge support from the UK Engineering and Physical Sciences Research Council (EPSRC) Grants EP/P01058X/1, EP/V047302/1, and EP/W00299X/1, UK Research and Innovation (UKRI) Frontier Research Grant EP/X023354/1, the Royal Society, and Durham University. 
R.G.F. gratefully acknowledges financial support by the Spanish projects PID2020-113390GB-I00 (MICIN), PY20-00082 (Junta de Andalucía) and A-FQM-52-UGR20 (ERDF-University of Granada), and the Andalusian Research Group FQM-207. H.R.S. acknowledges support from the NSF through a grant for ITAMP at
Harvard University.
The data presented in this paper are available from \url{http://doi.org/10.15128/r2fj236215f}.

\end{acknowledgments}
\
\appendix


\begin{thebibliography}{88}%
\makeatletter
\providecommand \@ifxundefined [1]{%
 \@ifx{#1\undefined}
}%
\providecommand \@ifnum [1]{%
 \ifnum #1\expandafter \@firstoftwo
 \else \expandafter \@secondoftwo
 \fi
}%
\providecommand \@ifx [1]{%
 \ifx #1\expandafter \@firstoftwo
 \else \expandafter \@secondoftwo
 \fi
}%
\providecommand \natexlab [1]{#1}%
\providecommand \enquote  [1]{``#1''}%
\providecommand \bibnamefont  [1]{#1}%
\providecommand \bibfnamefont [1]{#1}%
\providecommand \citenamefont [1]{#1}%
\providecommand \href@noop [0]{\@secondoftwo}%
\providecommand \href [0]{\begingroup \@sanitize@url \@href}%
\providecommand \@href[1]{\@@startlink{#1}\@@href}%
\providecommand \@@href[1]{\endgroup#1\@@endlink}%
\providecommand \@sanitize@url [0]{\catcode `\\12\catcode `\$12\catcode
  `\&12\catcode `\#12\catcode `\^12\catcode `\_12\catcode `\%12\relax}%
\providecommand \@@startlink[1]{}%
\providecommand \@@endlink[0]{}%
\providecommand \url  [0]{\begingroup\@sanitize@url \@url }%
\providecommand \@url [1]{\endgroup\@href {#1}{\urlprefix }}%
\providecommand \urlprefix  [0]{URL }%
\providecommand \Eprint [0]{\href }%
\providecommand \doibase [0]{https://doi.org/}%
\providecommand \selectlanguage [0]{\@gobble}%
\providecommand \bibinfo  [0]{\@secondoftwo}%
\providecommand \bibfield  [0]{\@secondoftwo}%
\providecommand \translation [1]{[#1]}%
\providecommand \BibitemOpen [0]{}%
\providecommand \bibitemStop [0]{}%
\providecommand \bibitemNoStop [0]{.\EOS\space}%
\providecommand \EOS [0]{\spacefactor3000\relax}%
\providecommand \BibitemShut  [1]{\csname bibitem#1\endcsname}%
\let\auto@bib@innerbib\@empty
\bibitem [{\citenamefont {Barnett}\ \emph {et~al.}(2006)\citenamefont
  {Barnett}, \citenamefont {Petrov}, \citenamefont {Lukin},\ and\ \citenamefont
  {Demler}}]{Barnett2006}%
  \BibitemOpen
  \bibfield  {author} {\bibinfo {author} {\bibfnamefont {R.}~\bibnamefont
  {Barnett}}, \bibinfo {author} {\bibfnamefont {D.}~\bibnamefont {Petrov}},
  \bibinfo {author} {\bibfnamefont {M.}~\bibnamefont {Lukin}},\ and\ \bibinfo
  {author} {\bibfnamefont {E.}~\bibnamefont {Demler}},\ }\bibfield  {title}
  {\bibinfo {title} {Quantum magnetism with multicomponent dipolar molecules in
  an optical lattice},\ }\href {https://doi.org/10.1103/PhysRevLett.96.190401}
  {\bibfield  {journal} {\bibinfo  {journal} {Phys. Rev. Lett.}\ }\textbf
  {\bibinfo {volume} {96}},\ \bibinfo {pages} {190401} (\bibinfo {year}
  {2006})}\BibitemShut {NoStop}%
\bibitem [{\citenamefont {Gorshkov}\ \emph {et~al.}(2011)\citenamefont
  {Gorshkov}, \citenamefont {Manmana}, \citenamefont {Chen}, \citenamefont
  {Ye}, \citenamefont {Demler}, \citenamefont {Lukin},\ and\ \citenamefont
  {Rey}}]{Gorshkov2011}%
  \BibitemOpen
  \bibfield  {author} {\bibinfo {author} {\bibfnamefont {A.~V.}\ \bibnamefont
  {Gorshkov}}, \bibinfo {author} {\bibfnamefont {S.~R.}\ \bibnamefont
  {Manmana}}, \bibinfo {author} {\bibfnamefont {G.}~\bibnamefont {Chen}},
  \bibinfo {author} {\bibfnamefont {J.}~\bibnamefont {Ye}}, \bibinfo {author}
  {\bibfnamefont {E.}~\bibnamefont {Demler}}, \bibinfo {author} {\bibfnamefont
  {M.~D.}\ \bibnamefont {Lukin}},\ and\ \bibinfo {author} {\bibfnamefont
  {A.~M.}\ \bibnamefont {Rey}},\ }\bibfield  {title} {\bibinfo {title} {Tunable
  superfluidity and quantum magnetism with ultracold polar molecules},\ }\href
  {https://doi.org/10.1103/PhysRevLett.107.115301} {\bibfield  {journal}
  {\bibinfo  {journal} {Phys. Rev. Lett.}\ }\textbf {\bibinfo {volume} {107}},\
  \bibinfo {pages} {115301} (\bibinfo {year} {2011})}\BibitemShut {NoStop}%
\bibitem [{\citenamefont {Baranov}\ \emph {et~al.}(2012)\citenamefont
  {Baranov}, \citenamefont {Dalmonte}, \citenamefont {Pupillo},\ and\
  \citenamefont {Zoller}}]{Baranov2012}%
  \BibitemOpen
  \bibfield  {author} {\bibinfo {author} {\bibfnamefont {M.}~\bibnamefont
  {Baranov}}, \bibinfo {author} {\bibfnamefont {M.}~\bibnamefont {Dalmonte}},
  \bibinfo {author} {\bibfnamefont {G.}~\bibnamefont {Pupillo}},\ and\ \bibinfo
  {author} {\bibfnamefont {P.}~\bibnamefont {Zoller}},\ }\bibfield  {title}
  {\bibinfo {title} {Condensed matter theory of dipolar quantum gases},\ }\href
  {https://doi.org/10.1021/cr2003568} {\bibfield  {journal} {\bibinfo
  {journal} {Chem. Rev.}\ }\textbf {\bibinfo {volume} {112}},\ \bibinfo {pages}
  {5012} (\bibinfo {year} {2012})}\BibitemShut {NoStop}%
\bibitem [{\citenamefont {Lechner}\ and\ \citenamefont
  {Zoller}(2013)}]{Lechner2013}%
  \BibitemOpen
  \bibfield  {author} {\bibinfo {author} {\bibfnamefont {W.}~\bibnamefont
  {Lechner}}\ and\ \bibinfo {author} {\bibfnamefont {P.}~\bibnamefont
  {Zoller}},\ }\bibfield  {title} {\bibinfo {title} {From classical to quantum
  glasses with ultracold polar molecules},\ }\href
  {https://doi.org/10.1103/PhysRevLett.111.185306} {\bibfield  {journal}
  {\bibinfo  {journal} {Phys. Rev. Lett.}\ }\textbf {\bibinfo {volume} {111}},\
  \bibinfo {pages} {185306} (\bibinfo {year} {2013})}\BibitemShut {NoStop}%
\bibitem [{\citenamefont {Wall}\ \emph {et~al.}(2015)\citenamefont {Wall},
  \citenamefont {Hazzard},\ and\ \citenamefont {Rey}}]{Wall2015}%
  \BibitemOpen
  \bibfield  {author} {\bibinfo {author} {\bibfnamefont {M.~L.}\ \bibnamefont
  {Wall}}, \bibinfo {author} {\bibfnamefont {K.~R.~A.}\ \bibnamefont
  {Hazzard}},\ and\ \bibinfo {author} {\bibfnamefont {A.~M.}\ \bibnamefont
  {Rey}},\ }\bibinfo {title} {Quantum magnetism with ultracold molecules},\ in\
  \href {https://doi.org/10.1142/9789814678704_0001} {\emph {\bibinfo
  {booktitle} {From Atomic to Mesoscale}}}\ (\bibinfo  {publisher} {World
  Scientific},\ \bibinfo {year} {2015})\ Chap.~\bibinfo {chapter} {1}, pp.\
  \bibinfo {pages} {3--37}\BibitemShut {NoStop}%
\bibitem [{\citenamefont {Bohn}\ \emph {et~al.}(2017)\citenamefont {Bohn},
  \citenamefont {Rey},\ and\ \citenamefont {Ye}}]{Bohn2017}%
  \BibitemOpen
  \bibfield  {author} {\bibinfo {author} {\bibfnamefont {J.~L.}\ \bibnamefont
  {Bohn}}, \bibinfo {author} {\bibfnamefont {A.~M.}\ \bibnamefont {Rey}},\ and\
  \bibinfo {author} {\bibfnamefont {J.}~\bibnamefont {Ye}},\ }\bibfield
  {title} {\bibinfo {title} {Cold molecules: Progress in quantum engineering of
  chemistry and quantum matter},\ }\href
  {https://doi.org/10.1126/science.aam6299} {\bibfield  {journal} {\bibinfo
  {journal} {Science}\ }\textbf {\bibinfo {volume} {357}},\ \bibinfo {pages}
  {1002} (\bibinfo {year} {2017})}\BibitemShut {NoStop}%
\bibitem [{\citenamefont {Yao}\ \emph {et~al.}(2018)\citenamefont {Yao},
  \citenamefont {Zaletel}, \citenamefont {Stamper-Kurn},\ and\ \citenamefont
  {Vishwanath}}]{Yao2018}%
  \BibitemOpen
  \bibfield  {author} {\bibinfo {author} {\bibfnamefont {N.~Y.}\ \bibnamefont
  {Yao}}, \bibinfo {author} {\bibfnamefont {M.~P.}\ \bibnamefont {Zaletel}},
  \bibinfo {author} {\bibfnamefont {D.~M.}\ \bibnamefont {Stamper-Kurn}},\ and\
  \bibinfo {author} {\bibfnamefont {A.}~\bibnamefont {Vishwanath}},\ }\bibfield
   {title} {\bibinfo {title} {A quantum dipolar spin liquid},\ }\href
  {https://doi.org/10.1038/s41567-017-0030-7} {\bibfield  {journal} {\bibinfo
  {journal} {Nat. Phys.}\ }\textbf {\bibinfo {volume} {14}},\ \bibinfo {pages}
  {405} (\bibinfo {year} {2018})}\BibitemShut {NoStop}%
\bibitem [{\citenamefont {Carr}\ \emph {et~al.}(2009)\citenamefont {Carr},
  \citenamefont {DeMille}, \citenamefont {Krems},\ and\ \citenamefont
  {Ye}}]{Carr2009}%
  \BibitemOpen
  \bibfield  {author} {\bibinfo {author} {\bibfnamefont {L.~D.}\ \bibnamefont
  {Carr}}, \bibinfo {author} {\bibfnamefont {D.}~\bibnamefont {DeMille}},
  \bibinfo {author} {\bibfnamefont {R.~V.}\ \bibnamefont {Krems}},\ and\
  \bibinfo {author} {\bibfnamefont {J.}~\bibnamefont {Ye}},\ }\bibfield
  {title} {\bibinfo {title} {Cold and ultracold molecules: science, technology
  and applications},\ }\href {https://doi.org/10.1088/1367-2630/11/5/055049}
  {\bibfield  {journal} {\bibinfo  {journal} {New J. Phys.}\ }\textbf {\bibinfo
  {volume} {11}},\ \bibinfo {pages} {055049} (\bibinfo {year}
  {2009})}\BibitemShut {NoStop}%
\bibitem [{\citenamefont {Saffman}(2016)}]{Saffman2016}%
  \BibitemOpen
  \bibfield  {author} {\bibinfo {author} {\bibfnamefont {M.}~\bibnamefont
  {Saffman}},\ }\bibfield  {title} {\bibinfo {title} {Quantum computing with
  atomic qubits and {R}ydberg interactions: progress and challenges},\ }\href
  {https://doi.org/10.1088/0953-4075/49/20/202001} {\bibfield  {journal}
  {\bibinfo  {journal} {J. Phys. B}\ }\textbf {\bibinfo {volume} {49}},\
  \bibinfo {pages} {202001} (\bibinfo {year} {2016})}\BibitemShut {NoStop}%
\bibitem [{\citenamefont {Adams}\ \emph {et~al.}(2019)\citenamefont {Adams},
  \citenamefont {Pritchard},\ and\ \citenamefont {Shaffer}}]{Adams2020}%
  \BibitemOpen
  \bibfield  {author} {\bibinfo {author} {\bibfnamefont {C.~S.}\ \bibnamefont
  {Adams}}, \bibinfo {author} {\bibfnamefont {J.~D.}\ \bibnamefont
  {Pritchard}},\ and\ \bibinfo {author} {\bibfnamefont {J.~P.}\ \bibnamefont
  {Shaffer}},\ }\bibfield  {title} {\bibinfo {title} {Rydberg atom quantum
  technologies},\ }\href {https://doi.org/10.1088/1361-6455/ab52ef} {\bibfield
  {journal} {\bibinfo  {journal} {J. Phys. B}\ }\textbf {\bibinfo {volume}
  {53}},\ \bibinfo {pages} {012002} (\bibinfo {year} {2019})}\BibitemShut
  {NoStop}%
\bibitem [{\citenamefont {Browaeys}\ and\ \citenamefont
  {Lahaye}(2020)}]{Browaeys2020}%
  \BibitemOpen
  \bibfield  {author} {\bibinfo {author} {\bibfnamefont {A.}~\bibnamefont
  {Browaeys}}\ and\ \bibinfo {author} {\bibfnamefont {T.}~\bibnamefont
  {Lahaye}},\ }\bibfield  {title} {\bibinfo {title} {Many-body physics with
  individually controlled {R}ydberg atoms},\ }\href
  {https://doi.org/10.1038/s41567-019-0733-z} {\bibfield  {journal} {\bibinfo
  {journal} {Nat. Phys.}\ }\textbf {\bibinfo {volume} {16}},\ \bibinfo {pages}
  {132} (\bibinfo {year} {2020})}\BibitemShut {NoStop}%
\bibitem [{\citenamefont {Morgado}\ and\ \citenamefont
  {Whitlock}(2021)}]{Morgado2021}%
  \BibitemOpen
  \bibfield  {author} {\bibinfo {author} {\bibfnamefont {M.}~\bibnamefont
  {Morgado}}\ and\ \bibinfo {author} {\bibfnamefont {S.}~\bibnamefont
  {Whitlock}},\ }\bibfield  {title} {\bibinfo {title} {Quantum simulation and
  computing with {Rydberg}-interacting qubits},\ }\href
  {https://doi.org/10.1116/5.0036562} {\bibfield  {journal} {\bibinfo
  {journal} {AVS Quantum Sci.}\ }\textbf {\bibinfo {volume} {3}},\ \bibinfo
  {pages} {023501} (\bibinfo {year} {2021})}\BibitemShut {NoStop}%
\bibitem [{\citenamefont {Wu}\ \emph {et~al.}(2021)\citenamefont {Wu},
  \citenamefont {Liang}, \citenamefont {Tian}, \citenamefont {Yang},
  \citenamefont {Chen}, \citenamefont {Liu}, \citenamefont {Tey},\ and\
  \citenamefont {You}}]{Wu2021}%
  \BibitemOpen
  \bibfield  {author} {\bibinfo {author} {\bibfnamefont {X.}~\bibnamefont
  {Wu}}, \bibinfo {author} {\bibfnamefont {X.}~\bibnamefont {Liang}}, \bibinfo
  {author} {\bibfnamefont {Y.}~\bibnamefont {Tian}}, \bibinfo {author}
  {\bibfnamefont {F.}~\bibnamefont {Yang}}, \bibinfo {author} {\bibfnamefont
  {C.}~\bibnamefont {Chen}}, \bibinfo {author} {\bibfnamefont {Y.-C.}\
  \bibnamefont {Liu}}, \bibinfo {author} {\bibfnamefont {M.~K.}\ \bibnamefont
  {Tey}},\ and\ \bibinfo {author} {\bibfnamefont {L.}~\bibnamefont {You}},\
  }\bibfield  {title} {\bibinfo {title} {A concise review of {R}ydberg atom
  based quantum computation and quantum simulation},\ }\href
  {https://doi.org/10.1088/1674-1056/abd76f} {\bibfield  {journal} {\bibinfo
  {journal} {Chin. Phys. B}\ }\textbf {\bibinfo {volume} {30}},\ \bibinfo
  {pages} {020305} (\bibinfo {year} {2021})}\BibitemShut {NoStop}%
\bibitem [{\citenamefont {Kaufman}\ and\ \citenamefont
  {Ni}(2021)}]{Kaufman2021}%
  \BibitemOpen
  \bibfield  {author} {\bibinfo {author} {\bibfnamefont {A.~M.}\ \bibnamefont
  {Kaufman}}\ and\ \bibinfo {author} {\bibfnamefont {K.-K.}\ \bibnamefont
  {Ni}},\ }\bibfield  {title} {\bibinfo {title} {Quantum science with optical
  tweezer arrays of ultracold atoms and molecules},\ }\href
  {https://doi.org/10.1038/s41567-021-01357-2} {\bibfield  {journal} {\bibinfo
  {journal} {Nat. Phys.}\ }\textbf {\bibinfo {volume} {17}},\ \bibinfo {pages}
  {1324} (\bibinfo {year} {2021})}\BibitemShut {NoStop}%
\bibitem [{\citenamefont {Jaksch}\ \emph {et~al.}(2000)\citenamefont {Jaksch},
  \citenamefont {Cirac}, \citenamefont {Zoller}, \citenamefont {Rolston},
  \citenamefont {C\^ot\'e},\ and\ \citenamefont {Lukin}}]{Jaksch2000}%
  \BibitemOpen
  \bibfield  {author} {\bibinfo {author} {\bibfnamefont {D.}~\bibnamefont
  {Jaksch}}, \bibinfo {author} {\bibfnamefont {J.~I.}\ \bibnamefont {Cirac}},
  \bibinfo {author} {\bibfnamefont {P.}~\bibnamefont {Zoller}}, \bibinfo
  {author} {\bibfnamefont {S.~L.}\ \bibnamefont {Rolston}}, \bibinfo {author}
  {\bibfnamefont {R.}~\bibnamefont {C\^ot\'e}},\ and\ \bibinfo {author}
  {\bibfnamefont {M.~D.}\ \bibnamefont {Lukin}},\ }\bibfield  {title} {\bibinfo
  {title} {Fast quantum gates for neutral atoms},\ }\href
  {https://doi.org/10.1103/PhysRevLett.85.2208} {\bibfield  {journal} {\bibinfo
   {journal} {Phys. Rev. Lett.}\ }\textbf {\bibinfo {volume} {85}},\ \bibinfo
  {pages} {2208} (\bibinfo {year} {2000})}\BibitemShut {NoStop}%
\bibitem [{\citenamefont {Lukin}\ \emph {et~al.}(2001)\citenamefont {Lukin},
  \citenamefont {Fleischhauer}, \citenamefont {Cote}, \citenamefont {Duan},
  \citenamefont {Jaksch}, \citenamefont {Cirac},\ and\ \citenamefont
  {Zoller}}]{Lukin2001}%
  \BibitemOpen
  \bibfield  {author} {\bibinfo {author} {\bibfnamefont {M.~D.}\ \bibnamefont
  {Lukin}}, \bibinfo {author} {\bibfnamefont {M.}~\bibnamefont {Fleischhauer}},
  \bibinfo {author} {\bibfnamefont {R.}~\bibnamefont {Cote}}, \bibinfo {author}
  {\bibfnamefont {L.~M.}\ \bibnamefont {Duan}}, \bibinfo {author}
  {\bibfnamefont {D.}~\bibnamefont {Jaksch}}, \bibinfo {author} {\bibfnamefont
  {J.~I.}\ \bibnamefont {Cirac}},\ and\ \bibinfo {author} {\bibfnamefont
  {P.}~\bibnamefont {Zoller}},\ }\bibfield  {title} {\bibinfo {title} {Dipole
  blockade and quantum information processing in mesoscopic atomic ensembles},\
  }\href {https://doi.org/10.1103/PhysRevLett.87.037901} {\bibfield  {journal}
  {\bibinfo  {journal} {Phys. Rev. Lett.}\ }\textbf {\bibinfo {volume} {87}},\
  \bibinfo {pages} {037901} (\bibinfo {year} {2001})}\BibitemShut {NoStop}%
\bibitem [{\citenamefont {Isenhower}\ \emph {et~al.}(2010)\citenamefont
  {Isenhower}, \citenamefont {Urban}, \citenamefont {Zhang}, \citenamefont
  {Gill}, \citenamefont {Henage}, \citenamefont {Johnson}, \citenamefont
  {Walker},\ and\ \citenamefont {Saffman}}]{Isenhower2010}%
  \BibitemOpen
  \bibfield  {author} {\bibinfo {author} {\bibfnamefont {L.}~\bibnamefont
  {Isenhower}}, \bibinfo {author} {\bibfnamefont {E.}~\bibnamefont {Urban}},
  \bibinfo {author} {\bibfnamefont {X.~L.}\ \bibnamefont {Zhang}}, \bibinfo
  {author} {\bibfnamefont {A.~T.}\ \bibnamefont {Gill}}, \bibinfo {author}
  {\bibfnamefont {T.}~\bibnamefont {Henage}}, \bibinfo {author} {\bibfnamefont
  {T.~A.}\ \bibnamefont {Johnson}}, \bibinfo {author} {\bibfnamefont {T.~G.}\
  \bibnamefont {Walker}},\ and\ \bibinfo {author} {\bibfnamefont
  {M.}~\bibnamefont {Saffman}},\ }\bibfield  {title} {\bibinfo {title}
  {Demonstration of a neutral atom controlled-\textsc{not} quantum gate},\
  }\href {https://doi.org/10.1103/PhysRevLett.104.010503} {\bibfield  {journal}
  {\bibinfo  {journal} {Phys. Rev. Lett.}\ }\textbf {\bibinfo {volume} {104}},\
  \bibinfo {pages} {010503} (\bibinfo {year} {2010})}\BibitemShut {NoStop}%
\bibitem [{\citenamefont {Wilk}\ \emph {et~al.}(2010)\citenamefont {Wilk},
  \citenamefont {Ga\"etan}, \citenamefont {Evellin}, \citenamefont {Wolters},
  \citenamefont {Miroshnychenko}, \citenamefont {Grangier},\ and\ \citenamefont
  {Browaeys}}]{Wilk2010}%
  \BibitemOpen
  \bibfield  {author} {\bibinfo {author} {\bibfnamefont {T.}~\bibnamefont
  {Wilk}}, \bibinfo {author} {\bibfnamefont {A.}~\bibnamefont {Ga\"etan}},
  \bibinfo {author} {\bibfnamefont {C.}~\bibnamefont {Evellin}}, \bibinfo
  {author} {\bibfnamefont {J.}~\bibnamefont {Wolters}}, \bibinfo {author}
  {\bibfnamefont {Y.}~\bibnamefont {Miroshnychenko}}, \bibinfo {author}
  {\bibfnamefont {P.}~\bibnamefont {Grangier}},\ and\ \bibinfo {author}
  {\bibfnamefont {A.}~\bibnamefont {Browaeys}},\ }\bibfield  {title} {\bibinfo
  {title} {Entanglement of two individual neutral atoms using {R}ydberg
  blockade},\ }\href {https://doi.org/10.1103/PhysRevLett.104.010502}
  {\bibfield  {journal} {\bibinfo  {journal} {Phys. Rev. Lett.}\ }\textbf
  {\bibinfo {volume} {104}},\ \bibinfo {pages} {010502} (\bibinfo {year}
  {2010})}\BibitemShut {NoStop}%
\bibitem [{\citenamefont {Saffman}\ \emph {et~al.}(2010)\citenamefont
  {Saffman}, \citenamefont {Walker},\ and\ \citenamefont
  {M\o{}lmer}}]{Saffman2010}%
  \BibitemOpen
  \bibfield  {author} {\bibinfo {author} {\bibfnamefont {M.}~\bibnamefont
  {Saffman}}, \bibinfo {author} {\bibfnamefont {T.~G.}\ \bibnamefont
  {Walker}},\ and\ \bibinfo {author} {\bibfnamefont {K.}~\bibnamefont
  {M\o{}lmer}},\ }\bibfield  {title} {\bibinfo {title} {Quantum information
  with {R}ydberg atoms},\ }\href {https://doi.org/10.1103/RevModPhys.82.2313}
  {\bibfield  {journal} {\bibinfo  {journal} {Rev. Mod. Phys.}\ }\textbf
  {\bibinfo {volume} {82}},\ \bibinfo {pages} {2313} (\bibinfo {year}
  {2010})}\BibitemShut {NoStop}%
\bibitem [{\citenamefont {Levine}\ \emph {et~al.}(2019)\citenamefont {Levine},
  \citenamefont {Keesling}, \citenamefont {Semeghini}, \citenamefont {Omran},
  \citenamefont {Wang}, \citenamefont {Ebadi}, \citenamefont {Bernien},
  \citenamefont {Greiner}, \citenamefont {Vuleti\ifmmode~\acute{c}\else
  \'{c}\fi{}}, \citenamefont {Pichler},\ and\ \citenamefont
  {Lukin}}]{Levine2019}%
  \BibitemOpen
  \bibfield  {author} {\bibinfo {author} {\bibfnamefont {H.}~\bibnamefont
  {Levine}}, \bibinfo {author} {\bibfnamefont {A.}~\bibnamefont {Keesling}},
  \bibinfo {author} {\bibfnamefont {G.}~\bibnamefont {Semeghini}}, \bibinfo
  {author} {\bibfnamefont {A.}~\bibnamefont {Omran}}, \bibinfo {author}
  {\bibfnamefont {T.~T.}\ \bibnamefont {Wang}}, \bibinfo {author}
  {\bibfnamefont {S.}~\bibnamefont {Ebadi}}, \bibinfo {author} {\bibfnamefont
  {H.}~\bibnamefont {Bernien}}, \bibinfo {author} {\bibfnamefont
  {M.}~\bibnamefont {Greiner}}, \bibinfo {author} {\bibfnamefont
  {V.}~\bibnamefont {Vuleti\ifmmode~\acute{c}\else \'{c}\fi{}}}, \bibinfo
  {author} {\bibfnamefont {H.}~\bibnamefont {Pichler}},\ and\ \bibinfo {author}
  {\bibfnamefont {M.~D.}\ \bibnamefont {Lukin}},\ }\bibfield  {title} {\bibinfo
  {title} {Parallel implementation of high-fidelity multiqubit gates with
  neutral atoms},\ }\href {https://doi.org/10.1103/PhysRevLett.123.170503}
  {\bibfield  {journal} {\bibinfo  {journal} {Phys. Rev. Lett.}\ }\textbf
  {\bibinfo {volume} {123}},\ \bibinfo {pages} {170503} (\bibinfo {year}
  {2019})}\BibitemShut {NoStop}%
\bibitem [{\citenamefont {Park}\ \emph {et~al.}(2017)\citenamefont {Park},
  \citenamefont {Yan}, \citenamefont {Loh}, \citenamefont {Will},\ and\
  \citenamefont {Zwierlein}}]{Park2017}%
  \BibitemOpen
  \bibfield  {author} {\bibinfo {author} {\bibfnamefont {J.~W.}\ \bibnamefont
  {Park}}, \bibinfo {author} {\bibfnamefont {Z.~Z.}\ \bibnamefont {Yan}},
  \bibinfo {author} {\bibfnamefont {H.}~\bibnamefont {Loh}}, \bibinfo {author}
  {\bibfnamefont {S.~A.}\ \bibnamefont {Will}},\ and\ \bibinfo {author}
  {\bibfnamefont {M.~W.}\ \bibnamefont {Zwierlein}},\ }\bibfield  {title}
  {\bibinfo {title} {Second-scale nuclear spin coherence time of ultracold
  $^{23}${N}a$^{40}${K} molecules},\ }\href
  {https://doi.org/10.1126/science.aal5066} {\bibfield  {journal} {\bibinfo
  {journal} {Science}\ }\textbf {\bibinfo {volume} {357}},\ \bibinfo {pages}
  {372} (\bibinfo {year} {2017})}\BibitemShut {NoStop}%
\bibitem [{\citenamefont {Gregory}\ \emph
  {et~al.}(2021{\natexlab{a}})\citenamefont {Gregory}, \citenamefont
  {Blackmore}, \citenamefont {Bromley}, \citenamefont {Hutson},\ and\
  \citenamefont {Cornish}}]{Gregory2021}%
  \BibitemOpen
  \bibfield  {author} {\bibinfo {author} {\bibfnamefont {P.~D.}\ \bibnamefont
  {Gregory}}, \bibinfo {author} {\bibfnamefont {J.~A.}\ \bibnamefont
  {Blackmore}}, \bibinfo {author} {\bibfnamefont {S.~L.}\ \bibnamefont
  {Bromley}}, \bibinfo {author} {\bibfnamefont {J.~M.}\ \bibnamefont
  {Hutson}},\ and\ \bibinfo {author} {\bibfnamefont {S.~L.}\ \bibnamefont
  {Cornish}},\ }\bibfield  {title} {\bibinfo {title} {Robust storage qubits in
  ultracold polar molecules},\ }\href
  {https://doi.org/10.1038/s41567-021-01328-7} {\bibfield  {journal} {\bibinfo
  {journal} {Nat. Phys.}\ }\textbf {\bibinfo {volume} {17}},\ \bibinfo {pages}
  {1149} (\bibinfo {year} {2021}{\natexlab{a}})}\BibitemShut {NoStop}%
\bibitem [{\citenamefont {Burchesky}\ \emph {et~al.}(2021)\citenamefont
  {Burchesky}, \citenamefont {Anderegg}, \citenamefont {Bao}, \citenamefont
  {Yu}, \citenamefont {Chae}, \citenamefont {Ketterle}, \citenamefont {Ni},\
  and\ \citenamefont {Doyle}}]{Burchesky2021}%
  \BibitemOpen
  \bibfield  {author} {\bibinfo {author} {\bibfnamefont {S.}~\bibnamefont
  {Burchesky}}, \bibinfo {author} {\bibfnamefont {L.}~\bibnamefont {Anderegg}},
  \bibinfo {author} {\bibfnamefont {Y.}~\bibnamefont {Bao}}, \bibinfo {author}
  {\bibfnamefont {S.~S.}\ \bibnamefont {Yu}}, \bibinfo {author} {\bibfnamefont
  {E.}~\bibnamefont {Chae}}, \bibinfo {author} {\bibfnamefont {W.}~\bibnamefont
  {Ketterle}}, \bibinfo {author} {\bibfnamefont {K.-K.}\ \bibnamefont {Ni}},\
  and\ \bibinfo {author} {\bibfnamefont {J.~M.}\ \bibnamefont {Doyle}},\
  }\bibfield  {title} {\bibinfo {title} {Rotational coherence times of polar
  molecules in optical tweezers},\ }\href
  {https://doi.org/10.1103/PhysRevLett.127.123202} {\bibfield  {journal}
  {\bibinfo  {journal} {Phys. Rev. Lett.}\ }\textbf {\bibinfo {volume} {127}},\
  \bibinfo {pages} {123202} (\bibinfo {year} {2021})}\BibitemShut {NoStop}%
\bibitem [{\citenamefont {DeMille}(2002)}]{DeMille2002}%
  \BibitemOpen
  \bibfield  {author} {\bibinfo {author} {\bibfnamefont {D.}~\bibnamefont
  {DeMille}},\ }\bibfield  {title} {\bibinfo {title} {Quantum computation with
  trapped polar molecules},\ }\href
  {https://doi.org/10.1103/PhysRevLett.88.067901} {\bibfield  {journal}
  {\bibinfo  {journal} {Phys. Rev. Lett.}\ }\textbf {\bibinfo {volume} {88}},\
  \bibinfo {pages} {067901} (\bibinfo {year} {2002})}\BibitemShut {NoStop}%
\bibitem [{\citenamefont {Yelin}\ \emph {et~al.}(2006)\citenamefont {Yelin},
  \citenamefont {Kirby},\ and\ \citenamefont {C\^ot\'e}}]{Yelin2006}%
  \BibitemOpen
  \bibfield  {author} {\bibinfo {author} {\bibfnamefont {S.~F.}\ \bibnamefont
  {Yelin}}, \bibinfo {author} {\bibfnamefont {K.}~\bibnamefont {Kirby}},\ and\
  \bibinfo {author} {\bibfnamefont {R.}~\bibnamefont {C\^ot\'e}},\ }\bibfield
  {title} {\bibinfo {title} {Schemes for robust quantum computation with polar
  molecules},\ }\href {https://doi.org/10.1103/PhysRevA.74.050301} {\bibfield
  {journal} {\bibinfo  {journal} {Phys. Rev. A}\ }\textbf {\bibinfo {volume}
  {74}},\ \bibinfo {pages} {050301(R)} (\bibinfo {year} {2006})}\BibitemShut
  {NoStop}%
\bibitem [{\citenamefont {Ni}\ \emph {et~al.}(2018)\citenamefont {Ni},
  \citenamefont {Rosenband},\ and\ \citenamefont {Grimes}}]{Ni2018}%
  \BibitemOpen
  \bibfield  {author} {\bibinfo {author} {\bibfnamefont {K.-K.}\ \bibnamefont
  {Ni}}, \bibinfo {author} {\bibfnamefont {T.}~\bibnamefont {Rosenband}},\ and\
  \bibinfo {author} {\bibfnamefont {D.~D.}\ \bibnamefont {Grimes}},\ }\bibfield
   {title} {\bibinfo {title} {Dipolar exchange quantum logic gate with polar
  molecules},\ }\href {https://doi.org/10.1039/C8SC02355G} {\bibfield
  {journal} {\bibinfo  {journal} {Chem. Sci.}\ }\textbf {\bibinfo {volume}
  {9}},\ \bibinfo {pages} {6830} (\bibinfo {year} {2018})}\BibitemShut
  {NoStop}%
\bibitem [{\citenamefont {Hughes}\ \emph {et~al.}(2020)\citenamefont {Hughes},
  \citenamefont {Frye}, \citenamefont {Sawant}, \citenamefont {Bhole},
  \citenamefont {Jones}, \citenamefont {Cornish}, \citenamefont {Tarbutt},
  \citenamefont {Hutson}, \citenamefont {Jaksch},\ and\ \citenamefont
  {Mur-Petit}}]{Hughes2020}%
  \BibitemOpen
  \bibfield  {author} {\bibinfo {author} {\bibfnamefont {M.}~\bibnamefont
  {Hughes}}, \bibinfo {author} {\bibfnamefont {M.~D.}\ \bibnamefont {Frye}},
  \bibinfo {author} {\bibfnamefont {R.}~\bibnamefont {Sawant}}, \bibinfo
  {author} {\bibfnamefont {G.}~\bibnamefont {Bhole}}, \bibinfo {author}
  {\bibfnamefont {J.~A.}\ \bibnamefont {Jones}}, \bibinfo {author}
  {\bibfnamefont {S.~L.}\ \bibnamefont {Cornish}}, \bibinfo {author}
  {\bibfnamefont {M.~R.}\ \bibnamefont {Tarbutt}}, \bibinfo {author}
  {\bibfnamefont {J.~M.}\ \bibnamefont {Hutson}}, \bibinfo {author}
  {\bibfnamefont {D.}~\bibnamefont {Jaksch}},\ and\ \bibinfo {author}
  {\bibfnamefont {J.}~\bibnamefont {Mur-Petit}},\ }\bibfield  {title} {\bibinfo
  {title} {Robust entangling gate for polar molecules using magnetic and
  microwave fields},\ }\href {https://doi.org/10.1103/PhysRevA.101.062308}
  {\bibfield  {journal} {\bibinfo  {journal} {Phys. Rev. A}\ }\textbf {\bibinfo
  {volume} {101}},\ \bibinfo {pages} {062308} (\bibinfo {year}
  {2020})}\BibitemShut {NoStop}%
\bibitem [{\citenamefont {Sawant}\ \emph {et~al.}(2020)\citenamefont {Sawant},
  \citenamefont {Blackmore}, \citenamefont {Gregory}, \citenamefont
  {Mur-Petit}, \citenamefont {Jaksch}, \citenamefont {Aldegunde}, \citenamefont
  {Hutson}, \citenamefont {Tarbutt},\ and\ \citenamefont
  {Cornish}}]{Sawant2020}%
  \BibitemOpen
  \bibfield  {author} {\bibinfo {author} {\bibfnamefont {R.}~\bibnamefont
  {Sawant}}, \bibinfo {author} {\bibfnamefont {J.~A.}\ \bibnamefont
  {Blackmore}}, \bibinfo {author} {\bibfnamefont {P.~D.}\ \bibnamefont
  {Gregory}}, \bibinfo {author} {\bibfnamefont {J.}~\bibnamefont {Mur-Petit}},
  \bibinfo {author} {\bibfnamefont {D.}~\bibnamefont {Jaksch}}, \bibinfo
  {author} {\bibfnamefont {J.}~\bibnamefont {Aldegunde}}, \bibinfo {author}
  {\bibfnamefont {J.~M.}\ \bibnamefont {Hutson}}, \bibinfo {author}
  {\bibfnamefont {M.~R.}\ \bibnamefont {Tarbutt}},\ and\ \bibinfo {author}
  {\bibfnamefont {S.~L.}\ \bibnamefont {Cornish}},\ }\bibfield  {title}
  {\bibinfo {title} {Ultracold polar molecules as qudits},\ }\href
  {https://doi.org/10.1088/1367-2630/ab60f4} {\bibfield  {journal} {\bibinfo
  {journal} {New J. Phys.}\ }\textbf {\bibinfo {volume} {22}},\ \bibinfo
  {pages} {013027} (\bibinfo {year} {2020})}\BibitemShut {NoStop}%
\bibitem [{\citenamefont {Albert}\ \emph {et~al.}(2020)\citenamefont {Albert},
  \citenamefont {Covey},\ and\ \citenamefont {Preskill}}]{Albert2020}%
  \BibitemOpen
  \bibfield  {author} {\bibinfo {author} {\bibfnamefont {V.~V.}\ \bibnamefont
  {Albert}}, \bibinfo {author} {\bibfnamefont {J.~P.}\ \bibnamefont {Covey}},\
  and\ \bibinfo {author} {\bibfnamefont {J.}~\bibnamefont {Preskill}},\
  }\bibfield  {title} {\bibinfo {title} {Robust encoding of a qubit in a
  molecule},\ }\href {https://doi.org/10.1103/PhysRevX.10.031050} {\bibfield
  {journal} {\bibinfo  {journal} {Phys. Rev. X}\ }\textbf {\bibinfo {volume}
  {10}},\ \bibinfo {pages} {031050} (\bibinfo {year} {2020})}\BibitemShut
  {NoStop}%
\bibitem [{\citenamefont {Liu}\ \emph {et~al.}(2018)\citenamefont {Liu},
  \citenamefont {Hood}, \citenamefont {Yu}, \citenamefont {Zhang},
  \citenamefont {Hutzler}, \citenamefont {Rosenband},\ and\ \citenamefont
  {Ni}}]{Liu2018}%
  \BibitemOpen
  \bibfield  {author} {\bibinfo {author} {\bibfnamefont {L.~R.}\ \bibnamefont
  {Liu}}, \bibinfo {author} {\bibfnamefont {J.~D.}\ \bibnamefont {Hood}},
  \bibinfo {author} {\bibfnamefont {Y.}~\bibnamefont {Yu}}, \bibinfo {author}
  {\bibfnamefont {J.~T.}\ \bibnamefont {Zhang}}, \bibinfo {author}
  {\bibfnamefont {N.~R.}\ \bibnamefont {Hutzler}}, \bibinfo {author}
  {\bibfnamefont {T.}~\bibnamefont {Rosenband}},\ and\ \bibinfo {author}
  {\bibfnamefont {K.-K.}\ \bibnamefont {Ni}},\ }\bibfield  {title} {\bibinfo
  {title} {Building one molecule from a reservoir of two atoms},\ }\href
  {https://doi.org/10.1126/science.aar7797} {\bibfield  {journal} {\bibinfo
  {journal} {Science}\ }\textbf {\bibinfo {volume} {360}},\ \bibinfo {pages}
  {900} (\bibinfo {year} {2018})}\BibitemShut {NoStop}%
\bibitem [{\citenamefont {Anderegg}\ \emph {et~al.}(2019)\citenamefont
  {Anderegg}, \citenamefont {Cheuk}, \citenamefont {Bao}, \citenamefont
  {Burchesky}, \citenamefont {Ketterle}, \citenamefont {Ni},\ and\
  \citenamefont {Doyle}}]{Anderegg2019}%
  \BibitemOpen
  \bibfield  {author} {\bibinfo {author} {\bibfnamefont {L.}~\bibnamefont
  {Anderegg}}, \bibinfo {author} {\bibfnamefont {L.~W.}\ \bibnamefont {Cheuk}},
  \bibinfo {author} {\bibfnamefont {Y.}~\bibnamefont {Bao}}, \bibinfo {author}
  {\bibfnamefont {S.}~\bibnamefont {Burchesky}}, \bibinfo {author}
  {\bibfnamefont {W.}~\bibnamefont {Ketterle}}, \bibinfo {author}
  {\bibfnamefont {K.-K.}\ \bibnamefont {Ni}},\ and\ \bibinfo {author}
  {\bibfnamefont {J.~M.}\ \bibnamefont {Doyle}},\ }\bibfield  {title} {\bibinfo
  {title} {An optical tweezer array of ultracold molecules},\ }\href
  {https://doi.org/10.1126/science.aax1265} {\bibfield  {journal} {\bibinfo
  {journal} {Science}\ }\textbf {\bibinfo {volume} {365}},\ \bibinfo {pages}
  {1156} (\bibinfo {year} {2019})}\BibitemShut {NoStop}%
\bibitem [{\citenamefont {He}\ \emph {et~al.}(2020)\citenamefont {He},
  \citenamefont {Wang}, \citenamefont {Zhuang}, \citenamefont {Xu},
  \citenamefont {Gao}, \citenamefont {Guo}, \citenamefont {Sheng},
  \citenamefont {Liu}, \citenamefont {Wang}, \citenamefont {Li}, \citenamefont
  {Shlyapnikov},\ and\ \citenamefont {Zhan}}]{He2020}%
  \BibitemOpen
  \bibfield  {author} {\bibinfo {author} {\bibfnamefont {X.}~\bibnamefont
  {He}}, \bibinfo {author} {\bibfnamefont {K.}~\bibnamefont {Wang}}, \bibinfo
  {author} {\bibfnamefont {J.}~\bibnamefont {Zhuang}}, \bibinfo {author}
  {\bibfnamefont {P.}~\bibnamefont {Xu}}, \bibinfo {author} {\bibfnamefont
  {X.}~\bibnamefont {Gao}}, \bibinfo {author} {\bibfnamefont {R.}~\bibnamefont
  {Guo}}, \bibinfo {author} {\bibfnamefont {C.}~\bibnamefont {Sheng}}, \bibinfo
  {author} {\bibfnamefont {M.}~\bibnamefont {Liu}}, \bibinfo {author}
  {\bibfnamefont {J.}~\bibnamefont {Wang}}, \bibinfo {author} {\bibfnamefont
  {J.}~\bibnamefont {Li}}, \bibinfo {author} {\bibfnamefont {G.~V.}\
  \bibnamefont {Shlyapnikov}},\ and\ \bibinfo {author} {\bibfnamefont
  {M.}~\bibnamefont {Zhan}},\ }\bibfield  {title} {\bibinfo {title} {Coherently
  forming a single molecule in an optical trap},\ }\href
  {https://doi.org/10.1126/science.aba7468} {\bibfield  {journal} {\bibinfo
  {journal} {Science}\ }\textbf {\bibinfo {volume} {370}},\ \bibinfo {pages}
  {331} (\bibinfo {year} {2020})}\BibitemShut {NoStop}%
\bibitem [{\citenamefont {Cairncross}\ \emph {et~al.}(2021)\citenamefont
  {Cairncross}, \citenamefont {Zhang}, \citenamefont {Picard}, \citenamefont
  {Yu}, \citenamefont {Wang},\ and\ \citenamefont {Ni}}]{Cairncross2021}%
  \BibitemOpen
  \bibfield  {author} {\bibinfo {author} {\bibfnamefont {W.~B.}\ \bibnamefont
  {Cairncross}}, \bibinfo {author} {\bibfnamefont {J.~T.}\ \bibnamefont
  {Zhang}}, \bibinfo {author} {\bibfnamefont {L.~R.}\ \bibnamefont {Picard}},
  \bibinfo {author} {\bibfnamefont {Y.}~\bibnamefont {Yu}}, \bibinfo {author}
  {\bibfnamefont {K.}~\bibnamefont {Wang}},\ and\ \bibinfo {author}
  {\bibfnamefont {K.-K.}\ \bibnamefont {Ni}},\ }\bibfield  {title} {\bibinfo
  {title} {Assembly of a rovibrational ground state molecule in an optical
  tweezer},\ }\href {https://doi.org/10.1103/physrevlett.126.123402} {\bibfield
   {journal} {\bibinfo  {journal} {Phys. Rev. Lett.}\ }\textbf {\bibinfo
  {volume} {126}},\ \bibinfo {pages} {123402} (\bibinfo {year}
  {2021})}\BibitemShut {NoStop}%
\bibitem [{\citenamefont {Zhang}\ \emph {et~al.}(2022)\citenamefont {Zhang},
  \citenamefont {Picard}, \citenamefont {Cairncross}, \citenamefont {Wang},
  \citenamefont {Yu}, \citenamefont {Fang},\ and\ \citenamefont
  {Ni}}]{Zhang2022a}%
  \BibitemOpen
  \bibfield  {author} {\bibinfo {author} {\bibfnamefont {J.~T.}\ \bibnamefont
  {Zhang}}, \bibinfo {author} {\bibfnamefont {L.~R.~B.}\ \bibnamefont
  {Picard}}, \bibinfo {author} {\bibfnamefont {W.~B.}\ \bibnamefont
  {Cairncross}}, \bibinfo {author} {\bibfnamefont {K.}~\bibnamefont {Wang}},
  \bibinfo {author} {\bibfnamefont {Y.}~\bibnamefont {Yu}}, \bibinfo {author}
  {\bibfnamefont {F.}~\bibnamefont {Fang}},\ and\ \bibinfo {author}
  {\bibfnamefont {K.-K.}\ \bibnamefont {Ni}},\ }\bibfield  {title} {\bibinfo
  {title} {An optical tweezer array of ground-state polar molecules},\ }\href
  {https://doi.org/10.1088/2058-9565/ac676c} {\bibfield  {journal} {\bibinfo
  {journal} {Quantum Sci. Technol.}\ }\textbf {\bibinfo {volume} {7}},\
  \bibinfo {pages} {035006} (\bibinfo {year} {2022})}\BibitemShut {NoStop}%
\bibitem [{\citenamefont {Ruttley}\ \emph {et~al.}()\citenamefont {Ruttley},
  \citenamefont {Guttridge}, \citenamefont {Spence}, \citenamefont {Bird},
  \citenamefont {Sueur}, \citenamefont {Hutson},\ and\ \citenamefont
  {Cornish}}]{Ruttley2023}%
  \BibitemOpen
  \bibfield  {author} {\bibinfo {author} {\bibfnamefont {D.~K.}\ \bibnamefont
  {Ruttley}}, \bibinfo {author} {\bibfnamefont {A.}~\bibnamefont {Guttridge}},
  \bibinfo {author} {\bibfnamefont {S.}~\bibnamefont {Spence}}, \bibinfo
  {author} {\bibfnamefont {R.~C.}\ \bibnamefont {Bird}}, \bibinfo {author}
  {\bibfnamefont {C.~R.~L.}\ \bibnamefont {Sueur}}, \bibinfo {author}
  {\bibfnamefont {J.~M.}\ \bibnamefont {Hutson}},\ and\ \bibinfo {author}
  {\bibfnamefont {S.~L.}\ \bibnamefont {Cornish}},\ }\bibfield  {title}
  {\bibinfo {title} {Formation of ultracold molecules by merging optical
  tweezers},\ }\href@noop {} {\ }\Eprint {https://arxiv.org/abs/2302.07296}
  {arXiv:2302.07296 [physics.atom-ph]} \BibitemShut {NoStop}%
\bibitem [{\citenamefont {Kuznetsova}\ \emph {et~al.}(2011)\citenamefont
  {Kuznetsova}, \citenamefont {Rittenhouse}, \citenamefont {Sadeghpour},\ and\
  \citenamefont {Yelin}}]{Kuznetsova2011}%
  \BibitemOpen
  \bibfield  {author} {\bibinfo {author} {\bibfnamefont {E.}~\bibnamefont
  {Kuznetsova}}, \bibinfo {author} {\bibfnamefont {S.~T.}\ \bibnamefont
  {Rittenhouse}}, \bibinfo {author} {\bibfnamefont {H.~R.}\ \bibnamefont
  {Sadeghpour}},\ and\ \bibinfo {author} {\bibfnamefont {S.~F.}\ \bibnamefont
  {Yelin}},\ }\bibfield  {title} {\bibinfo {title} {Rydberg atom mediated polar
  molecule interactions: a tool for molecular-state conditional quantum gates
  and individual addressability},\ }\href {https://doi.org/10.1039/C1CP21476D}
  {\bibfield  {journal} {\bibinfo  {journal} {Phys. Chem. Chem. Phys.}\
  }\textbf {\bibinfo {volume} {13}},\ \bibinfo {pages} {17115} (\bibinfo {year}
  {2011})}\BibitemShut {NoStop}%
\bibitem [{\citenamefont {Kuznetsova}\ \emph {et~al.}(2016)\citenamefont
  {Kuznetsova}, \citenamefont {Rittenhouse}, \citenamefont {Sadeghpour},\ and\
  \citenamefont {Yelin}}]{Kuznetsova2016}%
  \BibitemOpen
  \bibfield  {author} {\bibinfo {author} {\bibfnamefont {E.}~\bibnamefont
  {Kuznetsova}}, \bibinfo {author} {\bibfnamefont {S.~T.}\ \bibnamefont
  {Rittenhouse}}, \bibinfo {author} {\bibfnamefont {H.~R.}\ \bibnamefont
  {Sadeghpour}},\ and\ \bibinfo {author} {\bibfnamefont {S.~F.}\ \bibnamefont
  {Yelin}},\ }\bibfield  {title} {\bibinfo {title} {Rydberg-atom-mediated
  nondestructive readout of collective rotational states in polar-molecule
  arrays},\ }\href {https://doi.org/10.1103/PhysRevA.94.032325} {\bibfield
  {journal} {\bibinfo  {journal} {Phys. Rev. A}\ }\textbf {\bibinfo {volume}
  {94}},\ \bibinfo {pages} {032325} (\bibinfo {year} {2016})}\BibitemShut
  {NoStop}%
\bibitem [{\citenamefont {Wang}\ \emph {et~al.}(2022)\citenamefont {Wang},
  \citenamefont {Williams}, \citenamefont {Picard}, \citenamefont {Yao},\ and\
  \citenamefont {Ni}}]{Wang2022}%
  \BibitemOpen
  \bibfield  {author} {\bibinfo {author} {\bibfnamefont {K.}~\bibnamefont
  {Wang}}, \bibinfo {author} {\bibfnamefont {C.~P.}\ \bibnamefont {Williams}},
  \bibinfo {author} {\bibfnamefont {L.~R.~B.}\ \bibnamefont {Picard}}, \bibinfo
  {author} {\bibfnamefont {N.~Y.}\ \bibnamefont {Yao}},\ and\ \bibinfo {author}
  {\bibfnamefont {K.-K.}\ \bibnamefont {Ni}},\ }\bibfield  {title} {\bibinfo
  {title} {Enriching the quantum toolbox of ultracold molecules with {R}ydberg
  atoms},\ }\href {https://doi.org/10.1103/PRXQuantum.3.030339} {\bibfield
  {journal} {\bibinfo  {journal} {PRX Quantum}\ }\textbf {\bibinfo {volume}
  {3}},\ \bibinfo {pages} {030339} (\bibinfo {year} {2022})}\BibitemShut
  {NoStop}%
\bibitem [{\citenamefont {Zhang}\ and\ \citenamefont
  {Tarbutt}(2022)}]{Zhang2022}%
  \BibitemOpen
  \bibfield  {author} {\bibinfo {author} {\bibfnamefont {C.}~\bibnamefont
  {Zhang}}\ and\ \bibinfo {author} {\bibfnamefont {M.~R.}\ \bibnamefont
  {Tarbutt}},\ }\bibfield  {title} {\bibinfo {title} {Quantum computation in a
  hybrid array of molecules and {R}ydberg atoms},\ }\href
  {https://doi.org/10.1103/PRXQuantum.3.030340} {\bibfield  {journal} {\bibinfo
   {journal} {PRX Quantum}\ }\textbf {\bibinfo {volume} {3}},\ \bibinfo {pages}
  {030340} (\bibinfo {year} {2022})}\BibitemShut {NoStop}%
\bibitem [{\citenamefont {Graham}\ \emph {et~al.}(2019)\citenamefont {Graham},
  \citenamefont {Kwon}, \citenamefont {Grinkemeyer}, \citenamefont {Marra},
  \citenamefont {Jiang}, \citenamefont {Lichtman}, \citenamefont {Sun},
  \citenamefont {Ebert},\ and\ \citenamefont {Saffman}}]{Graham2019}%
  \BibitemOpen
  \bibfield  {author} {\bibinfo {author} {\bibfnamefont {T.~M.}\ \bibnamefont
  {Graham}}, \bibinfo {author} {\bibfnamefont {M.}~\bibnamefont {Kwon}},
  \bibinfo {author} {\bibfnamefont {B.}~\bibnamefont {Grinkemeyer}}, \bibinfo
  {author} {\bibfnamefont {Z.}~\bibnamefont {Marra}}, \bibinfo {author}
  {\bibfnamefont {X.}~\bibnamefont {Jiang}}, \bibinfo {author} {\bibfnamefont
  {M.~T.}\ \bibnamefont {Lichtman}}, \bibinfo {author} {\bibfnamefont
  {Y.}~\bibnamefont {Sun}}, \bibinfo {author} {\bibfnamefont {M.}~\bibnamefont
  {Ebert}},\ and\ \bibinfo {author} {\bibfnamefont {M.}~\bibnamefont
  {Saffman}},\ }\bibfield  {title} {\bibinfo {title} {Rydberg-mediated
  entanglement in a two-dimensional neutral atom qubit array},\ }\href
  {https://doi.org/10.1103/PhysRevLett.123.230501} {\bibfield  {journal}
  {\bibinfo  {journal} {Phys. Rev. Lett.}\ }\textbf {\bibinfo {volume} {123}},\
  \bibinfo {pages} {230501} (\bibinfo {year} {2019})}\BibitemShut {NoStop}%
\bibitem [{\citenamefont {Jarisch}\ and\ \citenamefont
  {Zeppenfeld}(2018)}]{Jarisch2018}%
  \BibitemOpen
  \bibfield  {author} {\bibinfo {author} {\bibfnamefont {F.}~\bibnamefont
  {Jarisch}}\ and\ \bibinfo {author} {\bibfnamefont {M.}~\bibnamefont
  {Zeppenfeld}},\ }\bibfield  {title} {\bibinfo {title} {State resolved
  investigation of {F}örster resonant energy transfer in collisions between
  polar molecules and {R}ydberg atoms},\ }\href
  {http://dx.doi.org/10.1088/1367-2630/aaf02e} {\bibfield  {journal} {\bibinfo
  {journal} {New J. Phys.}\ }\textbf {\bibinfo {volume} {20}},\ \bibinfo
  {pages} {113044} (\bibinfo {year} {2018})}\BibitemShut {NoStop}%
\bibitem [{\citenamefont {Gawlas}\ and\ \citenamefont
  {Hogan}(2020)}]{Gawlas2020}%
  \BibitemOpen
  \bibfield  {author} {\bibinfo {author} {\bibfnamefont {K.}~\bibnamefont
  {Gawlas}}\ and\ \bibinfo {author} {\bibfnamefont {S.~D.}\ \bibnamefont
  {Hogan}},\ }\bibfield  {title} {\bibinfo {title} {Rydberg-state-resolved
  resonant energy transfer in cold electric-field-controlled intrabeam
  collisions of {NH$_{3}$} with {R}ydberg {He} atoms},\ }\href
  {https://doi.org/10.1021/acs.jpclett.9b03290} {\bibfield  {journal} {\bibinfo
   {journal} {J. Phys. Chem. Lett.}\ }\textbf {\bibinfo {volume} {11}},\
  \bibinfo {pages} {83} (\bibinfo {year} {2020})}\BibitemShut {NoStop}%
\bibitem [{\citenamefont {Patsch}\ \emph {et~al.}(2022)\citenamefont {Patsch},
  \citenamefont {Zeppenfeld},\ and\ \citenamefont {Koch}}]{Patsch2022}%
  \BibitemOpen
  \bibfield  {author} {\bibinfo {author} {\bibfnamefont {S.}~\bibnamefont
  {Patsch}}, \bibinfo {author} {\bibfnamefont {M.}~\bibnamefont {Zeppenfeld}},\
  and\ \bibinfo {author} {\bibfnamefont {C.~P.}\ \bibnamefont {Koch}},\
  }\bibfield  {title} {\bibinfo {title} {Rydberg atom-enabled spectroscopy of
  polar molecules via {F}{\"o}rster resonance energy transfer},\ }\href
  {https://doi.org/10.1021/acs.jpclett.2c02521} {\bibfield  {journal} {\bibinfo
   {journal} {J. Phys. Chem. Lett.}\ }\textbf {\bibinfo {volume} {13}},\
  \bibinfo {pages} {10728} (\bibinfo {year} {2022})}\BibitemShut {NoStop}%
\bibitem [{\citenamefont {Zhao}\ \emph {et~al.}(2012)\citenamefont {Zhao},
  \citenamefont {Glaetzle}, \citenamefont {Pupillo},\ and\ \citenamefont
  {Zoller}}]{Zhao2012}%
  \BibitemOpen
  \bibfield  {author} {\bibinfo {author} {\bibfnamefont {B.}~\bibnamefont
  {Zhao}}, \bibinfo {author} {\bibfnamefont {A.~W.}\ \bibnamefont {Glaetzle}},
  \bibinfo {author} {\bibfnamefont {G.}~\bibnamefont {Pupillo}},\ and\ \bibinfo
  {author} {\bibfnamefont {P.}~\bibnamefont {Zoller}},\ }\bibfield  {title}
  {\bibinfo {title} {Atomic {Rydberg} reservoirs for polar molecules},\ }\href
  {https://link.aps.org/doi/10.1103/PhysRevLett.108.193007} {\bibfield
  {journal} {\bibinfo  {journal} {Phys. Rev. Lett.}\ }\textbf {\bibinfo
  {volume} {108}},\ \bibinfo {pages} {193007} (\bibinfo {year}
  {2012})}\BibitemShut {NoStop}%
\bibitem [{\citenamefont {Huber}\ and\ \citenamefont
  {B\"{u}chler}(2012)}]{Huber2012}%
  \BibitemOpen
  \bibfield  {author} {\bibinfo {author} {\bibfnamefont {S.~D.}\ \bibnamefont
  {Huber}}\ and\ \bibinfo {author} {\bibfnamefont {H.~P.}\ \bibnamefont
  {B\"{u}chler}},\ }\bibfield  {title} {\bibinfo {title}
  {Dipole-interaction-mediated laser cooling of polar molecules to ultracold
  temperatures},\ }\href {https://doi.org/10.1103/PhysRevLett.108.193006}
  {\bibfield  {journal} {\bibinfo  {journal} {Phys. Rev. Lett.}\ }\textbf
  {\bibinfo {volume} {108}},\ \bibinfo {pages} {193006} (\bibinfo {year}
  {2012})}\BibitemShut {NoStop}%
\bibitem [{\citenamefont {Rittenhouse}\ and\ \citenamefont
  {Sadeghpour}(2010)}]{Rittenhouse2010}%
  \BibitemOpen
  \bibfield  {author} {\bibinfo {author} {\bibfnamefont {S.~T.}\ \bibnamefont
  {Rittenhouse}}\ and\ \bibinfo {author} {\bibfnamefont {H.~R.}\ \bibnamefont
  {Sadeghpour}},\ }\bibfield  {title} {\bibinfo {title} {Ultracold giant
  polyatomic {R}ydberg molecules: Coherent control of molecular orientation},\
  }\href {https://doi.org/10.1103/PhysRevLett.104.243002} {\bibfield  {journal}
  {\bibinfo  {journal} {Phys. Rev. Lett.}\ }\textbf {\bibinfo {volume} {104}},\
  \bibinfo {pages} {243002} (\bibinfo {year} {2010})}\BibitemShut {NoStop}%
\bibitem [{\citenamefont {Rittenhouse}\ \emph {et~al.}(2011)\citenamefont
  {Rittenhouse}, \citenamefont {Mayle}, \citenamefont {Schmelcher},\ and\
  \citenamefont {Sadeghpour}}]{Rittenhouse2011}%
  \BibitemOpen
  \bibfield  {author} {\bibinfo {author} {\bibfnamefont {S.~T.}\ \bibnamefont
  {Rittenhouse}}, \bibinfo {author} {\bibfnamefont {M.}~\bibnamefont {Mayle}},
  \bibinfo {author} {\bibfnamefont {P.}~\bibnamefont {Schmelcher}},\ and\
  \bibinfo {author} {\bibfnamefont {H.~R.}\ \bibnamefont {Sadeghpour}},\
  }\bibfield  {title} {\bibinfo {title} {Ultralong-range polyatomic {Rydberg}
  molecules formed by a polar perturber},\ }\href
  {https://doi.org/10.1088/0953-4075/44/18/184005} {\bibfield  {journal}
  {\bibinfo  {journal} {J. Phys. B}\ }\textbf {\bibinfo {volume} {44}},\
  \bibinfo {pages} {184005} (\bibinfo {year} {2011})}\BibitemShut {NoStop}%
\bibitem [{\citenamefont {González-Férez}\ \emph {et~al.}(2020)\citenamefont
  {González-Férez}, \citenamefont {Rittenhouse}, \citenamefont {Schmelcher},\
  and\ \citenamefont {Sadeghpour}}]{GonzalezFerez2020}%
  \BibitemOpen
  \bibfield  {author} {\bibinfo {author} {\bibfnamefont {R.}~\bibnamefont
  {González-Férez}}, \bibinfo {author} {\bibfnamefont {S.~T.}\ \bibnamefont
  {Rittenhouse}}, \bibinfo {author} {\bibfnamefont {P.}~\bibnamefont
  {Schmelcher}},\ and\ \bibinfo {author} {\bibfnamefont {H.~R.}\ \bibnamefont
  {Sadeghpour}},\ }\bibfield  {title} {\bibinfo {title} {A protocol to realize
  triatomic ultralong range {Rydberg} molecules in an ultracold {KRb} gas},\
  }\href {https://doi.org/10.1088/1361-6455/ab68b8} {\bibfield  {journal}
  {\bibinfo  {journal} {J. Phys. B}\ }\textbf {\bibinfo {volume} {53}},\
  \bibinfo {pages} {074002} (\bibinfo {year} {2020})}\BibitemShut {NoStop}%
\bibitem [{\citenamefont {Yan}\ \emph {et~al.}(2013)\citenamefont {Yan},
  \citenamefont {Moses}, \citenamefont {Gadway}, \citenamefont {Covey},
  \citenamefont {Hazzard}, \citenamefont {Rey}, \citenamefont {Jin},\ and\
  \citenamefont {Ye}}]{Yan2013}%
  \BibitemOpen
  \bibfield  {author} {\bibinfo {author} {\bibfnamefont {B.}~\bibnamefont
  {Yan}}, \bibinfo {author} {\bibfnamefont {S.~A.}\ \bibnamefont {Moses}},
  \bibinfo {author} {\bibfnamefont {B.}~\bibnamefont {Gadway}}, \bibinfo
  {author} {\bibfnamefont {J.~P.}\ \bibnamefont {Covey}}, \bibinfo {author}
  {\bibfnamefont {K.~R.~A.}\ \bibnamefont {Hazzard}}, \bibinfo {author}
  {\bibfnamefont {A.~M.}\ \bibnamefont {Rey}}, \bibinfo {author} {\bibfnamefont
  {D.~S.}\ \bibnamefont {Jin}},\ and\ \bibinfo {author} {\bibfnamefont
  {J.}~\bibnamefont {Ye}},\ }\bibfield  {title} {\bibinfo {title} {Observation
  of dipolar spin-exchange interactions with lattice-confined polar
  molecules},\ }\href {https://doi.org/10.1038/nature12483} {\bibfield
  {journal} {\bibinfo  {journal} {Nature}\ }\textbf {\bibinfo {volume} {501}},\
  \bibinfo {pages} {521} (\bibinfo {year} {2013})}\BibitemShut {NoStop}%
\bibitem [{\citenamefont {Christakis}\ \emph {et~al.}(2023)\citenamefont
  {Christakis}, \citenamefont {Rosenberg}, \citenamefont {Raj}, \citenamefont
  {Chi}, \citenamefont {Morningstar}, \citenamefont {Huse}, \citenamefont
  {Yan},\ and\ \citenamefont {Bakr}}]{Christakis2023}%
  \BibitemOpen
  \bibfield  {author} {\bibinfo {author} {\bibfnamefont {L.}~\bibnamefont
  {Christakis}}, \bibinfo {author} {\bibfnamefont {J.~S.}\ \bibnamefont
  {Rosenberg}}, \bibinfo {author} {\bibfnamefont {R.}~\bibnamefont {Raj}},
  \bibinfo {author} {\bibfnamefont {S.}~\bibnamefont {Chi}}, \bibinfo {author}
  {\bibfnamefont {A.}~\bibnamefont {Morningstar}}, \bibinfo {author}
  {\bibfnamefont {D.~A.}\ \bibnamefont {Huse}}, \bibinfo {author}
  {\bibfnamefont {Z.~Z.}\ \bibnamefont {Yan}},\ and\ \bibinfo {author}
  {\bibfnamefont {W.~S.}\ \bibnamefont {Bakr}},\ }\bibfield  {title} {\bibinfo
  {title} {Probing site-resolved correlations in a spin system of ultracold
  molecules},\ }\href {https://doi.org/10.1038/s41586-022-05558-4} {\bibfield
  {journal} {\bibinfo  {journal} {Nature}\ }\textbf {\bibinfo {volume} {614}},\
  \bibinfo {pages} {64} (\bibinfo {year} {2023})}\BibitemShut {NoStop}%
\bibitem [{\citenamefont {Holland}\ \emph {et~al.}()\citenamefont {Holland},
  \citenamefont {Lu},\ and\ \citenamefont {Cheuk}}]{Holland2022}%
  \BibitemOpen
  \bibfield  {author} {\bibinfo {author} {\bibfnamefont {C.~M.}\ \bibnamefont
  {Holland}}, \bibinfo {author} {\bibfnamefont {Y.}~\bibnamefont {Lu}},\ and\
  \bibinfo {author} {\bibfnamefont {L.~W.}\ \bibnamefont {Cheuk}},\ }\bibfield
  {title} {\bibinfo {title} {On-demand entanglement of molecules in a
  reconfigurable optical tweezer array},\ }\href@noop {} {\ }\Eprint
  {https://arxiv.org/abs/2210.06309} {arXiv:2210.06309 [cond-mat.quant-gas]}
  \BibitemShut {NoStop}%
\bibitem [{\citenamefont {Bao}\ \emph {et~al.}()\citenamefont {Bao},
  \citenamefont {Yu}, \citenamefont {Anderegg}, \citenamefont {Chae},
  \citenamefont {Ketterle}, \citenamefont {Ni},\ and\ \citenamefont
  {Doyle}}]{Bao2022}%
  \BibitemOpen
  \bibfield  {author} {\bibinfo {author} {\bibfnamefont {Y.}~\bibnamefont
  {Bao}}, \bibinfo {author} {\bibfnamefont {S.~S.}\ \bibnamefont {Yu}},
  \bibinfo {author} {\bibfnamefont {L.}~\bibnamefont {Anderegg}}, \bibinfo
  {author} {\bibfnamefont {E.}~\bibnamefont {Chae}}, \bibinfo {author}
  {\bibfnamefont {W.}~\bibnamefont {Ketterle}}, \bibinfo {author}
  {\bibfnamefont {K.-K.}\ \bibnamefont {Ni}},\ and\ \bibinfo {author}
  {\bibfnamefont {J.~M.}\ \bibnamefont {Doyle}},\ }\bibfield  {title} {\bibinfo
  {title} {Dipolar spin-exchange and entanglement between molecules in an
  optical tweezer array},\ }\href@noop {} {\ }\Eprint
  {https://arxiv.org/abs/2211.09780} {arXiv:2211.09780 [physics.atom-ph]}
  \BibitemShut {NoStop}%
\bibitem [{\citenamefont {Shaffer}\ \emph {et~al.}(2018)\citenamefont
  {Shaffer}, \citenamefont {Rittenhouse},\ and\ \citenamefont
  {Sadeghpour}}]{Shaffer2018}%
  \BibitemOpen
  \bibfield  {author} {\bibinfo {author} {\bibfnamefont {J.~P.}\ \bibnamefont
  {Shaffer}}, \bibinfo {author} {\bibfnamefont {S.~T.}\ \bibnamefont
  {Rittenhouse}},\ and\ \bibinfo {author} {\bibfnamefont {H.~R.}\ \bibnamefont
  {Sadeghpour}},\ }\bibfield  {title} {\bibinfo {title} {Ultracold rydberg
  molecules},\ }\href {https://doi.org/10.1038/s41467-018-04135-6} {\bibfield
  {journal} {\bibinfo  {journal} {Nat. Commun.}\ }\textbf {\bibinfo {volume}
  {9}},\ \bibinfo {pages} {1965} (\bibinfo {year} {2018})}\BibitemShut
  {NoStop}%
\bibitem [{\citenamefont {Sadeghpour}(2023)}]{AMOHandbook2023}%
  \BibitemOpen
  \bibfield  {author} {\bibinfo {author} {\bibfnamefont {H.~R.}\ \bibnamefont
  {Sadeghpour}},\ }\bibinfo {title} {Ultracold rydberg atom--atom
  interaction},\ in\ \href {https://doi.org/10.1007/978-3-030-73893-8_54}
  {\emph {\bibinfo {booktitle} {Springer Handbook of Atomic, Molecular, and
  Optical Physics}}},\ \bibinfo {editor} {edited by\ \bibinfo {editor}
  {\bibfnamefont {G.~W.~F.}\ \bibnamefont {Drake}}}\ (\bibinfo  {publisher}
  {Springer International Publishing},\ \bibinfo {address} {Cham},\ \bibinfo
  {year} {2023})\ pp.\ \bibinfo {pages} {795--803}\BibitemShut {NoStop}%
\bibitem [{sup()}]{supplement}%
  \BibitemOpen
  \href@noop {} {}\bibinfo {note} {See Supplemental Material at [link to be
  inserted] for details of the theoretical calculations, STIRAP transfer, trap
  lifetimes, post-selection scheme, and Rydberg excitation scheme. Supplemental
  Material includes
  Refs.~\cite{Omont,FANO198661,EATON1992141,marinescu94,Black2001,Weber2017,sibalic2017,Leseleuc2018,Levine2018,Zhang2020,Blackmore2020}.}\BibitemShut
  {Stop}%
\bibitem [{\citenamefont {Fermi}\ and\ \citenamefont {Teller}(1947)}]{fermi47}%
  \BibitemOpen
  \bibfield  {author} {\bibinfo {author} {\bibfnamefont {E.}~\bibnamefont
  {Fermi}}\ and\ \bibinfo {author} {\bibfnamefont {E.}~\bibnamefont {Teller}},\
  }\bibfield  {title} {\bibinfo {title} {The capture of negative mesotrons in
  matter},\ }\href {https://doi.org/10.1103/PhysRev.72.399} {\bibfield
  {journal} {\bibinfo  {journal} {Phys. Rev.}\ }\textbf {\bibinfo {volume}
  {72}},\ \bibinfo {pages} {399} (\bibinfo {year} {1947})}\BibitemShut
  {NoStop}%
\bibitem [{\citenamefont {Molony}\ \emph {et~al.}(2014)\citenamefont {Molony},
  \citenamefont {Gregory}, \citenamefont {Ji}, \citenamefont {Lu},
  \citenamefont {K\"{o}ppinger}, \citenamefont {{Le Sueur}}, \citenamefont
  {Blackley}, \citenamefont {Hutson},\ and\ \citenamefont
  {Cornish}}]{Molony2014}%
  \BibitemOpen
  \bibfield  {author} {\bibinfo {author} {\bibfnamefont {P.~K.}\ \bibnamefont
  {Molony}}, \bibinfo {author} {\bibfnamefont {P.~D.}\ \bibnamefont {Gregory}},
  \bibinfo {author} {\bibfnamefont {Z.}~\bibnamefont {Ji}}, \bibinfo {author}
  {\bibfnamefont {B.}~\bibnamefont {Lu}}, \bibinfo {author} {\bibfnamefont
  {M.~P.}\ \bibnamefont {K\"{o}ppinger}}, \bibinfo {author} {\bibfnamefont
  {C.~R.}\ \bibnamefont {{Le Sueur}}}, \bibinfo {author} {\bibfnamefont
  {C.~L.}\ \bibnamefont {Blackley}}, \bibinfo {author} {\bibfnamefont {J.~M.}\
  \bibnamefont {Hutson}},\ and\ \bibinfo {author} {\bibfnamefont {S.~L.}\
  \bibnamefont {Cornish}},\ }\bibfield  {title} {\bibinfo {title} {Creation of
  ultracold $^{87} \mathrm{Rb}$$^{133}\mathrm{Cs}$ molecules in the
  rovibrational ground state},\ }\href
  {https://doi.org/10.1103/PhysRevLett.113.255301} {\bibfield  {journal}
  {\bibinfo  {journal} {Phys. Rev. Lett.}\ }\textbf {\bibinfo {volume} {113}},\
  \bibinfo {pages} {255301} (\bibinfo {year} {2014})}\BibitemShut {NoStop}%
\bibitem [{\citenamefont {Gregory}\ \emph {et~al.}(2016)\citenamefont
  {Gregory}, \citenamefont {Aldegunde}, \citenamefont {Hutson},\ and\
  \citenamefont {Cornish}}]{Gregory2016}%
  \BibitemOpen
  \bibfield  {author} {\bibinfo {author} {\bibfnamefont {P.~D.}\ \bibnamefont
  {Gregory}}, \bibinfo {author} {\bibfnamefont {J.}~\bibnamefont {Aldegunde}},
  \bibinfo {author} {\bibfnamefont {J.~M.}\ \bibnamefont {Hutson}},\ and\
  \bibinfo {author} {\bibfnamefont {S.~L.}\ \bibnamefont {Cornish}},\
  }\bibfield  {title} {\bibinfo {title} {Controlling the rotational and
  hyperfine state of ultracold $^{87}\mathrm{Rb}^{133}\mathrm{Cs}$ molecules},\
  }\href {https://doi.org/10.1103/PhysRevA.94.041403} {\bibfield  {journal}
  {\bibinfo  {journal} {Phys. Rev. A}\ }\textbf {\bibinfo {volume} {94}},\
  \bibinfo {pages} {041403} (\bibinfo {year} {2016})}\BibitemShut {NoStop}%
\bibitem [{\citenamefont {Brooks}\ \emph {et~al.}(2021)\citenamefont {Brooks},
  \citenamefont {Spence}, \citenamefont {Guttridge}, \citenamefont
  {Alampounti}, \citenamefont {Rakonjac}, \citenamefont {McArd}, \citenamefont
  {Hutson},\ and\ \citenamefont {Cornish}}]{Brooks21}%
  \BibitemOpen
  \bibfield  {author} {\bibinfo {author} {\bibfnamefont {R.~V.}\ \bibnamefont
  {Brooks}}, \bibinfo {author} {\bibfnamefont {S.}~\bibnamefont {Spence}},
  \bibinfo {author} {\bibfnamefont {A.}~\bibnamefont {Guttridge}}, \bibinfo
  {author} {\bibfnamefont {A.}~\bibnamefont {Alampounti}}, \bibinfo {author}
  {\bibfnamefont {A.}~\bibnamefont {Rakonjac}}, \bibinfo {author}
  {\bibfnamefont {L.~A.}\ \bibnamefont {McArd}}, \bibinfo {author}
  {\bibfnamefont {J.~M.}\ \bibnamefont {Hutson}},\ and\ \bibinfo {author}
  {\bibfnamefont {S.~L.}\ \bibnamefont {Cornish}},\ }\bibfield  {title}
  {\bibinfo {title} {Preparation of one $^{87}${R}b and one $^{133}${C}s atom
  in a single optical tweezer},\ }\href
  {https://doi.org/10.1088/1367-2630/ac0000} {\bibfield  {journal} {\bibinfo
  {journal} {New J. Phys.}\ }\textbf {\bibinfo {volume} {23}},\ \bibinfo
  {pages} {065002} (\bibinfo {year} {2021})}\BibitemShut {NoStop}%
\bibitem [{\citenamefont {Kaufman}\ \emph {et~al.}(2012)\citenamefont
  {Kaufman}, \citenamefont {Lester},\ and\ \citenamefont
  {Regal}}]{Kaufman2012}%
  \BibitemOpen
  \bibfield  {author} {\bibinfo {author} {\bibfnamefont {A.~M.}\ \bibnamefont
  {Kaufman}}, \bibinfo {author} {\bibfnamefont {B.~J.}\ \bibnamefont
  {Lester}},\ and\ \bibinfo {author} {\bibfnamefont {C.~A.}\ \bibnamefont
  {Regal}},\ }\bibfield  {title} {\bibinfo {title} {Cooling a single atom in an
  optical tweezer to its quantum ground state},\ }\href
  {https://doi.org/10.1103/PhysRevX.2.041014} {\bibfield  {journal} {\bibinfo
  {journal} {Phys. Rev. X}\ }\textbf {\bibinfo {volume} {2}},\ \bibinfo {pages}
  {041014} (\bibinfo {year} {2012})}\BibitemShut {NoStop}%
\bibitem [{\citenamefont {Thompson}\ \emph {et~al.}(2013)\citenamefont
  {Thompson}, \citenamefont {Tiecke}, \citenamefont {Zibrov}, \citenamefont
  {Vuleti\ifmmode~\acute{c}\else \'{c}\fi{}},\ and\ \citenamefont
  {Lukin}}]{Thompson2013}%
  \BibitemOpen
  \bibfield  {author} {\bibinfo {author} {\bibfnamefont {J.~D.}\ \bibnamefont
  {Thompson}}, \bibinfo {author} {\bibfnamefont {T.~G.}\ \bibnamefont
  {Tiecke}}, \bibinfo {author} {\bibfnamefont {A.~S.}\ \bibnamefont {Zibrov}},
  \bibinfo {author} {\bibfnamefont {V.}~\bibnamefont
  {Vuleti\ifmmode~\acute{c}\else \'{c}\fi{}}},\ and\ \bibinfo {author}
  {\bibfnamefont {M.~D.}\ \bibnamefont {Lukin}},\ }\bibfield  {title} {\bibinfo
  {title} {Coherence and raman sideband cooling of a single atom in an optical
  tweezer},\ }\href {https://doi.org/10.1103/PhysRevLett.110.133001} {\bibfield
   {journal} {\bibinfo  {journal} {Phys. Rev. Lett.}\ }\textbf {\bibinfo
  {volume} {110}},\ \bibinfo {pages} {133001} (\bibinfo {year}
  {2013})}\BibitemShut {NoStop}%
\bibitem [{\citenamefont {Spence}\ \emph {et~al.}(2022)\citenamefont {Spence},
  \citenamefont {Brooks}, \citenamefont {Ruttley}, \citenamefont {Guttridge},\
  and\ \citenamefont {Cornish}}]{Spence22}%
  \BibitemOpen
  \bibfield  {author} {\bibinfo {author} {\bibfnamefont {S.}~\bibnamefont
  {Spence}}, \bibinfo {author} {\bibfnamefont {R.~V.}\ \bibnamefont {Brooks}},
  \bibinfo {author} {\bibfnamefont {D.~K.}\ \bibnamefont {Ruttley}}, \bibinfo
  {author} {\bibfnamefont {A.}~\bibnamefont {Guttridge}},\ and\ \bibinfo
  {author} {\bibfnamefont {S.~L.}\ \bibnamefont {Cornish}},\ }\bibfield
  {title} {\bibinfo {title} {Preparation of \(^{87}\){R}b and \(^{133}\){C}s in
  the motional ground state of a single optical tweezer},\ }\href
  {https://doi.org/10.1088/1367-2630/ac95b9} {\bibfield  {journal} {\bibinfo
  {journal} {New J. Phys.}\ }\textbf {\bibinfo {volume} {24}},\ \bibinfo
  {pages} {103022} (\bibinfo {year} {2022})}\BibitemShut {NoStop}%
\bibitem [{\citenamefont {Pilch}\ \emph {et~al.}(2009)\citenamefont {Pilch},
  \citenamefont {Lange}, \citenamefont {Prantner}, \citenamefont {Kerner},
  \citenamefont {Ferlaino}, \citenamefont {N\"agerl},\ and\ \citenamefont
  {Grimm}}]{Pilch2009}%
  \BibitemOpen
  \bibfield  {author} {\bibinfo {author} {\bibfnamefont {K.}~\bibnamefont
  {Pilch}}, \bibinfo {author} {\bibfnamefont {A.~D.}\ \bibnamefont {Lange}},
  \bibinfo {author} {\bibfnamefont {A.}~\bibnamefont {Prantner}}, \bibinfo
  {author} {\bibfnamefont {G.}~\bibnamefont {Kerner}}, \bibinfo {author}
  {\bibfnamefont {F.}~\bibnamefont {Ferlaino}}, \bibinfo {author}
  {\bibfnamefont {H.-C.}\ \bibnamefont {N\"agerl}},\ and\ \bibinfo {author}
  {\bibfnamefont {R.}~\bibnamefont {Grimm}},\ }\bibfield  {title} {\bibinfo
  {title} {Observation of interspecies {F}eshbach resonances in an ultracold
  {R}b-{C}s mixture},\ }\href {https://doi.org/10.1103/PhysRevA.79.042718}
  {\bibfield  {journal} {\bibinfo  {journal} {Phys. Rev. A}\ }\textbf {\bibinfo
  {volume} {79}},\ \bibinfo {pages} {042718} (\bibinfo {year}
  {2009})}\BibitemShut {NoStop}%
\bibitem [{\citenamefont {Takekoshi}\ \emph {et~al.}(2012)\citenamefont
  {Takekoshi}, \citenamefont {Debatin}, \citenamefont {Rameshan}, \citenamefont
  {Ferlaino}, \citenamefont {Grimm}, \citenamefont {N\"agerl}, \citenamefont
  {Le~Sueur}, \citenamefont {Hutson}, \citenamefont {Julienne}, \citenamefont
  {Kotochigova},\ and\ \citenamefont {Tiemann}}]{Takekoshi2012}%
  \BibitemOpen
  \bibfield  {author} {\bibinfo {author} {\bibfnamefont {T.}~\bibnamefont
  {Takekoshi}}, \bibinfo {author} {\bibfnamefont {M.}~\bibnamefont {Debatin}},
  \bibinfo {author} {\bibfnamefont {R.}~\bibnamefont {Rameshan}}, \bibinfo
  {author} {\bibfnamefont {F.}~\bibnamefont {Ferlaino}}, \bibinfo {author}
  {\bibfnamefont {R.}~\bibnamefont {Grimm}}, \bibinfo {author} {\bibfnamefont
  {H.-C.}\ \bibnamefont {N\"agerl}}, \bibinfo {author} {\bibfnamefont {C.~R.}\
  \bibnamefont {Le~Sueur}}, \bibinfo {author} {\bibfnamefont {J.~M.}\
  \bibnamefont {Hutson}}, \bibinfo {author} {\bibfnamefont {P.~S.}\
  \bibnamefont {Julienne}}, \bibinfo {author} {\bibfnamefont {S.}~\bibnamefont
  {Kotochigova}},\ and\ \bibinfo {author} {\bibfnamefont {E.}~\bibnamefont
  {Tiemann}},\ }\bibfield  {title} {\bibinfo {title} {Towards the production of
  ultracold ground-state {R}b{C}s molecules: {F}eshbach resonances, weakly
  bound states, and the coupled-channel model},\ }\href
  {https://doi.org/10.1103/PhysRevA.85.032506} {\bibfield  {journal} {\bibinfo
  {journal} {Phys. Rev. A}\ }\textbf {\bibinfo {volume} {85}},\ \bibinfo
  {pages} {032506} (\bibinfo {year} {2012})}\BibitemShut {NoStop}%
\bibitem [{\citenamefont {K{\"o}ppinger}\ \emph {et~al.}(2014)\citenamefont
  {K{\"o}ppinger}, \citenamefont {McCarron}, \citenamefont {Jenkin},
  \citenamefont {Molony}, \citenamefont {Cho}, \citenamefont {Cornish},
  \citenamefont {Le~Sueur}, \citenamefont {Blackley},\ and\ \citenamefont
  {Hutson}}]{Koeppinger2014}%
  \BibitemOpen
  \bibfield  {author} {\bibinfo {author} {\bibfnamefont {M.~P.}\ \bibnamefont
  {K{\"o}ppinger}}, \bibinfo {author} {\bibfnamefont {D.~J.}\ \bibnamefont
  {McCarron}}, \bibinfo {author} {\bibfnamefont {D.~L.}\ \bibnamefont
  {Jenkin}}, \bibinfo {author} {\bibfnamefont {P.~K.}\ \bibnamefont {Molony}},
  \bibinfo {author} {\bibfnamefont {H.-W.}\ \bibnamefont {Cho}}, \bibinfo
  {author} {\bibfnamefont {S.~L.}\ \bibnamefont {Cornish}}, \bibinfo {author}
  {\bibfnamefont {C.~R.}\ \bibnamefont {Le~Sueur}}, \bibinfo {author}
  {\bibfnamefont {C.~L.}\ \bibnamefont {Blackley}},\ and\ \bibinfo {author}
  {\bibfnamefont {J.~M.}\ \bibnamefont {Hutson}},\ }\bibfield  {title}
  {\bibinfo {title} {Production of optically trapped $^{87}${R}b{C}s {F}eshbach
  molecules},\ }\href {https://doi.org/10.1103/PhysRevA.89.033604} {\bibfield
  {journal} {\bibinfo  {journal} {Phys. Rev. A}\ }\textbf {\bibinfo {volume}
  {89}},\ \bibinfo {pages} {033604} (\bibinfo {year} {2014})}\BibitemShut
  {NoStop}%
\bibitem [{\citenamefont {Gregory}\ \emph {et~al.}(2015)\citenamefont
  {Gregory}, \citenamefont {Molony}, \citenamefont {K\"{o}ppinger},
  \citenamefont {Kumar}, \citenamefont {Ji}, \citenamefont {Lu}, \citenamefont
  {Marchant},\ and\ \citenamefont {Cornish}}]{Gregory2015}%
  \BibitemOpen
  \bibfield  {author} {\bibinfo {author} {\bibfnamefont {P.~D.}\ \bibnamefont
  {Gregory}}, \bibinfo {author} {\bibfnamefont {P.~K.}\ \bibnamefont {Molony}},
  \bibinfo {author} {\bibfnamefont {M.~P.}\ \bibnamefont {K\"{o}ppinger}},
  \bibinfo {author} {\bibfnamefont {A.}~\bibnamefont {Kumar}}, \bibinfo
  {author} {\bibfnamefont {Z.}~\bibnamefont {Ji}}, \bibinfo {author}
  {\bibfnamefont {B.}~\bibnamefont {Lu}}, \bibinfo {author} {\bibfnamefont
  {A.~L.}\ \bibnamefont {Marchant}},\ and\ \bibinfo {author} {\bibfnamefont
  {S.~L.}\ \bibnamefont {Cornish}},\ }\bibfield  {title} {\bibinfo {title} {A
  simple, versatile laser system for the creation of ultracold ground state
  molecules},\ }\href {https://doi.org/10.1088/1367-2630/17/5/055006}
  {\bibfield  {journal} {\bibinfo  {journal} {New J. Phys.}\ }\textbf {\bibinfo
  {volume} {17}},\ \bibinfo {pages} {055006} (\bibinfo {year}
  {2015})}\BibitemShut {NoStop}%
\bibitem [{\citenamefont {Molony}\ \emph
  {et~al.}(2016{\natexlab{a}})\citenamefont {Molony}, \citenamefont {Kumar},
  \citenamefont {Gregory}, \citenamefont {Kliese}, \citenamefont {Puppe},
  \citenamefont {Le~Sueur}, \citenamefont {Aldegunde}, \citenamefont {Hutson},\
  and\ \citenamefont {Cornish}}]{Molony2016_2}%
  \BibitemOpen
  \bibfield  {author} {\bibinfo {author} {\bibfnamefont {P.~K.}\ \bibnamefont
  {Molony}}, \bibinfo {author} {\bibfnamefont {A.}~\bibnamefont {Kumar}},
  \bibinfo {author} {\bibfnamefont {P.~D.}\ \bibnamefont {Gregory}}, \bibinfo
  {author} {\bibfnamefont {R.}~\bibnamefont {Kliese}}, \bibinfo {author}
  {\bibfnamefont {T.}~\bibnamefont {Puppe}}, \bibinfo {author} {\bibfnamefont
  {C.~R.}\ \bibnamefont {Le~Sueur}}, \bibinfo {author} {\bibfnamefont
  {J.}~\bibnamefont {Aldegunde}}, \bibinfo {author} {\bibfnamefont {J.~M.}\
  \bibnamefont {Hutson}},\ and\ \bibinfo {author} {\bibfnamefont {S.~L.}\
  \bibnamefont {Cornish}},\ }\bibfield  {title} {\bibinfo {title} {Measurement
  of the binding energy of ultracold $^{87}\mathrm{Rb}^{133}\mathrm{Cs}$
  molecules using an offset-free optical frequency comb},\ }\href
  {https://doi.org/10.1103/PhysRevA.94.022507} {\bibfield  {journal} {\bibinfo
  {journal} {Phys. Rev. A}\ }\textbf {\bibinfo {volume} {94}},\ \bibinfo
  {pages} {022507} (\bibinfo {year} {2016}{\natexlab{a}})}\BibitemShut
  {NoStop}%
\bibitem [{\citenamefont {Takekoshi}\ \emph {et~al.}(2014)\citenamefont
  {Takekoshi}, \citenamefont {Reichs\"{o}llner}, \citenamefont {Schindewolf},
  \citenamefont {Hutson}, \citenamefont {Le~Sueur}, \citenamefont {Dulieu},
  \citenamefont {Ferlaino}, \citenamefont {Grimm},\ and\ \citenamefont
  {N{\"a}gerl}}]{Takekoshi2014}%
  \BibitemOpen
  \bibfield  {author} {\bibinfo {author} {\bibfnamefont {T.}~\bibnamefont
  {Takekoshi}}, \bibinfo {author} {\bibfnamefont {L.}~\bibnamefont
  {Reichs\"{o}llner}}, \bibinfo {author} {\bibfnamefont {A.}~\bibnamefont
  {Schindewolf}}, \bibinfo {author} {\bibfnamefont {J.~M.}\ \bibnamefont
  {Hutson}}, \bibinfo {author} {\bibfnamefont {C.~R.}\ \bibnamefont
  {Le~Sueur}}, \bibinfo {author} {\bibfnamefont {O.}~\bibnamefont {Dulieu}},
  \bibinfo {author} {\bibfnamefont {F.}~\bibnamefont {Ferlaino}}, \bibinfo
  {author} {\bibfnamefont {R.}~\bibnamefont {Grimm}},\ and\ \bibinfo {author}
  {\bibfnamefont {H.-C.}\ \bibnamefont {N{\"a}gerl}},\ }\bibfield  {title}
  {\bibinfo {title} {Ultracold dense samples of dipolar {R}b{C}s molecules in
  the rovibrational and hyperfine ground state},\ }\href
  {https://doi.org/10.1103/physrevlett.113.205301} {\bibfield  {journal}
  {\bibinfo  {journal} {Phys. Rev. Lett.}\ }\textbf {\bibinfo {volume} {113}},\
  \bibinfo {pages} {205301} (\bibinfo {year} {2014})}\BibitemShut {NoStop}%
\bibitem [{\citenamefont {Bergmann}\ \emph {et~al.}(1998)\citenamefont
  {Bergmann}, \citenamefont {Theuer},\ and\ \citenamefont
  {Shore}}]{Bergmann1998}%
  \BibitemOpen
  \bibfield  {author} {\bibinfo {author} {\bibfnamefont {K.}~\bibnamefont
  {Bergmann}}, \bibinfo {author} {\bibfnamefont {H.}~\bibnamefont {Theuer}},\
  and\ \bibinfo {author} {\bibfnamefont {B.}~\bibnamefont {Shore}},\ }\bibfield
   {title} {\bibinfo {title} {Coherent population transfer among quantum states
  of atoms and molecules},\ }\href {https://doi.org/10.1103/RevModPhys.70.1003}
  {\bibfield  {journal} {\bibinfo  {journal} {Rev. Mod. Phys.}\ }\textbf
  {\bibinfo {volume} {70}},\ \bibinfo {pages} {1003} (\bibinfo {year}
  {1998})}\BibitemShut {NoStop}%
\bibitem [{\citenamefont {Vitanov}\ \emph {et~al.}(2017)\citenamefont
  {Vitanov}, \citenamefont {Rangelov}, \citenamefont {Shore},\ and\
  \citenamefont {Bergmann}}]{Vitanov2017}%
  \BibitemOpen
  \bibfield  {author} {\bibinfo {author} {\bibfnamefont {N.~V.}\ \bibnamefont
  {Vitanov}}, \bibinfo {author} {\bibfnamefont {A.~A.}\ \bibnamefont
  {Rangelov}}, \bibinfo {author} {\bibfnamefont {B.~W.}\ \bibnamefont
  {Shore}},\ and\ \bibinfo {author} {\bibfnamefont {K.}~\bibnamefont
  {Bergmann}},\ }\bibfield  {title} {\bibinfo {title} {Stimulated {R}aman
  adiabatic passage in physics, chemistry, and beyond},\ }\href
  {https://doi.org/10.1103/RevModPhys.89.015006} {\bibfield  {journal}
  {\bibinfo  {journal} {Rev. Mod. Phys.}\ }\textbf {\bibinfo {volume} {89}},\
  \bibinfo {pages} {015006} (\bibinfo {year} {2017})}\BibitemShut {NoStop}%
\bibitem [{\citenamefont {Molony}\ \emph
  {et~al.}(2016{\natexlab{b}})\citenamefont {Molony}, \citenamefont {Gregory},
  \citenamefont {Kumar}, \citenamefont {{Le Sueur}}, \citenamefont {Hutson},\
  and\ \citenamefont {Cornish}}]{Molony2016}%
  \BibitemOpen
  \bibfield  {author} {\bibinfo {author} {\bibfnamefont {P.~K.}\ \bibnamefont
  {Molony}}, \bibinfo {author} {\bibfnamefont {P.~D.}\ \bibnamefont {Gregory}},
  \bibinfo {author} {\bibfnamefont {A.}~\bibnamefont {Kumar}}, \bibinfo
  {author} {\bibfnamefont {C.~R.}\ \bibnamefont {{Le Sueur}}}, \bibinfo
  {author} {\bibfnamefont {J.~M.}\ \bibnamefont {Hutson}},\ and\ \bibinfo
  {author} {\bibfnamefont {S.~L.}\ \bibnamefont {Cornish}},\ }\bibfield
  {title} {\bibinfo {title} {Production of ultracold $^{87}
  \mathrm{Rb}$$^{133}\mathrm{Cs}$ in the absolute ground state : {Complete}
  characterisation of the {STIRAP} transfer},\ }\href
  {https://doi.org/10.1002/cphc.201600501} {\bibfield  {journal} {\bibinfo
  {journal} {{ChemPhysChem}}\ }\textbf {\bibinfo {volume} {17}},\ \bibinfo
  {pages} {3811} (\bibinfo {year} {2016}{\natexlab{b}})}\BibitemShut {NoStop}%
\bibitem [{\citenamefont {Gregory}\ \emph
  {et~al.}(2021{\natexlab{b}})\citenamefont {Gregory}, \citenamefont
  {Blackmore}, \citenamefont {D}, \citenamefont {Fernley}, \citenamefont
  {Bromley}, \citenamefont {Hutson},\ and\ \citenamefont
  {Cornish}}]{Gregory2021a}%
  \BibitemOpen
  \bibfield  {author} {\bibinfo {author} {\bibfnamefont {P.~D.}\ \bibnamefont
  {Gregory}}, \bibinfo {author} {\bibfnamefont {J.~A.}\ \bibnamefont
  {Blackmore}}, \bibinfo {author} {\bibfnamefont {F.~M.}\ \bibnamefont {D}},
  \bibinfo {author} {\bibfnamefont {L.~M.}\ \bibnamefont {Fernley}}, \bibinfo
  {author} {\bibfnamefont {S.~L.}\ \bibnamefont {Bromley}}, \bibinfo {author}
  {\bibfnamefont {J.~M.}\ \bibnamefont {Hutson}},\ and\ \bibinfo {author}
  {\bibfnamefont {S.~L.}\ \bibnamefont {Cornish}},\ }\bibfield  {title}
  {\bibinfo {title} {Molecule–molecule and atom–molecule collisions with
  ultracold {RbCs} molecules},\ }\href
  {https://doi.org/10.1088/1367-2630/ac3c63} {\bibfield  {journal} {\bibinfo
  {journal} {New J. Phys.}\ }\textbf {\bibinfo {volume} {23}},\ \bibinfo
  {pages} {125004} (\bibinfo {year} {2021}{\natexlab{b}})}\BibitemShut
  {NoStop}%
\bibitem [{\citenamefont {Barakhshan}\ \emph {et~al.}(2022)\citenamefont
  {Barakhshan}, \citenamefont {Marrs}, \citenamefont {Bhosale}, \citenamefont
  {Arora}, \citenamefont {Eigenmann},\ and\ \citenamefont
  {Safronova}}]{UDportal}%
  \BibitemOpen
  \bibfield  {author} {\bibinfo {author} {\bibfnamefont {P.}~\bibnamefont
  {Barakhshan}}, \bibinfo {author} {\bibfnamefont {A.}~\bibnamefont {Marrs}},
  \bibinfo {author} {\bibfnamefont {A.}~\bibnamefont {Bhosale}}, \bibinfo
  {author} {\bibfnamefont {B.}~\bibnamefont {Arora}}, \bibinfo {author}
  {\bibfnamefont {R.}~\bibnamefont {Eigenmann}},\ and\ \bibinfo {author}
  {\bibfnamefont {M.~S.}\ \bibnamefont {Safronova}},\ }\href
  {https://www.udel.edu/atom} {\bibinfo {title} {Portal for high-precision
  atomic data and computation (version 2.0)}},\ \bibinfo {howpublished}
  {[Online]} (\bibinfo {year} {2022})\BibitemShut {NoStop}%
\bibitem [{\citenamefont {Vexiau}\ \emph {et~al.}(2017)\citenamefont {Vexiau},
  \citenamefont {Borsalino}, \citenamefont {Lepers}, \citenamefont {Orbán},
  \citenamefont {Aymar}, \citenamefont {Dulieu},\ and\ \citenamefont
  {Bouloufa-Maafa}}]{Vexiau2017}%
  \BibitemOpen
  \bibfield  {author} {\bibinfo {author} {\bibfnamefont {R.}~\bibnamefont
  {Vexiau}}, \bibinfo {author} {\bibfnamefont {D.}~\bibnamefont {Borsalino}},
  \bibinfo {author} {\bibfnamefont {M.}~\bibnamefont {Lepers}}, \bibinfo
  {author} {\bibfnamefont {A.}~\bibnamefont {Orbán}}, \bibinfo {author}
  {\bibfnamefont {M.}~\bibnamefont {Aymar}}, \bibinfo {author} {\bibfnamefont
  {O.}~\bibnamefont {Dulieu}},\ and\ \bibinfo {author} {\bibfnamefont
  {N.}~\bibnamefont {Bouloufa-Maafa}},\ }\bibfield  {title} {\bibinfo {title}
  {Dynamic dipole polarizabilities of heteronuclear alkali dimers: optical
  response, trapping and control of ultracold molecules},\ }\href
  {https://doi.org/10.1080/0144235X.2017.1351821} {\bibfield  {journal}
  {\bibinfo  {journal} {Int. Rev. Phys. Chem.}\ }\textbf {\bibinfo {volume}
  {36}},\ \bibinfo {pages} {709} (\bibinfo {year} {2017})}\BibitemShut
  {NoStop}%
\bibitem [{\citenamefont {Olaya}\ \emph {et~al.}(2020)\citenamefont {Olaya},
  \citenamefont {P\'erez-R\'{\i}os},\ and\ \citenamefont
  {Herrera}}]{Olaya2020}%
  \BibitemOpen
  \bibfield  {author} {\bibinfo {author} {\bibfnamefont {V.}~\bibnamefont
  {Olaya}}, \bibinfo {author} {\bibfnamefont {J.}~\bibnamefont
  {P\'erez-R\'{\i}os}},\ and\ \bibinfo {author} {\bibfnamefont
  {F.}~\bibnamefont {Herrera}},\ }\bibfield  {title} {\bibinfo {title}
  {${C}_{6}$ coefficients for interacting {R}ydberg atoms and alkali-metal
  dimers},\ }\href {https://doi.org/10.1103/PhysRevA.101.032705} {\bibfield
  {journal} {\bibinfo  {journal} {Phys. Rev. A}\ }\textbf {\bibinfo {volume}
  {101}},\ \bibinfo {pages} {032705} (\bibinfo {year} {2020})}\BibitemShut
  {NoStop}%
\bibitem [{\citenamefont {Kuznetsova}\ \emph {et~al.}(2018)\citenamefont
  {Kuznetsova}, \citenamefont {Rittenhouse}, \citenamefont {Beterov},
  \citenamefont {Scully}, \citenamefont {Yelin},\ and\ \citenamefont
  {Sadeghpour}}]{Kuznetsova2018}%
  \BibitemOpen
  \bibfield  {author} {\bibinfo {author} {\bibfnamefont {E.}~\bibnamefont
  {Kuznetsova}}, \bibinfo {author} {\bibfnamefont {S.~T.}\ \bibnamefont
  {Rittenhouse}}, \bibinfo {author} {\bibfnamefont {I.~I.}\ \bibnamefont
  {Beterov}}, \bibinfo {author} {\bibfnamefont {M.~O.}\ \bibnamefont {Scully}},
  \bibinfo {author} {\bibfnamefont {S.~F.}\ \bibnamefont {Yelin}},\ and\
  \bibinfo {author} {\bibfnamefont {H.~R.}\ \bibnamefont {Sadeghpour}},\
  }\bibfield  {title} {\bibinfo {title} {Effective spin-spin interactions in
  bilayers of {R}ydberg atoms and polar molecules},\ }\href
  {https://doi.org/10.1103/PhysRevA.98.043609} {\bibfield  {journal} {\bibinfo
  {journal} {Phys. Rev. A}\ }\textbf {\bibinfo {volume} {98}},\ \bibinfo
  {pages} {043609} (\bibinfo {year} {2018})}\BibitemShut {NoStop}%
\bibitem [{\citenamefont {Gonz\'alez-F\'erez}\ \emph
  {et~al.}(2015)\citenamefont {Gonz\'alez-F\'erez}, \citenamefont
  {Sadeghpour},\ and\ \citenamefont {Schmelcher}}]{gonzalez15}%
  \BibitemOpen
  \bibfield  {author} {\bibinfo {author} {\bibfnamefont {R.}~\bibnamefont
  {Gonz\'alez-F\'erez}}, \bibinfo {author} {\bibfnamefont {H.~R.}\ \bibnamefont
  {Sadeghpour}},\ and\ \bibinfo {author} {\bibfnamefont {P.}~\bibnamefont
  {Schmelcher}},\ }\bibfield  {title} {\bibinfo {title} {Rotational
  hybridization, and control of alignment and orientation in triatomic
  ultralong-range {R}ydberg molecules},\ }\href
  {https://doi.org/10.1088/1367-2630/17/1/013021} {\bibfield  {journal}
  {\bibinfo  {journal} {New J. Phys.}\ }\textbf {\bibinfo {volume} {17}},\
  \bibinfo {pages} {013021} (\bibinfo {year} {2015})}\BibitemShut {NoStop}%
\bibitem [{\citenamefont {Omont}(1977)}]{Omont}%
  \BibitemOpen
  \bibfield  {author} {\bibinfo {author} {\bibfnamefont {A.}~\bibnamefont
  {Omont}},\ }\bibfield  {title} {\bibinfo {title} {On the theory of collisions
  of atoms in {R}ydberg states with neutral particles},\ }\href
  {https://doi.org/10.1051/jphys:0197700380110134300} {\bibfield  {journal}
  {\bibinfo  {journal} {J. Phys. France}\ }\textbf {\bibinfo {volume} {38}},\
  \bibinfo {pages} {1343} (\bibinfo {year} {1977})}\BibitemShut {NoStop}%
\bibitem [{\citenamefont {Fano}\ and\ \citenamefont {Rau}(1986)}]{FANO198661}%
  \BibitemOpen
  \bibfield  {author} {\bibinfo {author} {\bibfnamefont {U.}~\bibnamefont
  {Fano}}\ and\ \bibinfo {author} {\bibfnamefont {A.}~\bibnamefont {Rau}},\
  }\bibfield  {title} {\bibinfo {title} {4 - elastic scattering by a
  short-range central potential},\ }in\ \href
  {https://doi.org/https://doi.org/10.1016/B978-0-12-248460-5.50007-1} {\emph
  {\bibinfo {booktitle} {Atomic Collisions and Spectra}}},\ \bibinfo {editor}
  {edited by\ \bibinfo {editor} {\bibfnamefont {U.}~\bibnamefont {Fano}}\ and\
  \bibinfo {editor} {\bibfnamefont {A.}~\bibnamefont {Rau}}}\ (\bibinfo
  {publisher} {Academic Press},\ \bibinfo {year} {1986})\ pp.\ \bibinfo {pages}
  {61--80}\BibitemShut {NoStop}%
\bibitem [{\citenamefont {Eaton}\ \emph {et~al.}(1992)\citenamefont {Eaton},
  \citenamefont {Sarkas}, \citenamefont {Arnold}, \citenamefont {McHugh},\ and\
  \citenamefont {Bowen}}]{EATON1992141}%
  \BibitemOpen
  \bibfield  {author} {\bibinfo {author} {\bibfnamefont {J.}~\bibnamefont
  {Eaton}}, \bibinfo {author} {\bibfnamefont {H.}~\bibnamefont {Sarkas}},
  \bibinfo {author} {\bibfnamefont {S.}~\bibnamefont {Arnold}}, \bibinfo
  {author} {\bibfnamefont {K.}~\bibnamefont {McHugh}},\ and\ \bibinfo {author}
  {\bibfnamefont {K.}~\bibnamefont {Bowen}},\ }\bibfield  {title} {\bibinfo
  {title} {Negative ion photoelectron spectroscopy of the heteronuclear
  alkali-metal dimer and trimer anions: {N}a{K}$^-$, {KRb}$^-$, {RbCs}$^-$,
  {KCs}$^-$, {Na}$_2${K}$^-$, and {K}$_2${Cs}$^-$},\ }\href
  {https://doi.org/10.1016/0009-2614(92)85697-9} {\bibfield  {journal}
  {\bibinfo  {journal} {Chem. Phys. Lett.}\ }\textbf {\bibinfo {volume}
  {193}},\ \bibinfo {pages} {141} (\bibinfo {year} {1992})}\BibitemShut
  {NoStop}%
\bibitem [{\citenamefont {Marinescu}\ \emph {et~al.}(1994)\citenamefont
  {Marinescu}, \citenamefont {Sadeghpour},\ and\ \citenamefont
  {Dalgarno}}]{marinescu94}%
  \BibitemOpen
  \bibfield  {author} {\bibinfo {author} {\bibfnamefont {M.}~\bibnamefont
  {Marinescu}}, \bibinfo {author} {\bibfnamefont {H.~R.}\ \bibnamefont
  {Sadeghpour}},\ and\ \bibinfo {author} {\bibfnamefont {A.}~\bibnamefont
  {Dalgarno}},\ }\bibfield  {title} {\bibinfo {title} {Dispersion coefficients
  for alkali-metal dimers},\ }\href {https://doi.org/10.1103/PhysRevA.49.982}
  {\bibfield  {journal} {\bibinfo  {journal} {Phys. Rev. A}\ }\textbf {\bibinfo
  {volume} {49}},\ \bibinfo {pages} {982} (\bibinfo {year} {1994})}\BibitemShut
  {NoStop}%
\bibitem [{\citenamefont {Black}(2001)}]{Black2001}%
  \BibitemOpen
  \bibfield  {author} {\bibinfo {author} {\bibfnamefont {E.~D.}\ \bibnamefont
  {Black}},\ }\bibfield  {title} {\bibinfo {title} {An introduction to
  {P}ound–{D}rever–{H}all laser frequency stabilization},\ }\href
  {https://doi.org/10.1119/1.1286663} {\bibfield  {journal} {\bibinfo
  {journal} {Am. J. Phys.}\ }\textbf {\bibinfo {volume} {69}},\ \bibinfo
  {pages} {79} (\bibinfo {year} {2001})}\BibitemShut {NoStop}%
\bibitem [{\citenamefont {Weber}\ \emph {et~al.}(2017)\citenamefont {Weber},
  \citenamefont {Tresp}, \citenamefont {Menke}, \citenamefont {Urvoy},
  \citenamefont {Firstenberg}, \citenamefont {Büchler},\ and\ \citenamefont
  {Hofferberth}}]{Weber2017}%
  \BibitemOpen
  \bibfield  {author} {\bibinfo {author} {\bibfnamefont {S.}~\bibnamefont
  {Weber}}, \bibinfo {author} {\bibfnamefont {C.}~\bibnamefont {Tresp}},
  \bibinfo {author} {\bibfnamefont {H.}~\bibnamefont {Menke}}, \bibinfo
  {author} {\bibfnamefont {A.}~\bibnamefont {Urvoy}}, \bibinfo {author}
  {\bibfnamefont {O.}~\bibnamefont {Firstenberg}}, \bibinfo {author}
  {\bibfnamefont {H.~P.}\ \bibnamefont {Büchler}},\ and\ \bibinfo {author}
  {\bibfnamefont {S.}~\bibnamefont {Hofferberth}},\ }\bibfield  {title}
  {\bibinfo {title} {Calculation of {R}ydberg interaction potentials},\ }\href
  {https://doi.org/10.1088/1361-6455/aa743a} {\bibfield  {journal} {\bibinfo
  {journal} {J. Phys. B}\ }\textbf {\bibinfo {volume} {50}},\ \bibinfo {pages}
  {133001} (\bibinfo {year} {2017})}\BibitemShut {NoStop}%
\bibitem [{\citenamefont {{{\v{S}}}ibali{\'{c}}}\ \emph {et~al.}(2017)\citenamefont
  {{{\v{S}}}ibali{\'{c}}}, \citenamefont {Pritchard}, \citenamefont {Adams},\ and\
  \citenamefont {Weatherill}}]{sibalic2017}%
  \BibitemOpen
  \bibfield  {author} {\bibinfo {author} {\bibfnamefont {N.}~\bibnamefont
  {Šibalić}}, \bibinfo {author} {\bibfnamefont {J.~D.}\ \bibnamefont
  {Pritchard}}, \bibinfo {author} {\bibfnamefont {C.~S.}\ \bibnamefont
  {Adams}},\ and\ \bibinfo {author} {\bibfnamefont {K.~J.}\ \bibnamefont
  {Weatherill}},\ }\bibfield  {title} {\bibinfo {title} {{ARC}: An open-source
  library for calculating properties of alkali {R}ydberg atoms},\ }\href
  {https://doi.org/10.1016/j.cpc.2017.06.015} {\bibfield  {journal} {\bibinfo
  {journal} {Comput. Phys. Commun.}\ }\textbf {\bibinfo {volume} {220}},\
  \bibinfo {pages} {319} (\bibinfo {year} {2017})}\BibitemShut {NoStop}%
\bibitem [{\citenamefont {de~L\'es\'eleuc}\ \emph {et~al.}(2018)\citenamefont
  {de~L\'es\'eleuc}, \citenamefont {Barredo}, \citenamefont {Lienhard},
  \citenamefont {Browaeys},\ and\ \citenamefont {Lahaye}}]{Leseleuc2018}%
  \BibitemOpen
  \bibfield  {author} {\bibinfo {author} {\bibfnamefont {S.}~\bibnamefont
  {de~L\'es\'eleuc}}, \bibinfo {author} {\bibfnamefont {D.}~\bibnamefont
  {Barredo}}, \bibinfo {author} {\bibfnamefont {V.}~\bibnamefont {Lienhard}},
  \bibinfo {author} {\bibfnamefont {A.}~\bibnamefont {Browaeys}},\ and\
  \bibinfo {author} {\bibfnamefont {T.}~\bibnamefont {Lahaye}},\ }\bibfield
  {title} {\bibinfo {title} {Analysis of imperfections in the coherent optical
  excitation of single atoms to {R}ydberg states},\ }\href
  {https://doi.org/10.1103/PhysRevA.97.053803} {\bibfield  {journal} {\bibinfo
  {journal} {Phys. Rev. A}\ }\textbf {\bibinfo {volume} {97}},\ \bibinfo
  {pages} {053803} (\bibinfo {year} {2018})}\BibitemShut {NoStop}%
\bibitem [{\citenamefont {Levine}\ \emph {et~al.}(2018)\citenamefont {Levine},
  \citenamefont {Keesling}, \citenamefont {Omran}, \citenamefont {Bernien},
  \citenamefont {Schwartz}, \citenamefont {Zibrov}, \citenamefont {Endres},
  \citenamefont {Greiner}, \citenamefont {Vuleti\ifmmode~\acute{c}\else
  \'{c}\fi{}},\ and\ \citenamefont {Lukin}}]{Levine2018}%
  \BibitemOpen
  \bibfield  {author} {\bibinfo {author} {\bibfnamefont {H.}~\bibnamefont
  {Levine}}, \bibinfo {author} {\bibfnamefont {A.}~\bibnamefont {Keesling}},
  \bibinfo {author} {\bibfnamefont {A.}~\bibnamefont {Omran}}, \bibinfo
  {author} {\bibfnamefont {H.}~\bibnamefont {Bernien}}, \bibinfo {author}
  {\bibfnamefont {S.}~\bibnamefont {Schwartz}}, \bibinfo {author}
  {\bibfnamefont {A.~S.}\ \bibnamefont {Zibrov}}, \bibinfo {author}
  {\bibfnamefont {M.}~\bibnamefont {Endres}}, \bibinfo {author} {\bibfnamefont
  {M.}~\bibnamefont {Greiner}}, \bibinfo {author} {\bibfnamefont
  {V.}~\bibnamefont {Vuleti\ifmmode~\acute{c}\else \'{c}\fi{}}},\ and\ \bibinfo
  {author} {\bibfnamefont {M.~D.}\ \bibnamefont {Lukin}},\ }\bibfield  {title}
  {\bibinfo {title} {High-fidelity control and entanglement of {R}ydberg-atom
  qubits},\ }\href {https://doi.org/10.1103/PhysRevLett.121.123603} {\bibfield
  {journal} {\bibinfo  {journal} {Phys. Rev. Lett.}\ }\textbf {\bibinfo
  {volume} {121}},\ \bibinfo {pages} {123603} (\bibinfo {year}
  {2018})}\BibitemShut {NoStop}%
\bibitem [{\citenamefont {Zhang}\ \emph {et~al.}(2020)\citenamefont {Zhang},
  \citenamefont {Yu}, \citenamefont {Cairncross}, \citenamefont {Wang},
  \citenamefont {Picard}, \citenamefont {Hood}, \citenamefont {Lin},
  \citenamefont {Hutson},\ and\ \citenamefont {Ni}}]{Zhang2020}%
  \BibitemOpen
  \bibfield  {author} {\bibinfo {author} {\bibfnamefont {J.~T.}\ \bibnamefont
  {Zhang}}, \bibinfo {author} {\bibfnamefont {Y.}~\bibnamefont {Yu}}, \bibinfo
  {author} {\bibfnamefont {W.~B.}\ \bibnamefont {Cairncross}}, \bibinfo
  {author} {\bibfnamefont {K.}~\bibnamefont {Wang}}, \bibinfo {author}
  {\bibfnamefont {L.~R.~B.}\ \bibnamefont {Picard}}, \bibinfo {author}
  {\bibfnamefont {J.~D.}\ \bibnamefont {Hood}}, \bibinfo {author}
  {\bibfnamefont {Y.-W.}\ \bibnamefont {Lin}}, \bibinfo {author} {\bibfnamefont
  {J.~M.}\ \bibnamefont {Hutson}},\ and\ \bibinfo {author} {\bibfnamefont
  {K.-K.}\ \bibnamefont {Ni}},\ }\bibfield  {title} {\bibinfo {title} {Forming
  a single molecule by magnetoassociation in an optical tweezer},\ }\href
  {https://doi.org/10.1103/PhysRevLett.124.253401} {\bibfield  {journal}
  {\bibinfo  {journal} {Phys. Rev. Lett.}\ }\textbf {\bibinfo {volume} {124}},\
  \bibinfo {pages} {253401} (\bibinfo {year} {2020})}\BibitemShut {NoStop}%
\bibitem [{\citenamefont {Blackmore}\ \emph {et~al.}(2020)\citenamefont
  {Blackmore}, \citenamefont {Sawant}, \citenamefont {Gregory}, \citenamefont
  {Bromley}, \citenamefont {Aldegunde}, \citenamefont {Hutson},\ and\
  \citenamefont {Cornish}}]{Blackmore2020}%
  \BibitemOpen
  \bibfield  {author} {\bibinfo {author} {\bibfnamefont {J.~A.}\ \bibnamefont
  {Blackmore}}, \bibinfo {author} {\bibfnamefont {R.}~\bibnamefont {Sawant}},
  \bibinfo {author} {\bibfnamefont {P.~D.}\ \bibnamefont {Gregory}}, \bibinfo
  {author} {\bibfnamefont {S.~L.}\ \bibnamefont {Bromley}}, \bibinfo {author}
  {\bibfnamefont {J.}~\bibnamefont {Aldegunde}}, \bibinfo {author}
  {\bibfnamefont {J.~M.}\ \bibnamefont {Hutson}},\ and\ \bibinfo {author}
  {\bibfnamefont {S.~L.}\ \bibnamefont {Cornish}},\ }\bibfield  {title}
  {\bibinfo {title} {Controlling the ac {Stark} effect of {RbCs} with dc
  electric and magnetic fields},\ }\href
  {https://doi.org/10.1103/PhysRevA.102.053316} {\bibfield  {journal} {\bibinfo
   {journal} {Phys. Rev. A}\ }\textbf {\bibinfo {volume} {102}},\ \bibinfo
  {pages} {053316} (\bibinfo {year} {2020})}\BibitemShut {NoStop}%
\end{thebibliography}
%

\ifarXiv
    

\section*{Supplementary material}

# Supplementary Material:
# Observation of Rydberg blockade due to the charge-dipole interaction between an atom and a polar molecule

Alexander Guttridge,[1,2,*] Daniel K. Ruttley,[1,2,*] Archie C. Baldock,[1] Rosario González-Férez,[3] H. R. Sadeghpour,[4] C. S. Adams,[1,2,†] and Simon L. Cornish[1,2,‡]

[1]*Department of Physics, Durham University, South Road, Durham, DH1 3LE, United Kingdom*
[2]*Joint Quantum Centre (JQC) Durham-Newcastle, Durham University, South Road, Durham, DH1 3LE, United Kingdom*
[3]*Instituto Carlos I de Física Teórica y Computacional, and Departamento de Física Atómica, Molecular y Nuclear, Universidad de Granada, 18071 Granada, Spain*
[4]*ITAMP, Center for Astrophysics | Harvard & Smithsonian, Cambridge, MA 02138 USA*

## RYDBERG ATOM-POLAR MOLECULE INTERACTION

The Born-Oppenheimer Hamiltonian at large separations between the Rb atom and the RbCs molecule, reads as [1, 2]

$$H = H_{\rm R}(\mathbf{r}) + V_{\rm cd}(\mathbf{r}, \mathbf{R}_{\rm am}) + V_{\rm em}(\mathbf{r}, \mathbf{R}_{\rm am}) \tag{1}$$

where $\mathbf{r}$ and $\mathbf{R}_{\rm am}$ are the positions of the Rydberg electron and diatomic molecule with respect to the atomic core $\rm Rb^+$, respectively. $H_{\rm R}$ represents the single electron Hamiltonian describing the Rydberg atom

$$H_{\rm R}(\mathbf{r}) = -\frac{\hbar^2}{2m_e}\nabla_r^2 + V_l(r) \tag{2}$$

with $V_l(r)$ being the $l$-dependent model potential [3], where $l$ is the angular momentum of the Rydberg electron. The second term includes the rotational energy and the Rydberg electron-RbCs dipole interaction, in the Born-Oppenheimer and rigid rotor approximations,

$$V_{\rm cd} = B\mathbf{N}^2 - \mathbf{d} \cdot \mathbf{F}(\mathbf{r}, \mathbf{R}_{\rm am}) \tag{3}$$

with $B$ being the rotational constant, $\mathbf{N}$ the molecular angular momentum operator, and $\mathbf{d}$ the permanent electric dipole moments of the diatomic molecule. The internal electric field due to the Rydberg electron, at the position of RbCs, $\mathbf{R}_{\rm am}$, is

$$\mathbf{F}(\mathbf{r}, \mathbf{R}_{\rm am}) = e\frac{\mathbf{r} - \mathbf{R}_{\rm am}}{|\mathbf{r} - \mathbf{R}_{\rm am}|^3} + e\frac{\mathbf{R}_{\rm am}}{\mathrm{R}_{\rm am}^3}, \tag{4}$$

the first term is due to the Rydberg electron and the second one to the atomic core $\rm Rb^+$, with $e$ being the electron charge. The charge-dipole term allows for scattering of electrons for diatomic molecules possessing permanent electric dipole moments less than a critical value $d_{\rm cr} = 1.639$ D [4]; at supercritical dipole moments, the Rydberg electron could bind.

The last term in the adiabatic Hamiltonian in Eq. (1) represents the scattering of the Rydberg electron from RbCs and is approximated as the Fermi pseudopotential [5]

$$V_{\rm em} = 2\pi a_S(k)\delta\left(\mathbf{r} - \mathbf{R}_{\rm am}\right), \tag{5}$$

accounting for the contribution to low-energy electron-molecule scattering in the $L = 0$ (S-wave) partial wave. The S-wave scattering length is $a_{\rm S}(k) = -\tan(\delta_{\rm S}(k))/k$, with $\delta_{\rm S}(k)$ being the $L = 0$ scattering phase shift at momentum $k$. The electron scattering from RbCs($\rm X^1\Sigma$) can form negative ions, and the electron affinity (EA) of RbCs, $E_A = 0.478 \pm 0.020$ eV [6], is used to estimate the S-wave scattering length. Ab initio calculations of electron scattering length in collision with molecules are notoriously difficult to converge at extreme low collision energies. When the electron affinity of the molecule is small, a reasonable measure of the size of the scattering length can be made using $E_A = -\hbar^2/(2ma_S^2)$, where $m$ is the electron mass [7]. This scattering becomes important when the molecule enters the Rydberg orbit. At the separations considered here, the dominant interaction is the electron-dipole interaction.

The Schrödinger equation associated with the Hamiltonian in Eq. (1) is solved by expanding in a basis set,

$$\Psi(\mathbf{r}, \Omega_d; \mathbf{R}_{\rm am}) = \sum_{n,l,N,J} C_{nlNJ}(\mathbf{R}_{\rm am})\Psi_{nlNJM_J}(\mathbf{r}, \Omega_d), \tag{6}$$

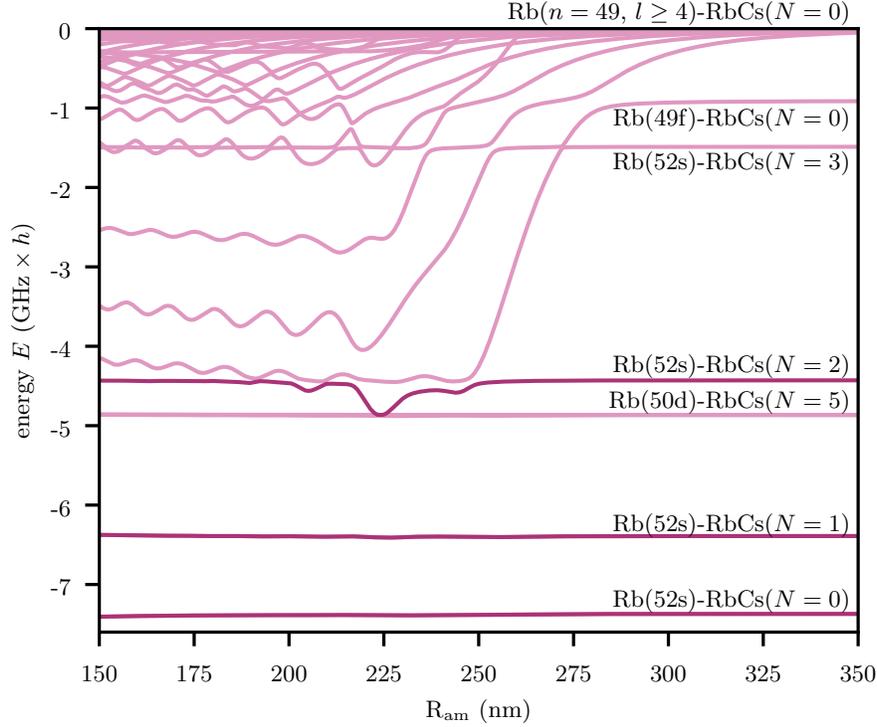

FIG. S1. Born-Oppenheimer electronic potential energy curves for the Rydberg polyatomic molecule, Rb-RbCs($M_J = 0$) evolving from the degenerate Rydberg Rb($n = 49, l \geq 4$)-RbCs($N = 0$) manifold. The other dissociation limits are: Rb(52s)-RbCs($N$), with $N = 0, 1, 2, 3$, Rb(50d)-RbCs($N = 5$), and Rb(49f)-RbCs($N = 0$).

with

$$\Psi_{nlNJM_J}(\mathbf{r}, \Omega_d) = \sum_{m_l=-l}^{m_l=l} \sum_{M_N=-N}^{M_N=N} \langle lm_l NM_N | JM_J \rangle \psi_{nlm}(\mathbf{r}) Y_{NM_N}(\Omega_d), \quad (7)$$

where $\langle lm_l NM_N | JM_J \rangle$ are the Clebsch-Gordan coefficients, $J = |l - N|, \ldots, l + N$, and $M_J = -J, \ldots, J$. $\psi_{nlm}(\mathbf{r})$ is the Rydberg electron wave function with $n$, $l$, and $m$ being the principal, orbital, and magnetic quantum numbers respectively. $Y_{NM_N}(\Omega_d)$ is the spherical harmonics, which represents the field-free rotational wave function of the diatomic molecule, with $N$ and $M_N$ being the rotational and magnetic quantum numbers. The internal rotational motion of the diatomic molecule is described by the Euler angles $\Omega_d = (\theta_d, \phi_d)$.

In the basis expansion in Eq. (6), we have assumed that the projection of the total angular momentum along the inter-species axis $M_J$ is a good quantum number. For $M_J = 0$, our calculations presented in Fig. 1(a) include the rotational excitations up to RbCs($N \leq 5$), and for the quantum defect Rydberg states Rb(($n + 3$)s), Rb(($n + 2$)p), Rb(($n + 1$)d), Rb($n$f) and the degenerate manifold Rb($n, 4 \leq l \leq n - 1$), with $n = 48, 49$, and $50$, satisfying $M_J = m_l + M_N$, and all possible values of the total angular momentum $J$. By using in Eq. (6), a reduced basis formed by Rydberg states, with non-zero quantum defects, Rb(52s), Rb(51p), Rb(50d), Rb(49f), and the degenerate manifold Rb($49, 4 \leq l \leq 48$), the largest difference of energy shift presented in Fig. 1(a) is $\sim 2.5$ MHz in the outermost minimum, whereas at 300 nm this difference is reduced to $\sim 75$ kHz.

To illustrate the electronic features of the ultralong-range Rydberg molecule, we present in Fig. S1 the adiabatic potential energy curves close to the dissociation threshold Rb($n = 49, l \geq 4$)-RbCs($N = 0$). These results have been computed using the reduced basis. Within the states evolving from this degenerate manifold, we encounter the potential energy curves Rb(52s)-RbCs($N = 2$) and Rb(52s)-RbCs($N = 3$), which are resonantly coupled to the zero quantum defect Rydberg levels due to the Rydberg atom-molecule anisotropic dipole interaction [8]. As a consequence, the adiabatic potential wells evolving from the hydrogenic like Rydberg manifold acquire $s$-wave admixture, which facilitates the creation of the Rydberg molecule Rb-RbCs via conventional two-photon Rydberg excitation schemes of Rb. The vibrational bound states of the potentials Rb(52s)-RbCs($N$), with $N = 0, 1$, and $2$ are presented in Fig. S2.

Finally we note that the Fermi pseudopotential term $V_{\text{em}}$ Eq. (3), which is included in the full Hamiltonian in Eq. (1), only has a small effect on the potential energy curves in the range of interest for the experiments. In Fig. S3

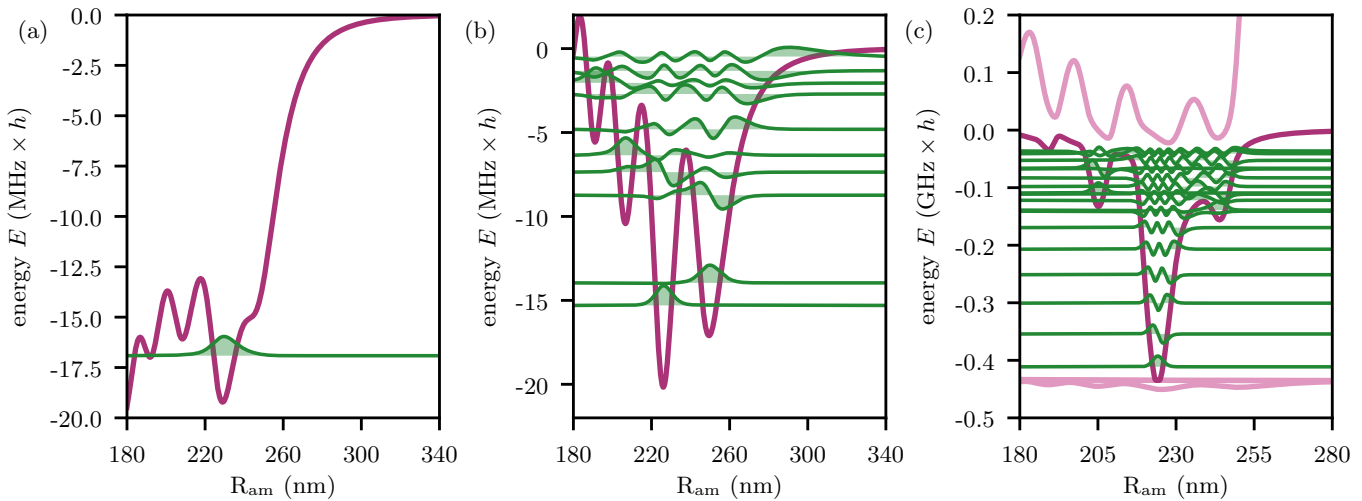

FIG. S2. Adiabatic electronic potential energy curves (purple) and vibrational wavefunctions (green) for the Rb-RbCs Rydberg molecule. Energies are defined relative to the corresponding asymptotes (a) Rb(52s)-RbCs($N = 0$), (b) (52s)-RbCs($N = 1$), and (c) (52s)-RbCs($N = 2$). The wavefunctions have been shifted to the corresponding vibrational energies.

we plot the potential energy curves with (purple) and without (blue) the Fermi pseudopotential term. The inset highlights the experimentally relevant region and at these separations the energy shift due to electron scattering is not significant.

## EXPERIMENTAL DETAILS

Each experimental run starts by loading two 1D arrays of optical tweezers which trap Rb and Cs atoms. Rearrangement is performed on each of these 1D arrays in order to prepare individually trapped Rb and Cs atoms in the designated optical tweezers. The rearrangement protocol prepares a single Cs atom in a single 1065.512 nm trap with beam waists $\{w_x^{1065}, w_y^{1065}\} = \{1.08(1), 1.21(1)\}$ µm. For atom+molecule measurements, the rearrangement protocol prepares single Rb atoms in two 816.848 nm tweezers with separation 4 µm along the $y$-axis and beam waists $\{w_x^{817}, w_y^{817}\} = \{0.82(1), 0.92(1)\}$ µm. For measurements whose main focus is to produce molecules (Fig. 2 of main text), the second 817 nm trap is not loaded. The 817 nm tweezers are intially offset from the 1065 nm tweezer by 4 µm along the $x$-axis.

Following rearrangement, the trap occupations are verified by fluorescence imaging. This image is later used to post-select on different experimental scenarios. For example, if only 2 Rb atoms and no Cs atoms are present in this image, no molecule will be created in the routine. Therefore, this data is used to calculate the 1-body Rb survival probability in Fig. 3 of the main text.

The Rb and Cs atoms are then cooled to the motional ground state of their traps with Raman sideband cooling before being prepared in the hyperfine states $|f = 1, m_f = 1\rangle_\text{Rb}$ and $|f = 3, m_f = 3\rangle_\text{Cs}$, as described in Ref. [9]. Rb+Cs atom pairs are prepared in the 1065 nm tweezer by translating the 817 nm tweezers in the $x$-direction such that one overlaps with the 1065 nm tweezer. This 817 nm tweezer is then switched off and molecule formation is performed as described in the main text.

For the collision measurements presented in Fig. 3 of the main text, the remaining 817 nm tweezer containing a Rb atom is swept along both the $x$- and $y$-axes. The sweeps are optimised to minimise heating during this process similarly to the sweeps used for molecule formation. We estimate these sweeps cause approximately 0.1 quanta of heating to the atom. The tweezer is first swept along the $x$-axis to a programmed tweezer separation which is plotted as the independent variable on the x-axis of Fig. 3. The tweezer is then swept along the $y$-axis such that it overlaps with the 1065 nm tweezer in this direction and is held at this position for 9.5 ms.

For measurements involving both an atom and a molecule, the Rb atom is cooled to the motional ground state of the 817 nm tweezer. For the trap parameters used in the Rydberg blockade measurements, we estimate the spread of the zero-point wavefunction of the atom $\sqrt{\hbar/2m\omega}$ is $\{\delta x, \delta y, \delta z\}_\text{Rb} = \{27, 30, 69\}$ nm. The RbCs molecule is formed from atoms cooled to the motional ground state and inherits the low motional excitation of the atoms. We estimate

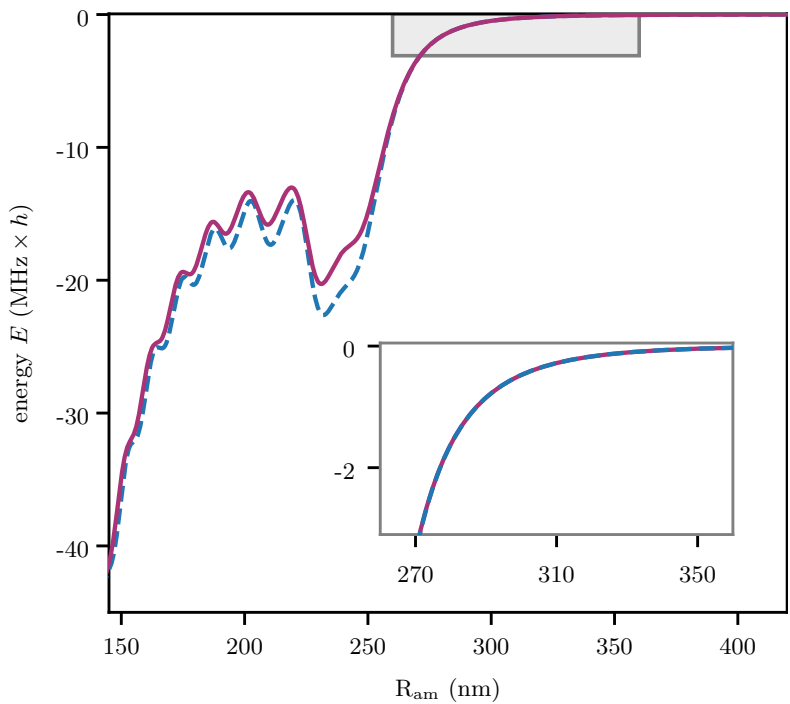

FIG. S3. Adiabatic electronic potential energy curves of Rb(52s)-RbCs($N = 0$). The purple line shows the calculated curve using the full Hamiltonian in Eq. (1), which includes the Fermi pseudopotential term $V_{em}$ in Eq. (3). The blue dashed line shows the calculated curves without $V_{em}$ in Eq. (3). The energies are defined relative to the Rb(52s)-RbCs($N = 0$) asymptote. The inset corresponds to the shaded region in the main figure and highlights the energy shift relevant to the blockade measurements.

the probability for the weakly bound RbCs molecule to occupy the motional ground state to be $\sim 0.66$. This estimate is based on Raman thermometry of the individual Rb and Cs atoms along with an estimate of the probability to occupy the lowest state of relative motion inferred from the molecule formation probability [10]. In comparison to the previous demonstration of two photon transfer of molecules in optical tweezers [11], the STIRAP transfer in our system is not expected to cause significant motional heating of the molecules. The tweezer wavelength used in our work is near-magic for the STIRAP transition [12]. By measuring the shift of the STIRAP transition as a function of the tweezer intensity, we measure the polarizability ratio between the states $|G\rangle$ and $|F\rangle$ to be $\alpha_G/\alpha_F = 1.07(2)$ in our 1065.512 nm optical tweezer. In addition, motional excitation from the momentum kick of the Raman beams is suppressed because the STIRAP beams propagate along the radial axis of the tweezers. The molecule experiences an optical potential with trapping frequency of 10 kHz during STIRAP and we estimate a probability of $> 0.99$ for the motional state to be unchanged during the STIRAP process. Assuming negligible heating from the movement of the 817 nm tweezer towards the 1065 nm tweezer containing the molecule, we estimate the spread of the wavefunction of the RbCs molecule to be $\{\delta x, \delta y, \delta z\}_{\text{RbCs}} = \{34, 39, 87\}$ nm during the Rydberg blockade measurements.

## FESHBACH MOLECULE LIFETIME

Rb+Cs atom pairs which are prepared in the motional ground state of the 1065 nm tweezer are associated into the weakly bound molecular state $|-1(1,3)s(1,3)\rangle$ using magnetoassociation across a Feshbach resonance at 197.1 G [13]. Here the molecule state is labelled as $|\nu(f_{\text{Rb}}, f_{\text{Cs}})\ell(m_{\text{Rb}}, m_{\text{Cs}})\rangle$, where $\nu$ is the vibrational quantum number relative to the supporting threshold and $\ell$ is the rotational quantum number around the centre of mass. The weakly bound molecules are transferred to state $|F\rangle = |-6(2,4)d(2,4)\rangle$ by ramping the magnetic field down to 181.6 G. Prior to the STIRAP pulse sequence, we allow the magnetic field to settle for 10 ms to avoid the STIRAP transitions drifting significantly between the transfer to the ground state and the transfer back from the ground state. The state $|F\rangle$ is deeply bound with respect to its atomic threshold ($f_{\text{Rb}} = 2, f_{\text{Cs}} = 4$), resulting in a larger photon scattering rate in comparison to the weakly bound state $|-1(1,3)s(1,3)\rangle$.

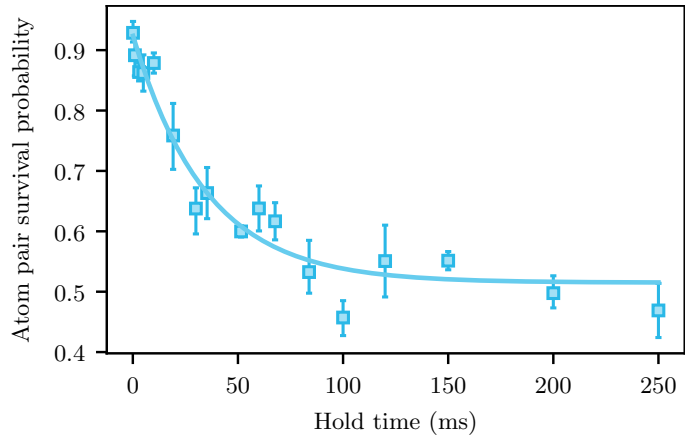

FIG. S4. Trap lifetime of molecules in state $|F\rangle$ immediately after magnetoassociation. The tweezer intensity is 12.7 kWcm$^{-2}$ and the extracted lifetime is 35(6) ms.

Figure S4 shows the measured trap lifetime of the molecules in $|F\rangle$ in the 1065 nm tweezer. Following a hold time in the tweezer, the magnetic field is ramped back over the Feshbach resonance to convert the molecule back into an atom pair which is reimaged. We extract a lifetime of 35(6) ms for a tweezer intensity of 12.7 kWcm$^{-2}$ and thus expect 25(4)% of molecules in $|F\rangle$ to be lost prior to STIRAP in the molecule formation sequence.

## STIRAP PULSE SEQUENCE

STIRAP transfer relies on the formation and adiabatic transfer of an eigenstate of the on-resonant lambda system from $|F\rangle$ to the rovibrational ground state $|G\rangle$. These states are coupled via a common excited state $|E\rangle$ with two lasers: the pump laser ($|F\rangle \to |E\rangle$; $\lambda = 1557$ nm) and the Stokes laser ($|E\rangle \to |G\rangle$; $\lambda = 977$ nm). The eigenstate (the "dark state") takes the form

$$|D\rangle = \cos\theta\,|F\rangle + \sin\theta\,|G\rangle, \tag{8}$$

where the mixing angle $\theta = \arctan(\Omega_p/\Omega_S)$ for Rabi frequencies $\Omega_p$ and $\Omega_S$ of the pump and Stokes beams respectively. To initialise the dark state in $|F\rangle$ and evolve it to $|G\rangle$, a so called 'counter-intuitive' pulse sequence is used which begins with only the Stokes beam on and evolves to having only the pump beam on [14].

We adiabatically evolve $\theta$ by ramping the intensities of the STIRAP beams with a $\cos^4 t$ profile for time $t$. The time to complete a one-way transfer is 150 µs. The intensity of the beams are controlled with acousto-optic modulators (AOMs) (pump: ISOMET M1205-P80-L-0.6, Stokes: ISOMET 1205C-843) driven by a digital direct synthesiser (DDS) (homebuilt, based on Anologue Devices AD9910). These AOMs also provide tuneability of the frequencies of the STIRAP beams.

The pump and Stokes beams are frequency stabilised to an ultra-low expansion (ULE) cavity (Stable Laser Systems, custom) using a Pound-Drever-Hall (PDH) scheme [15]. The finesse $\mathcal{F}$ and free-spectral-range $\omega_{\text{fsr}}$ of this cavity at the different wavelengths used are shown in Table I. Two sets of frequency sidebands are added to the light going to the cavity by an electro-optic modulator (EOM) (pump: Thorlabs LN65S-FC, Stokes: EOSpace PM-0S5-10-PFA-PFA-980). Light reflected from the cavity is incident on fast photodiodes (Thorlabs PDA05CF2). One set of sidebands at $\omega_{\text{PDH}}$ [Table I] is used to derive the PDH error signal and the other set is used to tune the frequency offset of the error signal between the cavity transmission peaks to allow for stabilisation to an arbitrary frequency. Two independent function generators produce the driving tones for these sidebands (Rigol DG822 and Windfreak Technologies Synth HD respectively) before they are combined with a power splitter (Minicircuits ZAPD-2-252-S+) and sent to the modulator. The demodulation of the PDH error signal and laser feedback is performed by fast laser locking modules (Toptica FALC pro).

TABLE I. Quantities relevant to the frequency stabilisation of the STIRAP and Rydberg lasers.

| Laser | | ULE cavity | | PDH lock |
|---|---|---|---|---|
| | | $\mathcal{F}$ | $\omega_{\text{fsr}}$ (MHz) | $\omega_{\text{PDH}}$ (MHz) |
| STIRAP | Stokes ($\lambda = 977$ nm) | $2.00(8) \times 10^4$ | 1498.8123(9) | 20.21 |
| | pump ($\lambda = 1557$ nm) | $1.70(2) \times 10^4$ | 1498.796(3) | 15.51 |
| Rydberg | $\lambda = 420$ nm | $> 8 \times 10^2$ | $\sim 1500$ | 25.00 |
| | $\lambda = 1013$ nm | $2.8(1) \times 10^4$ | 1498.8117(6) | 11.31 |

## MOLECULE DETECTION SCHEME

All experimental data presented in the main text and here are obtained by repeating an experimental sequence multiple times (typically between 200 - 1000 times). The error bars shown in figures are obtained by block bootstrapping the acquired data. This technique incorporates uncertainties from the number of samples and variations in experimental conditions over time. These data are post-selected based on the initial and final occupancy of various traps to obtain statistics for different scenarios. The occupancy is measured with fluorescence imaging. For example, in Fig. 2(d) of the main text we show the atom pair survival probability $P_{11}$, this data was post-selected on the condition that a Rb and a Cs atom were initially loaded.

To post-select for experimental runs in which the formation of a weakly bound molecule was successful, such as those presented in Fig. 3 of the main text, we implement a molecule detection scheme. This detection scheme is required as direct fluorescence imaging of the molecule is precluded by the lack of closed cycling transitions in RbCs. We instead detect the successful formation of a weakly bound molecule by exploiting the species-specificity of our optical tweezers.

When the Rb+Cs atom pair occupies the relative motional ground state, the probability to form a weakly bound molecule using magnetoassociation is $> 0.99$. However, when there is heating during the merging process or imperfect cooling of the atoms to the 3D motional ground state, the atom pair can occupy an excited state of relative motion, prohibiting molecule formation using our chosen Feshbach resonance. This imperfect state preparation is the dominant error mechanism in the molecule formation process with $\sim 50\%$ of experimental runs producing a weakly bound molecule. When the atoms are prepared in hyperfine states $|1,1\rangle_{\text{Rb}}$ and $|3,3\rangle_{\text{Cs}}$ there are no hyperfine-changing inelastic collisions. Therefore, failure to associate a weakly-bound molecule leaves a Rb+Cs atom pair which occupies an excited state of relative motion of the 1065 tweezer.

The different polarizabilities of the tweezers for our species are shown in Table II; RbCs is more strongly attracted to the tweezer at 1065 nm, whereas Rb is more strongly attracted to the tweezers at 817 nm. In contrast, Cs is only attracted to the tweezer at 1065 nm and is repelled by the tweezers at 817 nm.

Our detection scheme is shown in detail in Fig. S5(a). The initial occupancy of all traps is measured at the start of a routine. After the molecule preparation stages (including STIRAP), we turn on an additional atom tweezer ($\lambda = 817$ nm) at the position of the molecule tweezer ($\lambda = 1065$ nm) and sweep it away from this trap. The occupancy of this "detection tweezer" is measured at the end of the routine alongside the occupancy of the other traps. The presence of an atom in the detection tweezer flags runs where the molecule tweezer contained an unassociated atom pair (i.e. molecule formation was unsuccessful). The intensity of the molecule tweezer is chosen such that its potential is deeper than that of the detection tweezer for a RbCs molecule or Cs atom, but shallower than the detection tweezer for a Rb atom. In Fig. S5(b) we present measurements showing that for 1065 nm tweezer powers between $0.2 - 4$ mW, the Rb atom is removed from the molecule tweezer and the RbCs molecule remains. Following removal of the Rb atom using the detection tweezer, light which is near resonant with the Cs atom is applied, resulting in rapid heating and loss of the Cs atom from the molecule tweezer. The RbCs molecule and Rb atom are unaffected by this light due to the large detuning from any allowed optical transitions.

TABLE II. Calculated polarizabilities of the different species in our experiment at our optical tweezer wavelengths. All polarizabilities are for atomic or molecular electronic ground states.

| Species | $\alpha_{817}$ ($a_0^3$) | $\alpha_{1065}$ ($a_0^3$) | $\alpha_{817}/\alpha_{1065}$ |
|---|---|---|---|
| RbCs [12] | $4.0 \times 10^2$ | $1.8 \times 10^3$ | $\sim 1/4.5$ |
| Rb [16] | 4307 | 687 | $\sim 6.3$ |
| Cs [16] | -3477 | 1163 | $\sim -3.0$ |

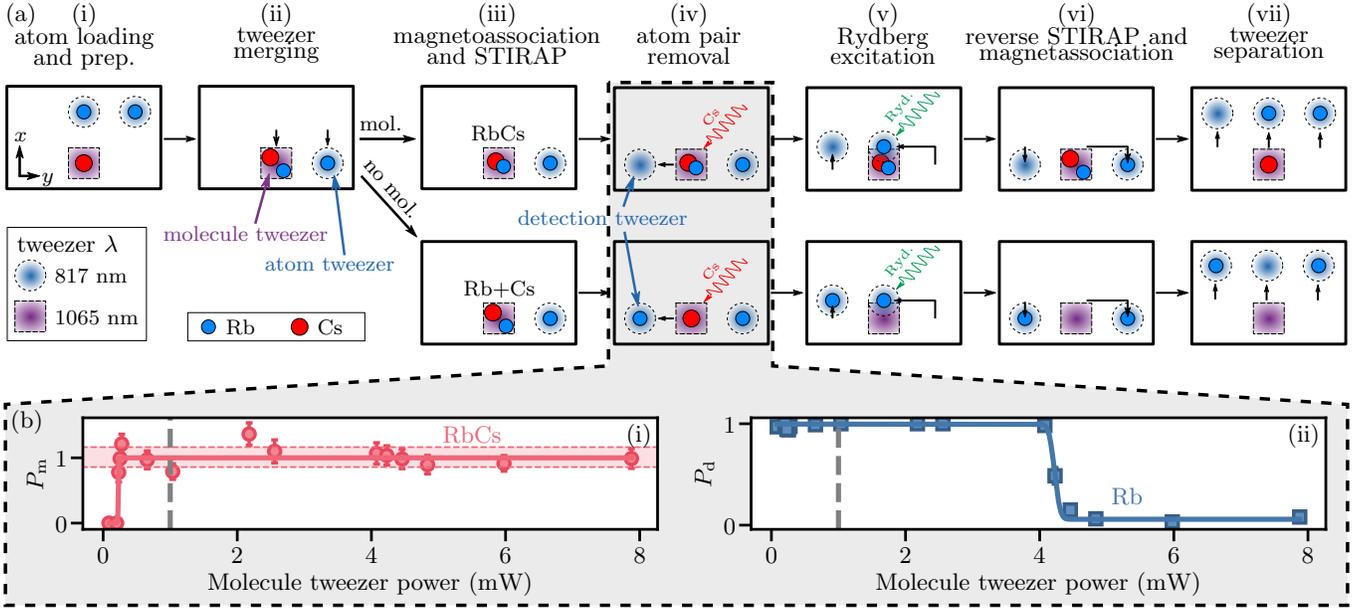

FIG. S5. (a) Experimental sequence for the study of RbCs and Rb interactions. (i) Two Rb atoms and one Cs atom are trapped 817 nm and 1065 nm optical tweezers respectively. (ii) The 817 nm traps are swept to prepare a Rb+Cs atom pair in the 1065 nm trap ("molecule tweezer"). The remaining Rb atom remains in the "atom tweezer". (iii) The Rb+Cs atom pair is magnetoassociated with ∼50% efficiency and then transferred to $|G\rangle$. The top (bottom) row of the figure shows the events that occur in the cases of (un)successful molecule formation. (iv) In the event of unsuccessful molecule formation, the Rb+Cs atom pair is removed from the molecule tweezer. An 817 nm tweezer ("detection tweezer") selectively removes the Rb atom and light near-resonant with the remaining Cs atom is then applied to eject Cs. These steps do not affect the RbCs molecule present if the molecular formation was successful. (v) The atom tweezer is swept to be close to the molecule tweezer. The Rydberg excitation light is applied here for the experiment presented in Fig. 4 of the main text. (vi) The tweezer sweep is reversed and remaining molecules are transferred back to $|F\rangle$ and disassociated into an atom pair. (vii) The tweezers are separated and atomic fluorescence imaging is performed. (b) The probability $P_\mathrm{m}$ of a RbCs molecule remaining in the molecule tweezer [panel (i)] and the probability $P_\mathrm{d}$ of a Rb atom being transferred to the detection tweezer [panel (ii)] as a function of the molecule tweezer power during step (a)(iv). Data in panel (i) has been rescaled to account for molecule loss in other stages of the sequence; the shaded region shows the uncertainty on $P_\mathrm{m} = 1$ due to this rescaling. Over a broad range of powers the detection tweezer can selectively remove Rb atoms and leave RbCs molecules. Therefore, occupation of the detection tweezer indicates unsuccessful molecule formation. We use a power of 1 mW (gray dashed lines) for the experiments presented in the main text.

As shown in the lower panel of Fig. S5(a)(iv), unsuccessful molecule formation results in a Rb atom occupying the detection tweezer and an empty molecule tweezer. The upper panel shows the alternative scenario where association into $|F\rangle$ was successful. In this case, the detection tweezer remains unoccupied as the molecule will have either been lost or transferred to $|G\rangle$ and thus remains in the 1065 nm trap.

For the experiment presented in Fig. 4 of the main text, data labelled "no molecule present" are from experimental runs where the detection tweezer was occupied. In contrast, data labelled "molecule present" are from runs where both the detection tweezer was unoccupied and the associated Rb+Cs atom pair was successfully recovered when reversing the molecule preparation stages. These post-selection criteria ignore runs in which the weakly-bound molecule was formed but subsequently lost.

## RYDBERG EXCITATION

Rb atoms are prepared in the hyperfine ground state $|g\rangle = |5^2\mathrm{S}_{1/2}, f = 1, m_f = 1\rangle$. They are excited to the Rydberg state $|r\rangle = |52^2\mathrm{S}_{1/2}, m_j = -1/2\rangle$ via the state $|e\rangle = |6^2\mathrm{P}_{3/2}, m_j = -1/2\rangle$ at 181.6 G; at this field the states $|e\rangle$ and $|r\rangle$ are in the Paschen-Back regime where $m_j$ is a good quantum number. This excitation requires two lasers at wavelengths 420 nm (Toptica DL PRO) and 1013 nm (Toptica TA PRO) for the first and second excitation stages respectively. The single-photon Rabi frequencies are approximately $\Omega_{420} \approx \Omega_{1013} = 28$ MHz; the single-photon Rabi frequencies are matched to minimise the transition light shift and excess scattering. A single-photon detuning

of $\Delta_{1p} = +800$ MHz from state $|e\rangle$ prevents significant population of this state. The two-photon Rabi frequency $\Omega = \Omega_{420}\Omega_{1013}/(2\Delta_{1p})$ is measured to be 500(3) kHz.

The Rydberg lasers are frequency stabilised to the same ULE cavity as the STIRAP lasers. The stabilisation setup is much the same; the only difference is the EOMs used to add frequency sidebands (420 nm: Photonics Technologies EOM-02-25-U, 1013 nm: EOSpace PM-0S5-10-PFA-PFA-1013) and the fast photodiode used for the light at 420 nm (Thorlabs PDA8A2). The light at 1013 nm is incident on the same photodiode as the Stokes light, the difference in $\omega_{PDH}$ between the two [Table I] allows their error signals to be extracted independently. The EOM for the light at 420 nm is a free-space EOM as opposed to the fibre EOMs for the other lasers; this precludes the stabilisation to an arbitrary frequency and we instead stabilise the laser to a cavity mode. For this reason, we cannot precisely measure $\mathcal{F}$ and $\omega_{fsr}$ for the light at 420 nm; we instead estimate these quantities based on the cavity linewidth and position of the PDH sidebands on the error signal.

Optical excitation to $|r\rangle$ is performed using square pulses produced by AOMs (420 nm: ISOMET M1250-T250L-0.45, 1013 nm: ISOMET 1205C-843) which are driven using a DDS. When performing a Rydberg pulse, the light at 1013 nm is switched on before and remains on after the light at 420 nm to ensure consistent two-photon pulse timing. The Rydberg pulse times presented in the main text are the times for which both lasers are applied to the atoms (i.e. the pulse duration of the light at 420 nm).

In runs where a weakly bound molecule is not formed, a Rb atom may be present in the detection tweezer as well as the atom tweezer. To prevent deleterious effects from Rydberg excitation of the second Rb atom, the detection tweezer depth is reduced such that the Rydberg transition shifts to lower energy. This is shown in Fig. S6.

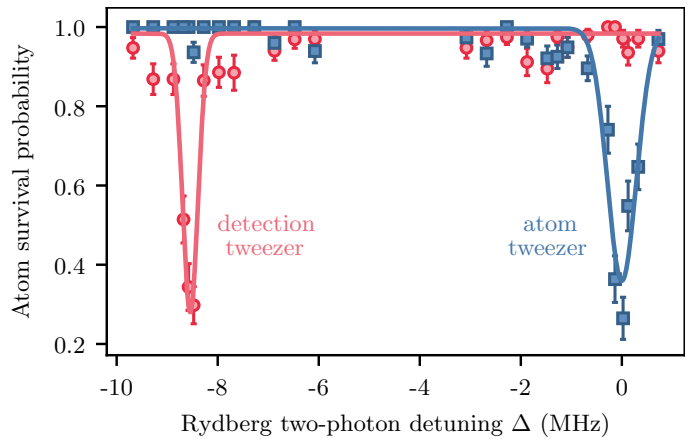

FIG. S6. Survival probability of a Rb atom trapped in the atom tweezer (blue squares) or detection tweezer (red circles) as a function of detuning from the Rydberg transition for the Rb atom in the atom tweezer. Reducing the detection tweezer depth shifts the Rydberg transition of a Rb atom in this trap by -8.52(2) MHz. This prevents unwanted Rydberg excitation of an atom in this trap.

## RYDBERG DETECTION

The optical tweezers in our experiment are anti-trapping for atoms in the state $|r\rangle$. As the trapping light remains on during Rydberg excitation, atoms excited to the Rydberg state are mapped onto atom loss, providing a convenient detection mechanism. We experimentally measure the trap lifetime of the state $|r\rangle$ using two $\pi$-pulses with a variable delay between the pulses. The results of this experiment are presented in in Fig. S7; the trap lifetime is measured to be 8.5(1.1) µs in an atom tweezer of intensity $35\,\text{kWcm}^{-2}$.

We calculate the zero-temperature lifetime of the state $|r\rangle$ with ARC [17] to be 146 µs. We therefore expect to map excitation to loss with a fidelity of 94(1)%, where the infidelity is from events where the state $|r\rangle$ decays before ejection of the atom from the tweezer.

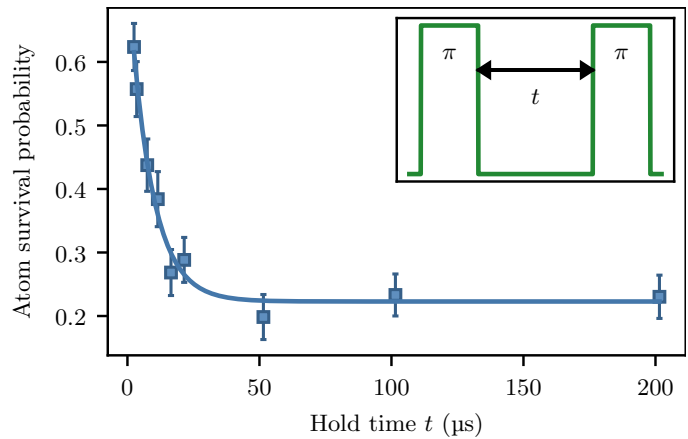

FIG. S7. Time taken for the Rydberg atom to be ejected from its tweezer. A $\pi$ pulse excites the atom to the Rydberg state. After a variable hold time a second $\pi$ pulse returns the atom to the ground state. The lifetime of the Rydberg atom in the tweezer is 8.5(1.1) µs.

## ATOM-MOLECULE SEPARATION

### Trap Separation in the Radial Axes

In order to estimate the separation of the Rb atom and RbCs molecule we use a combination of techniques which allows the separation to be estimated in three spatial dimensions. In the radial direction of the tweezers, perpendicular to the tweezer light propagation direction, we can directly infer the positions of the tweezers by imaging the fluorescence of the trapped atoms. However, the finite resolution of our imaging system ($\sim 1.5$ µm) precludes accurate measurement of the trap separations for sub-micron separations. We estimate the separation at sub-micron scales by careful calibration of the acousto-optic deflector (AOD) which is used to control the position of the atom tweezer ($\lambda = 817$ nm). We calibrate this device by recording multiple fluorescence images of a Rb atom trapped in the tweezer for a range of RF drive frequencies. This procedure along with knowledge of the magnification of our imaging system allows us to calibrate distance moved in the object plane as a function of the drive frequency of the AOD.

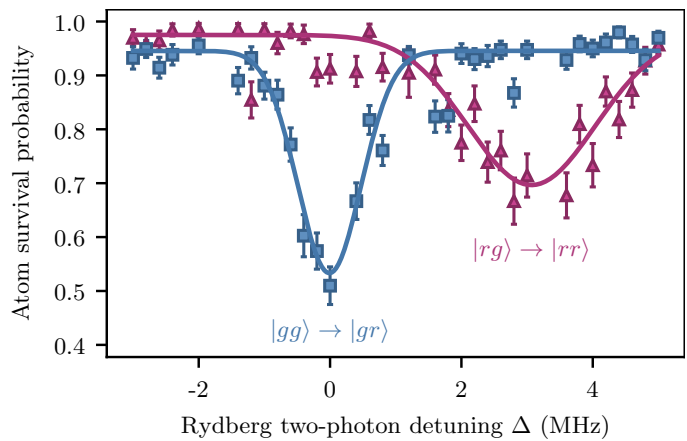

FIG. S8. Interaction shift of the Rydberg transition of one atom caused by the presence of another Rydberg atom. The blue squares are data from runs where the other atom is not successfully excited (i.e. remains in state $|g\rangle$); the purple triangles are data from runs where the other atom is excited to state $|r\rangle$ prior to the Rydberg pulse shown here.

We have verified the accuracy of the AOD calibration by performing measurements of the interaction shift of two Rb atoms in neighbouring tweezers excited to $|r\rangle$. In this experiment, two atom tweezers are generated using the AOD, bypassing any potential misalignment along the axial direction (tweezer light propagation direction) which is possible when using tweezers of a different wavelength. We excite a single atom to $|r\rangle$ using a $\pi$-pulse and then after

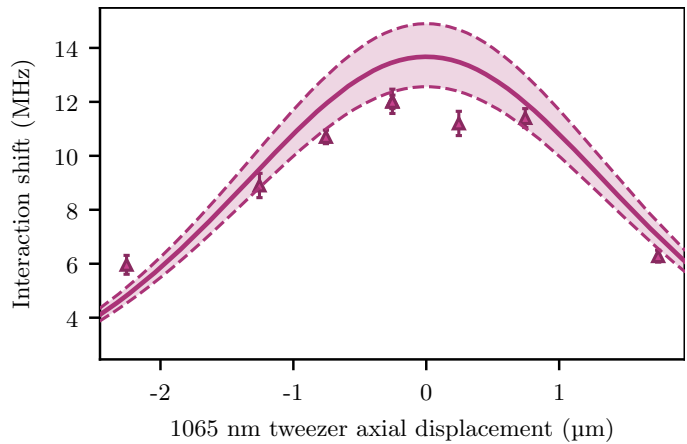

FIG. S9. Interaction shift of two Rydberg atoms as a function of the axial displacement between the tweezers. Before excitation, one atom is trapped in a 817 nm tweezer and the other is trapped in a 1065 nm tweezer. The tweezers are offset radially by 3.49(5) µm. When the tweezers are overlapped axially, the interaction shift is maximised. The solid line is the expected interaction shift as the axial displacement is varied; the shaded region results from the error on the radial displacement.

1 µs we perform a second pulse of the same duration but with a variable frequency. We show in Fig. S8 the probability of exciting the second Rb atom to $|r\rangle$. Two features appear conditional on the successful excitation of the first Rb atom to $|r\rangle$, which we measure as that atom being lost from the tweezer. The difference in frequency between the two peaks measures the shift in energy of the Rydberg transition. We measure an interaction shift of 3.07(14) MHz for a trap separation of 4.48(3) µm, as predicted by our AOD calibration. This is in good agreement with calculations of the interaction shift using the *pairinteraction* library [18] which predicts a separation of 4.48(4) µm for our measured interaction shift.

The RbCs molecule and Rb atom are trapped in optical tweezers of different wavelengths which are focussed through the same high numerical-aperture objective lens. With our AOD calibration we can predict the relative change in position of the 817 nm tweezer but we also need to identify the position where the 1065 nm and 817 nm are overlapped. To do this we perform atom-molecule loss measurements like those shown in Fig 3 of the main text. The inelastic collision rate is largest when the wavefunctions of the atom and molecule are well overlapped, allowing us to find the radial overlap position of the traps. We can then tune the trap separation to the desired value by changing the driving frequency of the AOD by the required amount.

### Trap Separation in the Axial Direction

The above procedure of using the AOD to estimate the trap separation is not possible in the axial direction as the 2D AOD does not control the trap position in this direction. Instead we control the trap separation in this direction by imprinting a phase pattern corresponding to a Fresnel lens onto a spatial light modulator (SLM) in the 1065 nm tweezer path. We cannot measure trap displacements in this direction directly using fluorescence imaging as this direction is along the imaging axis. Instead, we find the overlap position of the 1065 nm and 817 nm tweezers in the axial direction by measuring the interaction shift of two Rb Rydberg atoms. One atom is trapped in a 817 nm tweezer and the other in the 1065 nm tweezer. We are unable to observe the interaction-shifted peak for tweezer separations smaller than $\lesssim 3$ µm because the interaction shift for two Rb atoms in $|r\rangle$ becomes larger than the bandwidth of the AOM used for the Rydberg pulses. Therefore, we radially displace the tweezers by 3.49(5) µm such that the atomic separation is never smaller than this value.

Figure S9 shows the measured interaction shift as a function of the 1065 nm tweezer axial displacement. The relative change in axial position is estimated based on the focal length of the lens applied using the SLM. When the traps are aligned axially, the atom-atom separation is minimised and the interaction shift is maximised. We use the *pairinteraction* library to calculate the predicted interaction shift as a function of axial displacement given our radial displacement, and fit the measured data with this function. This fitting procedure returns a value for the overlap position with a $1\sigma$ uncertainty of 0.1 µm. We have verified the accuracy of this alignment procedure using RbCs molecule formation. Molecule formation is sensitive to the relative overlap of the tweezers as the formation probability is dependent on the probability of atom pairs to occupy the relative motional ground state. Misalignment

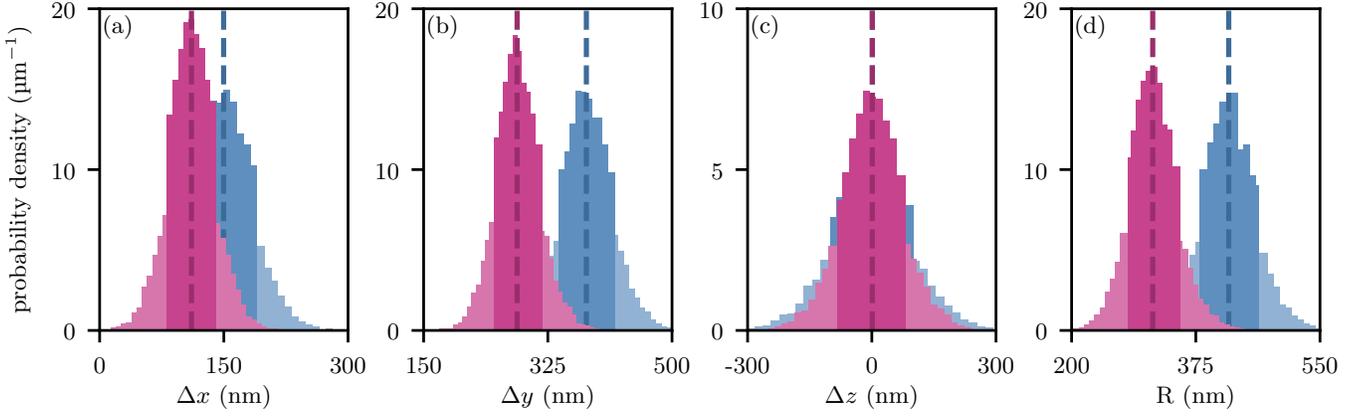

FIG. S10. Results of the atom-molecule separation simulation for the experiment shown in Fig. 4(a) of the main text. The probability distributions of the tweezer displacements (blue) and atom-molecule displacements (purple) are shown along the three experimental axes [panels (a)-(c)] and the inter-particle axis [panel (d)]. The best estimate for each parameter is taken to be equal to the mean of the distribution (indicated by the dashed lines) and the associated systematic uncertainty is read from the $1\sigma$ bounds (shown as the darker shaded regions).

of the optical tweezers causes heating during the merging process and hence a reduction in the molecule formation probability.

### Simulations of the atom-molecule separation in species-specific traps

For experiments where the atom and molecule tweezers are brought to sub-micron separations, such as the experiments presented in Fig. 4 of the main text, the optical potential experienced by one of the particles is significantly perturbed by the presence of the other tweezer. As both Rb and RbCs are attracted to both tweezers, these perturbations result in the atom-molecule separation $R_{am}$ being significantly smaller than the tweezer separation $R_t$. To account for this, we perform numerical simulations of the total potential experienced by each species. These simulations use the measured optical powers, trap waists and Rayleigh lengths of our tweezers. For a given tweezer displacement, the location of each species' potential minima is found; the separation between these minima is $R_{am}$. For example, using the techniques described in the previous sections, the tweezer displacements along each axis are estimated to be $\{\Delta x_t, \Delta y_t, \Delta z_t\} = \{150(40), 380(40), 0(100)\}$ nm for the experiment presented in Fig. 4(a) of the main text. Here the numbers in parentheses represent the systematic uncertainties resulting from the calibrations presented earlier. To estimate the corresponding atom-molecule displacements $\{\Delta x_{am}, \Delta y_{am}, \Delta z_{am}\}$ and their systematic uncertainties we use a Monte Carlo method. The tweezer displacements are sampled from normal distributions with the means equal to our best estimates and standard deviations equal to the associated systematic uncertainties on these values. Distributions of $R_t$ and $R_{am}$ are then calculated. With $10^4$ samples, we find $\{\Delta x_{am}, \Delta y_{am}, \Delta z_{am}\} = \{110(30), 280(40), 0(80)\}$ nm, $R_t = 420(40)$ nm, and $R_{am} = 310(40)$ nm. The probability distributions from this simulation are shown in Fig. S10.

In our calculations of the optical potential depth we use the calculated polarizabilities for each species (listed in Table II). We note that accurately calculating the polarizability of the RbCs molecule is more challenging compared to calculating the polarizabilities of alkali-metal atoms like Rb and Cs because of the more complex energy level structure of the molecule. In Ref. [19] the isotropic polarizability of the RbCs vibronic ground state was measured to be $2.02(4) \times 10^3$ at a wavelength of 1064.513 nm. This measured value is $\sim 10\%$ higher than the calculated value of $1.8 \times 10^3$. In our simulations we have incorporated this uncertainty in the molecule polarizability. For example, for the experiment presented in Fig. 4(a) of the main text, we find that when the RbCs polarizabilities at both wavelengths are scaled by 10% there is no significant change to $R_{am}$. When the ratio between the two polarizabilities is scaled by 10% the change to $R_{am}$ is on the order of $\sim 2\%$, which is much less than the uncertainty resulting from the systematic uncertainties on the tweezer displacements.

For the measurement presented in the upper panel of Fig. 4(b) of the main text where the atom and molecule are held further apart, the tweezer displacements are estimated to be $\{\Delta x_t, \Delta y_t, \Delta z_t\} = \{760(30), 40(100), 0(100)\}$ nm. Using the Monte Carlo simulation described above with $10^4$ samples, we find that for this measurement $\{\Delta x_{am}, \Delta y_{am}, \Delta z_{am}\} = \{690(40), 40(90), 0(110)\}$ nm, $R_t = 770(30)$ nm, and $R_{am} = 700(40)$ nm.

**Variation in atom-molecule separation**

In addition to the mean trap separation which is estimated using the techniques described above, the separation between the atom and molecule varies between experimental runs due to shot-to-shot variations in the positions of the optical tweezers. As the 817 nm and 1065 nm tweezers do not share the same optical path, we measure significant drifts in the relative position of the two traps. In the radial direction we have characterised this drift by fitting images of the atomic fluorescence during the experimental measurements reported in the main text. We find that drifts in the relative positions of the 1065 nm and 817 nm tweezers lead to a change in the trap separation with a standard deviation of 50 nm. We have also verified this measurement of the positional drift using a routine which ejects Cs atoms trapped in the 1065 nm tweezer by overlapping a 817 nm tweezer with this trap. The 817 nm tweezer produces a repulsive potential for Cs and measuring the position of maximum loss over time can track the tweezer overlap with a precision of $\sim 10$ nm [20]. From this measurement, we observed the drift was correlated with the change in humidity of the lab (roughly 10 nm/%). The humidity changes significantly during a single air-conditioning cycle which typically lasts 20 mins. For the blockade measurement, we anticipate that the trap separation will have varied in the radial direction by around 40-60 nm. Unfortunately, we were unable to perform a measurement in the axial direction with the same precision because of the much increased trap size in this direction. We therefore just take the estimate of the radial drift for this axis as well. Our estimate for the standard deviations in tweezer separations is therefore $\{\sigma_x, \sigma_y, \sigma_z\} = \{50, 50, 50\}$ nm.

## SIMULATIONS OF ATOM-MOLECULE RYDBERG BLOCKADE

The model Hamiltonian describing our atom-molecule system $\mathcal{H}$ is given by

$$\mathcal{H} = H_a + H_m + V(R_{am}), \quad (9)$$

where $H_a$ and $H_m$ are the uncoupled, single particle Hamiltonians corresponding to the atom and molecule respectively and $V(R_{am})$ is the interaction potential between the atom and the molecule. In our simulations of the system, we approximate the atom as a two-level system with Rabi frequency $\Omega$ and detuning $\Delta$ from the $|g\rangle \to |r\rangle$ transition. The interaction between an atom in state $|g\rangle$ and a molecule in state $|G\rangle$ is extremely small for the values of $R_{am}$ used in our experiments. The only significant interaction term involves the pair state $|r\rangle + |G\rangle$ (shown in Fig. 1(a) of the main text). This is the only non-zero interaction term in our simulations of the dynamics.

As the atom and molecule are predominantly prepared in the motional ground state of their respective tweezers, we average the interaction between $|r\rangle$ and $|G\rangle$ over the ground state wavefunction of relative motion. For the Rydberg blockade measurements, we calculate the root-mean-square width of the ground state wavefunction for relative motion, $\beta$, to be $\beta = 53$ nm [13]. The blue line in Fig. S11 shows the energy shift of the pair state after averaging the interaction potential using a Gaussian function with a standard deviation of 53 nm. For comparison, we also plot the energy shift for the unaveraged potential in purple.

We simulate the dynamics of the atom-molecule blockade measurements in Fig. 4, by solving the Lindblad master equation for the density matrix $\rho$:

$$\dot{\rho} = -i[\mathcal{H}, \rho] + \mathcal{L}[\rho], \quad (10)$$

where $\mathcal{L}$ is the Lindblad superoperator. We neglect spontaneous emission from the state $|r\rangle$ as this occurs on longer timescales compared to the timescale of the dynamics. To reproduce the observed dynamics in the absence of a molecule, we find it important to include dephasing from the Rydberg lasers. We achieve this using Lindblad operators $-\frac{\gamma}{2}\rho_{gr}$ and $-\frac{\gamma}{2}\rho_{rg}$ which incorporate the decay of the coherences between the ground and Rydberg states. Using the measured Rabi frequency $\Omega = 0.5$ MHz, we empirically find a value for $\gamma = 0.1$ MHz that matches the observed dynamics in the absence of a molecule. This corresponds to laser frequency noise resulting from the stabilisation of the lasers to the ULE cavity and in future could be suppressed using established techniques [21].

To incorporate the variation in the atom-molecule position over different iterations of the experiment, we randomly sample $\{\Delta x_{am}, \Delta y_{am}, \Delta z_{am}\}$ from a normal distribution. We construct the distribution using our best estimate of the atom-molecule separation as the mean and our estimate of the shot-to-shot fluctuations in the tweezer separations as the standard deviation of the distribution. For each iteration, we calculate $R_{am}$ and solve Eq. 10. We average the results over 200 iterations for each time or detuning.

To map the results of the simulations onto the results of the experiments, it is important to include experimental imperfections that result in state preparation and measurement errors. Following the approach of Ref. [22], we

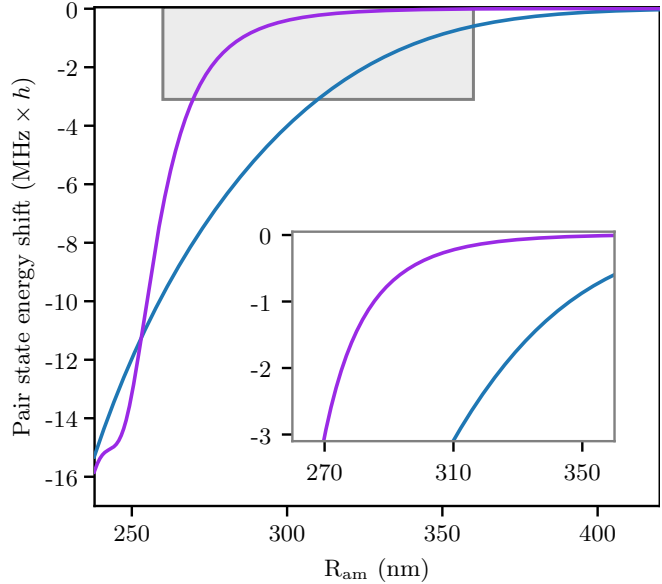

FIG. S11. Energy shift of the pair state $|r\rangle + |G\rangle$ as a function of the separation $R_{am}$ between the Rb Rydberg atom and RbCs molecule. The blue line shows the result obtained by averaging the interaction using a Gaussian potential with standard deviation of 53 nm. The purple line shows the energy shift for no averaging. The inset corresponds to the shaded region in the main figure and highlights the energy shift relevant to the blockade measurements.

include terms $\eta$, $\epsilon$, and $\epsilon'$ with each term measured experimentally. The term $\eta$ accounts for preparation errors and corresponds to the probability the atom was not prepared in $|g\rangle = |5^2S_{1/2}, f=1, m_f=1\rangle$. We find $\eta = 0.13$ which is predominantly limited by Raman scattering from the optical tweezers while the atom is held in the atom tweezer during the molecule formation stages of the experimental sequence. The term $\epsilon$ accounts for loss of atoms in state $|g\rangle$ which were not excited to state $|r\rangle$. We measure $\epsilon = 0.02$, limited by collisions with background gas particles in the vacuum chamber. Finally, $\epsilon'$ accounts for the probability that an atom in state $|r\rangle$ decays before it is lost from the optical tweezer. The measurements in Fig. S7 suggest $\epsilon' = 0.06$. We convert the probability to populate the state $|g\rangle$, $\tilde{P}_g$, (obtained from the simulations) into the experimentally measured atom survival probability $P_g$ using [22]

$$P_g = \eta(1-\varepsilon) + (1-\eta)(1-\varepsilon)\left[\tilde{P}_g + \varepsilon'\tilde{P}_r\right], \qquad (11)$$

where $\tilde{P}_r$ is the population of state $|r\rangle$ that we also obtain from solving Eq. (10).

---

* A. G. and D. K. R. contributed equally to this work.
† c.s.adams@durham.ac.uk
‡ s.l.cornish@durham.ac.uk
[1] S. T. Rittenhouse and H. R. Sadeghpour, Ultracold giant polyatomic Rydberg molecules: Coherent control of molecular orientation, Phys. Rev. Lett. **104**, 243002 (2010).
[2] R. González-Férez, H. R. Sadeghpour, and P. Schmelcher, Rotational hybridization, and control of alignment and orientation in triatomic ultralong-range Rydberg molecules, New J. Phys. **17**, 013021 (2015).
[3] M. Marinescu, H. R. Sadeghpour, and A. Dalgarno, Dispersion coefficients for alkali-metal dimers, Phys. Rev. A **49**, 982 (1994).
[4] E. Fermi and E. Teller, The capture of negative mesotrons in matter, Phys. Rev. **72**, 399 (1947).
[5] A. Omont, On the theory of collisions of atoms in Rydberg states with neutral particles, J. Phys. France **38**, 1343 (1977).
[6] J. Eaton, H. Sarkas, S. Arnold, K. McHugh, and K. Bowen, Negative ion photoelectron spectroscopy of the heteronuclear alkali-metal dimer and trimer anions: NaK⁻, KRb⁻, RbCs⁻, KCs⁻, Na₂K⁻, and K₂Cs⁻, Chem. Phys. Lett. **193**, 141 (1992).
[7] U. Fano and A. Rau, 4 - elastic scattering by a short-range central potential, in *Atomic Collisions and Spectra*, edited by U. Fano and A. Rau (Academic Press, 1986) pp. 61–80.

- [8] R. González-Férez, S. T. Rittenhouse, P. Schmelcher, and H. R. Sadeghpour, A protocol to realize triatomic ultralong range Rydberg molecules in an ultracold KRb gas, J. Phys. B **53**, 074002 (2020).
- [9] S. Spence, R. V. Brooks, D. K. Ruttley, A. Guttridge, and S. L. Cornish, Preparation of $^{87}$Rb and $^{133}$Cs in the motional ground state of a single optical tweezer, New J. Phys. **24**, 103022 (2022).
- [10] J. T. Zhang, Y. Yu, W. B. Cairncross, K. Wang, L. R. B. Picard, J. D. Hood, Y.-W. Lin, J. M. Hutson, and K.-K. Ni, Forming a single molecule by magnetoassociation in an optical tweezer, Phys. Rev. Lett. **124**, 253401 (2020).
- [11] W. B. Cairncross, J. T. Zhang, L. R. Picard, Y. Yu, K. Wang, and K.-K. Ni, Assembly of a rovibrational ground state molecule in an optical tweezer, Phys. Rev. Lett. **126**, 123402 (2021).
- [12] R. Vexiau, D. Borsalino, M. Lepers, A. Orbán, M. Aymar, O. Dulieu, and N. Bouloufa-Maafa, Dynamic dipole polarizabilities of heteronuclear alkali dimers: optical response, trapping and control of ultracold molecules, Int. Rev. Phys. Chem. **36**, 709 (2017).
- [13] D. K. Ruttley, A. Guttridge, S. Spence, R. C. Bird, C. R. L. Sueur, J. M. Hutson, and S. L. Cornish, Formation of ultracold molecules by merging optical tweezers, arXiv:2302.07296 [physics.atom-ph].
- [14] K. Bergmann, H. Theuer, and B. Shore, Coherent population transfer among quantum states of atoms and molecules, Rev. Mod. Phys. **70**, 1003 (1998).
- [15] E. D. Black, An introduction to Pound–Drever–Hall laser frequency stabilization, Am. J. Phys. **69**, 79 (2001).
- [16] P. Barakhshan, A. Marrs, A. Bhosale, B. Arora, R. Eigenmann, and M. S. Safronova, Portal for high-precision atomic data and computation (version 2.0), [Online] (2022).
- [17] N. Šibalić, J. D. Pritchard, C. S. Adams, and K. J. Weatherill, ARC: An open-source library for calculating properties of alkali Rydberg atoms, Comput. Phys. Commun. **220**, 319 (2017).
- [18] S. Weber, C. Tresp, H. Menke, A. Urvoy, O. Firstenberg, H. P. Büchler, and S. Hofferberth, Calculation of Rydberg interaction potentials, J. Phys. B **50**, 133001 (2017).
- [19] J. A. Blackmore, R. Sawant, P. D. Gregory, S. L. Bromley, J. Aldegunde, J. M. Hutson, and S. L. Cornish, Controlling the ac Stark effect of RbCs with dc electric and magnetic fields, Phys. Rev. A **102**, 053316 (2020).
- [20] R. V. Brooks, S. Spence, A. Guttridge, A. Alampounti, A. Rakonjac, L. A. McArd, J. M. Hutson, and S. L. Cornish, Preparation of one $^{87}$Rb and one $^{133}$Cs atom in a single optical tweezer, New J. Phys. **23**, 065002 (2021).
- [21] H. Levine, A. Keesling, A. Omran, H. Bernien, S. Schwartz, A. S. Zibrov, M. Endres, M. Greiner, V. Vuletić, and M. D. Lukin, High-fidelity control and entanglement of Rydberg-atom qubits, Phys. Rev. Lett. **121**, 123603 (2018).
- [22] S. de Léséleuc, D. Barredo, V. Lienhard, A. Browaeys, and T. Lahaye, Analysis of imperfections in the coherent optical excitation of single atoms to Rydberg states, Phys. Rev. A **97**, 053803 (2018).


\fi

\end{document}